\documentclass[a4paper,11pt]{article}

\usepackage{xcolor}
\usepackage{xspace}
\usepackage{subfigure}
\usepackage{lineno}
\usepackage{multirow}
\usepackage{footnote}
\usepackage{upgreek}
\usepackage[flushleft]{threeparttable}

\addtocontents{toc}{\setcounter{tocdepth}{2}}

\pdfoutput=1 

\usepackage{jcappub} 

\graphicspath{{Images/}}

\usepackage[T1]{fontenc} 

\title{Constraining the sources of ultra-high-energy cosmic rays across and above the ankle with the spectrum and composition data measured at the Pierre Auger Observatory } 
\author[13]{A.~Abdul Halim,}
\author[71]{P.~Abreu,}
\author[53,51]{M.~Aglietta,}
\author[1]{I.~Allekotte,}
\author[69]{K.~Almeida Cheminant,}
\author[8,12]{A.~Almela,}
\author[78]{J.~Alvarez-Mu\~niz,}
\author[78]{J.~Ammerman Yebra,}
\author[53,51]{G.A.~Anastasi,}
\author[85]{L.~Anchordoqui,}
% ...

\affiliation[1]{Centro At\'omico Bariloche and Instituto Balseiro (CNEA-UNCuyo-CONICET), San Carlos de Bariloche, Argentina}
% ...
\affiliation[8]{Instituto de Tecnolog\'\i{}as en Detecci\'on y Astropart\'\i{}culas (CNEA, CONICET, UNSAM), Buenos Aires, Argentina}
% ...
\affiliation[12]{Universidad Tecnol\'ogica Nacional -- Facultad Regional Buenos Aires, Buenos Aires, Argentina}
\affiliation[13]{University of Adelaide, Adelaide, S.A., Australia}
% ...
\affiliation[51]{INFN, Sezione di Torino, Torino, Italy}
% ...
\affiliation[53]{Osservatorio Astrofisico di Torino (INAF), Torino, Italy}
% ...
\affiliation[69]{Institute of Nuclear Physics PAN, Krakow, Poland}
% ...
\affiliation[71]{Laborat\'orio de Instrumenta\c{c}\~ao e F\'\i{}sica Experimental de Part\'\i{}culas -- LIP and Instituto Superior T\'ecnico -- IST, Universidade de Lisboa -- UL, Lisboa, Portugal}
% ...
\affiliation[78]{Instituto Galego de F\'\i{}sica de Altas Enerx\'\i{}as (IGFAE), Universidade de Santiago de Compostela, Santiago de Compostela, Spain}
% ...
\affiliation[85]{Department of Physics and Astronomy, Lehman College, City University of New York, Bronx, NY, USA}
% ...
\affiliation[c]{School of Physics and Astronomy, University of Leeds, Leeds, United Kingdom}
% ...

\usepackage[normalem]{ulem}

\DeclareMathOperator{\sech}{sech}
\newcommand{\xmax}{\ensuremath{X_\text{max}}\xspace}

\emailAdd{spokespersons@auger.org}

\abstract{In this work we present the interpretation of the energy spectrum and mass composition data as measured by the Pierre Auger Collaboration above $6{\times}10^{17}$\,eV. We use an astrophysical model with two extragalactic source populations to model the hardening of the cosmic-ray flux at around $5{\times}10^{18}$\,eV (the so-called ``ankle'' feature) as a transition between these two components. We find our data to be well reproduced if sources above the ankle emit a mixed composition with a hard spectrum and a low rigidity cutoff. The component below the ankle is required to have a very soft spectrum and a mix of protons and intermediate-mass nuclei. The origin of this intermediate-mass component is not well constrained and it could originate from either Galactic or extragalactic sources.
To the aim of evaluating our capability to constrain astrophysical models, we discuss the impact on the fit results of the main experimental systematic uncertainties and of the assumptions about quantities affecting the air shower development as well as the propagation and redshift distribution of injected ultra-high-energy cosmic rays (UHECRs).
}

\begin{document}
\maketitle
\flushbottom

\section{Introduction}
The quest for the sources of ultra-high-energy cosmic rays (UHECRs) is central in modern
astroparticle physics. While the bulk of Galactic cosmic rays (GCRs) is thought to be accelerated by diffusive shocks in supernova remnants~\cite{Gabici:2019jvz}, 
the origin and acceleration mechanism governing the most energetic particles is still under debate.
High data quality has been reached in the past decade from the experimental point of view, setting the basis for the development of theoretical models aiming at describing the observations.

The Pierre Auger Observatory~\cite{auger} has allowed us to study the features of the all-particle energy spectrum  with unprecedented precision~\cite{paospe20,PierreAuger:2021hun,paoICRC21}. 
Far from being described by a simple power law, in the highest-energy region the all-particle cosmic-ray spectrum shows several features. A sharp feature, known as the ankle, is observed at ${\sim}10^{18.7}$\,eV, corresponding to a hardening of the spectrum. 
A new feature, dubbed the instep, at ${\sim}10^{19.1}$\,eV, could reflect the interplay of light-to-intermediate nuclei~\cite{paospe20b}.
Finally, a suppression of the total flux above ${\sim}10^{19.7}$\,eV may be attributed to energy losses during the propagation of UHECRs~\cite{GZKa,GZKb}, to the limited maximum energy the sources can provide to particle acceleration, or possibly to a combination of both effects.
The spectrum measured by Telescope Array~\cite{TAspe} (TA) agrees in both shape and normalisation with the one measured by Auger within the systematic uncertainties (14\% and 21\% for Auger and TA respectively), with a noticeable difference only showing up at energies $\gtrsim10^{19.5}$\,eV~\cite{Auger-TAspe}.

The composition of the primary beam~\cite{bellido2017,PAOcomp}, as estimated by the distributions of depth of maximum development of the showers \xmax, appears to be given by
a mix of protons and medium-mass (e.g.\ nitrogen) nuclei at energies above the second knee,
gradually getting lighter with increasing energy up to $10^{18.3}$\,eV. From this energy up to the
ankle, the primaries are mainly mixed.
A study of the event-by-event correlation between two different observables, the depth of shower maximum and the ground-level signal, measured by the fluorescence detector (FD) and the surface detector (SD) respectively~\cite{PierreAuger:2016qzj,xmax_icrc19}, which is rather insensitive to the experimental systematic uncertainties and to the uncertainties in the modelling of air showers affecting composition estimates based on the \xmax distributions alone, confirms that the composition is mixed in the ankle region, excluding any pure elements or $(\text{p}+\text{He})$-only mixtures with ${>}6\sigma$~significance.
Above the ankle, the mass composition appears increasingly heavier and less mixed, suggesting that the total UHECR spectrum
is the superposition of alternating groups of elements with progressively
heavier mass each with a steep cutoff, though with increasingly sparse statistics towards the suppression region.  Such a sequence is analogous to the Peters cycle~\cite{peters} which has already been associated to the knee of the cosmic-ray spectrum. 
The composition  has also been measured by the TA Collaboration~\cite{TAcomp}; the comparison  between the \xmax moments of Auger with that of TA is not immediate because TA includes the detector effects in their result. By converting the Auger \xmax values into the values folded with the TA detector effects, both experiments appear to be compatible up to $10^{19}$\,eV~\cite{Yushkov:2019hoh}.

The energy region where GCRs give room to extragalactic cosmic rays (EGCRs), somewhere between the second knee and the ankle, is particularly important to draw a complete description of the origin of UHECRs. In the region immediately below and around the ankle, a dominance of Galactic protons and medium-mass nuclei can be excluded based on the measured low level of anisotropy in the distribution of arrival directions~\cite{PAOlsa,Gia2012}. On the other hand, a dominance of heavier nuclei, which would comply with the allowed limits, is disfavoured by the interpretation of \xmax measurements as mentioned above. These findings exclude the models, very popular in the past, which proposed that the GCR--EGCR transition occurrs at the ankle \cite{Aloisio:2006wv}. As a consequence, the large fraction of protons found in composition measurements around the ankle  must be of extragalactic origin, and the mixed composition visible just above the second knee should be provided by an additional component, whether Galactic or extragalactic~\cite{Hillas2005,Allard2008,Gaisser2013,Deligny2014,Giacinti2015,Aloisio:2013hya,Aloisio2019}. 
Recently it has been suggested~\cite{Globus:2015xga,UFA2015} that a fair amount of protons at and below the ankle might result from interactions of cosmic ray nuclei in the source environment (see also~\cite{Baerwald:2014zga,GAP,Biehl:2017zlw,TDE_ref,Guepin:2017abw,Fang:2017zjf,Kach2017,Supanitsky:2018jje,Zhang2018,Zhang:2018agl,Boncioli2019,Muzio:2019leu,Heinze:2020zqb,Rodrigues2021,Muzio:2021zud,Condorelli:2022vfa}), possibly with the addition of some contribution from GCRs. Comparison between the expected and measured~\cite{Auger-neu2019} neutrino limits have been extensively used to further check the viability of the different scenarios where the interactions in sources are taken into account~\cite{Biehl:2017zlw,TDE_ref,Boncioli2019,Muzio:2019leu,Heinze:2020zqb,Rodrigues2021,Muzio:2021zud,Muzio:2022bak,Condorelli:2022vfa}, as well as the ones where only cosmogenic neutrinos are considered~\cite{Heinze:2015hhp,Heinze2019,AlvesBatista:2018zui}. Above 8\,EeV, the extragalactic origin of UHECRs is clearly suggested by the observation of a dipolar anisotropy with amplitude of~7.3\% and phase pointing $115^\circ$ away from the Galactic centre, and by the evolution of its amplitude with energy, which is consistent with a shrinking horizon for the sources of the highest-energy particles~\cite{PAOdip,PAOlsa20,PierreAuger:2021dqp}.

\newcommand{\CF}{our previous work~\cite{combinedfit}}
In a previous publication~\cite{combinedfit}, in which we focused only on the energy region above the ankle, we exploited a combined fit of a simple astrophysical model of UHECR sources to both the energy spectrum and mass composition data measured by the Pierre Auger Observatory, to investigate the constraining power of the collected data on the source properties. In that paper, the possibility to extend the fit to lower energies without spoiling the above-ankle results was also considered by subtracting from data the extrapolation of the above-ankle best-fit results at lower energies. Even if an actual fit in the whole energy region had not been performed yet, we found first indications of the need of an additional light-to-intermediate component with a steeper generation spectrum with respect to the one of the above-ankle component. More recently, it was shown in Ref.~\cite{protonspectrum}, starting from the same baseline astrophysical model, that the inferred fraction of protons below the ankle can be described as an extragalactic component with a much softer energy spectrum with respect to the one of the high-energy population that describes the measured mixed composition above the ankle. This scenario calls for an additional component to fully describe the total flux of UHECRs.
Here, we assume from the beginning a two-population model, and perform a complete simultaneous fit of the different components. The novelties of this analysis lay in the assumption from the beginning of a two-population model, and in performing a complete simultaneous fit of the different components in the full energy range from below the ankle up to the highest energies. 
The study of the systematic uncertainties, both from measurements and models, is extended to the whole energy range. The careful evaluation of such uncertainties is performed thanks to a data-driven approach, which exploits the complete
knowledge on data available within the Pierre Auger Collaboration, whose statistics have been extended by six full years.

\section{The combined fit} \label{combfit}

\subsection{Astrophysical and propagation models}
\label{astromodel}
\subsubsection{Extragalactic and Galactic sources}

In this study we aim at constraining the physical parameters related to the energy spectrum and the mass composition of particles escaping the environments of extragalactic sources.  In \CF, a single population of identical extragalactic sources was fitted to the data above the ankle ($E>10^{18.7}$\,eV). In this work we adopt a similar baseline astrophysical model
but, since we also want to interpret the ankle region, we assume the presence of one (or more) additional contribution(s) at low energies, so that the ankle is produced by the superposition of different components. 

In our model, each extragalactic component is here
assumed to originate from a population of identical sources uniformly distributed in the
comoving volume. A correction, based on Ref.~\cite{biteau} as described in Appendix~\ref{app:overdensity}, takes into
account the higher densities for distances shorter than ${\sim}30$\,Mpc with a minimum source distance of 1\,Mpc. Such a correction allows to take into account the fact that the Milky Way belongs to a group of galaxies, itself embedded on the Local Sheet~\cite{mccall2014}. 
The effects of using different assumptions for the local overdensity are discussed in Appendix~\ref{sec:overdensity}, and those of assuming different evolutions of the source emissivity with cosmological time are discussed in Section~\ref{SourceEvolution}.

The starting basic assumption is that UHECRs are accelerated by electromagnetic
processes up to a maximum energy proportional to their electric charge. For each extragalactic population of sources the
spectrum of particles escaping from the source environment (after acceleration and in-source propagation) can be modelled as
the superposition of the contributions of $n\le 5$ representative stable nuclear species $A$, chosen among $^{1}$H, $^{4}$He, $^{14}$N, $^{28}$Si, $^{56}$Fe,\footnote{We have verified that considering also other intermediate nuclear species (e.g.\ $^{20}$Ne and $^{40}$Ca) escaping from the sources does not significantly change the fit results.}
each following a power-law spectrum with a broken exponential rigidity cutoff.   The generation rate $\widetilde{Q}_A(E)$, defined as the number of nuclei with mass $A$ ejected per unit of energy, volume and time, is given by
\begin{equation}
    \widetilde{Q}_A(E) 
    = \widetilde{Q}_{0A} \, \left( \frac{E}{E_0} \right)^{-\gamma}
    \, \begin{cases}
        1, & E \leq Z_A \, R_{\mathrm{cut}}; \\
        \exp\left(1-\frac{E}{Z_A \, R_\text{cut}}\right), & E > Z_A \, R_\text{cut},
    \end{cases}\label{eq1bis}
\end{equation}
where $Z_A$ is the atomic number of each species $A$, and $\widetilde{Q}_{0A}$ is the generation rate at a reference energy $E_0$, which is set to a value arbitrarily lower than the energy cutoff of protons; the total generation rate is then $\widetilde{Q}(E) = \sum_A \widetilde{Q}_A(E)$ and is expressed in units of erg$^{-1}$\,Mpc$^{-3}$\,yr$^{-1}$.  
These are of course simplifications, aiming at keeping the number of free parameters manageable during the fit procedure. For the same reason, we neglect the differences among sources within the same population (see~\cite{arXiv:2207.10691} for a discussion of the effect of the population variance on the combined fit), so all the estimated parameters are the effective ones which characterise the total escape spectrum from all sources in the population.
For each extragalactic population, there are then $2+n$ free parameters: the spectral index $\gamma$, the rigidity cutoff $R_\text{cut}$, and $n$~partial normalisations $\widetilde{Q}_{0A}$. 
To compare the estimated compositions corresponding to different $\gamma$ values and to immediately get a physically more meaningful information about the nuclear species at the sources from the fit results, it is thus useful to express the mass fractions in terms of 
fractions~$I_A$ of the total source emissivity~$\mathcal{L}_0$ of each population, defined as the total energy ejected per unit of comoving volume per unit of time at redshift~$z=0$,
\begin{equation}
    I_{A} = \frac{\int_{E_\text{min}}^\infty E \, \widetilde{Q}_A(E)\,\mathrm{d}E}{\mathcal{L}_0},
    \quad\text{where}\quad
    \mathcal{L}_0 = \sum_A \textstyle\int_{E_\text{min}}^\infty E \, \widetilde{Q}_A(E)\,\mathrm{d}E,
    \label{intenergydensity}
\end{equation}
starting from the fit energy threshold $E_\text{min}=10^{17.8}$\,eV. 
The emissivity~$\mathcal{L}_0$ is thus expressed in units of erg\,Mpc$^{-3}$\,yr$^{-1}$.

In Section~\ref{scenarioA}, we also consider the possible presence of a Galactic component at Earth, which is modelled as a power law  with~$\gamma_\text{Gal}=3.2$ modified by a simple exponential cutoff.\footnote{This value for the slope of the spectrum of the Galactic component was chosen based on the slope of the high-energy tail of the spectrum as estimated for example from the measurements of KASCADE-Grande electron-poor (heavy) events at $E \ge 10^{16.7}$\,eV~\cite{KASCADE}. We checked that different choices would not affect the result: given the narrowness of the energy range in which this component is non-negligible, the spectral index and the cutoff energy are nearly degenerate with each other.}  
As for its mass composition, we considered the cases of pure~Fe, a mix of Fe+Si, pure~Si, a mix of Si+N, pure~N and a mix of N+He.
The normalisation~$J_0^\text{Gal}$ at $E_0^\text{Gal}=10^{16.85}$\,eV,
the rigidity cutoff~$R_\text{cut}^\text{Gal}$,
and (in the cases with two elements) the fraction of the heavier element
are free parameters of the fit.

\subsubsection{Propagation in intergalactic space}\label{propa}
\newcommand{\TALYS}{\textsc{Talys}}
The energy spectrum and mass composition of the particles escaping from extragalactic source environments are modified during the propagation in the intergalactic medium by the adiabatic energy losses and the interactions with background photons. Assuming standard cosmology, the adiabatic energy losses due to the expansion of the Universe are given by the relationship between time and redshift $(\mathrm{d}t/\mathrm{d}z)^{-1}=-H_{0}(1+z)\sqrt{\smash[b]{\Omega_\text{m}(1+z)^3+\Omega_\Lambda}}$, where we use the values $H_0=70$\,km\,s$^{-1}$\,Mpc$^{-1}$ for the Hubble constant at present time, $\Omega_\text{m}=0.3$ for the matter density, and $\Omega_\Lambda=0.7$ for the dark energy density.\footnote{The effects of uncertainties in $H_0$, $\Omega_\text{m}$ and $\Omega_\Lambda$ on
   predicted propagated UHECR fluxes are negligible~\cite{AlvesBatista:2019rhs}.} The effect of the interactions with background photons is described by $\eta_{A'A}(E',E,z)$, the fraction of particles with energy $E'$ and mass number $A'$ at Earth produced by a nucleus escaping the source environment at a redshift~$z$ with energy~$E$ and mass number~$A$. The relevant interaction processes taken into account are the electron--positron pair photoproduction, the pion photoproduction, and the photodisintegration of nuclei. The photon fields playing a role in the propagation of UHECRs are the ones from the cosmic microwave background (CMB) and the ones from the infrared/visible/ultraviolet extragalactic background light (EBL).

The observed energy spectrum $J_\text{obs}(E')$ is thus obtained by integrating the contributions of all the sources weighted by the redshift and modified by the effects of interactions with radiation photons,
\begin{equation}
J_\text{obs}(E')=\frac{c}{4\pi}\sum_A\sum_{A'} \iint\mathrm{d}E\,\mathrm{d}z\left|\frac{\mathrm{d}t}{\mathrm{d}z}\right|\, S(z)\,\widetilde{Q}_A(E)\,\frac{\mathrm{d}\eta_{A'A}(E',E,z)}{\mathrm{d}E'}
\end{equation}
where $c$ is the speed of light and $S(z)$ is the evolution of the luminosity density of UHECRs; in the simplest case of a flat evolution $S(z)=1$. 

We take into account the propagation effects by using SimProp~\cite{simprop} simulations.  A direct comparison between CRPropa \cite{AlvesBatista:2016vpy} and SimProp has been reported in Ref.~\cite{AlvesBatista:2015jem}, showing consistent results for the same model assumptions.  The uncertain quantities are treated with phenomenological models. More specifically, the photodisintegration cross sections $\sigma_\text{pd}$ are much less known than the pair photoproduction and pion photoproduction ones, as shown also in~\cite{Boncioli:2016lkt}. There are also large uncertainties in the spectrum and evolution of the EBL, unlike for the CMB.
In this work, we model photodisintegrations via the cross sections computed by \TALYS~\cite{talys, talys2,talys3}
with the settings described in Ref.~\cite{AlvesBatista:2015jem},
or the ones from the Puget, Stecker and Bredekamp (PSB)~\cite{psb,psb2} model. The EBL is described using the Gilmore~\cite{gilmore} or Dom\'inguez~\cite{dominguez} model.
The differences induced by the employment of different models, studied in Ref.~\cite{AlvesBatista:2015jem}, are used to evaluate the corresponding systematic uncertainties in Section~\ref{sysunc}.

We neglect the effects of intergalactic magnetic fields on the UHECR energy spectrum and mass composition.  According to the propagation theorem~\cite{Aloisio:2004jda}, such effects are negligible in the limit that the distances between sources are much less than all other relevant length scales, most notably the Larmor radius~$r_\text{L} \approx 1.08 \, (E/\text{EeV}) \, Z^{-1} \, (B_\perp/\text{nG})^{-1}\,\text{Mpc}$, where~$B_\perp$ is the magnetic field strength in the direction perpendicular to the propagation.  In our model, the lowest relevant magnetic rigidity~$E/Z$ is that of nitrogen ($Z=7$) at $10^{17.8}$\,eV and typical distances between sources are ${\lesssim}10$\,Mpc, hence the theorem is applicable for $B_\perp \ll 10^{-11}$\,G. For stronger IGMFs a modification of the spectrum at low energies could appear because of the magnetic horizon effect, as discussed in Refs.~\cite{Lemoine2005,Aloisio:2004jda,Mollerach2013}. However, in the present work, in order to follow a data-driven approach with simple model assumptions, we assume $B_\perp \ll 10^{-11}$\,G and defer the treatment of the possible magnetic effects to future studies.

\subsubsection{Development of air showers}
\label{gumbel}
\newcommand{\EPOS}{\textsc{Epos}-LHC}           %
\newcommand{\QGSJET}{\textsc{QGSJet}\,II-04}    %
\newcommand{\SIBYLL}{\textsc{Sibyll}\,2.3d}     %

Since a direct measurement of the mass composition is not possible on an event-by-event basis, we use the distribution of \xmax\ as an estimator of the mass distribution in each energy bin. Such a conversion depends on the choice of hadronic interaction model (HIM), which is thus another source of uncertainty.  In this work, we use the HIMs \EPOS~\cite{eposlhc}, \QGSJET~\cite{Ostapchenko:2010vb} and \SIBYLL~\cite{Riehn:2019jet}.

We first modelled the true \xmax distributions as generalised Gumbel distribution functions~$g(\xmax|E,A)$, with parameters depending on the HIM and on the mass and energy of the primary cosmic ray, as described in Ref.~\cite{gumbel}; a discussion on the effect of using different parameterisations for the \xmax distributions can be found in Ref.~\cite{desouza}. The Gumbel parameter values were fitted to CONEX~\cite{conex} simulations. We computed the total predicted \xmax distribution in each energy bin as $g_\text{tot}(\xmax|E)$, considering the contribution of all the 
simulated events in that bin. To take detector effects into account, these distributions were then multiplied by a function describing the acceptance and convolved by the resolution. The model prediction $G^\text{mod}$ was thus obtained. Further details about the Gumbel parameterisation can be found in Appendix~\ref{gumbel_par}.

\subsection{The data sets}
We use the recently published measurement of the UHECR energy spectrum obtained from events detected using the SD array of the Pierre Auger Observatory up to August 2018, including both the original stations with 1500\,m spacing (SD-1500) and the low-energy extension with 750\,m spacing (SD-750), fully corrected for detector acceptance and resolution effects~\cite{PierreAuger:2021hun}.  
The energy range $10^{17.8}\,\text{eV}\le E<10^{20.2}$\,eV is subdivided in 24 bins of $\log_{10}(E/\text{eV})=0.1$.
Each bin up to $10^{20.0}$\,eV contains more than 20 events, and the second-to-last and last bins contain 9 and 6 events, respectively.

The \xmax\ distributions 
measured using the FD telescopes up to December 2017~\cite{xmax_icrc19} are used
as an estimator of the mass distribution in each energy bin.
They are divided in eighteen bins of $\log_{10}(E/\text{eV})=0.1$ from~$10^{17.8}$\,eV to $10^{19.6}$\,eV (the same binning chosen for the energy spectrum) plus one additional larger bin containing events with energies above $10^{19.6}$\,eV.
In this last bin,
the median energy is $10^{19.70}$\,eV and that of the most energetic event is $10^{20.02}$\,eV, hence we effectively only have composition information up to the suppression energy.
The total number of collected events is 31\,085; it ranges from 5476 in the first energy bin to 35 in the last.
In each of these energy bins, the \xmax distribution is binned in intervals of 20\,g/cm$^2$. 
There is a total of 329 non-empty bins in the whole dataset, which extends by about six years the one used in the previous combined fit analysis~\cite{combinedfit}.

\subsection{Fit procedure}
\label{fit_procedure}
In the fit we minimise the deviance $D=-2\ln(L/L_\text{sat})$, a generalised $\chi^2$, where $L$ is the likelihood of our model and $L_\text{sat}$ that of a model which perfectly describes the data; thus minimising $D$ is equivalent to maximising $L$ (see e.g.\ Ref.~\cite{baker} for further details).  The deviance consists of two terms, $D_J$ and $D_{X_\text{max}}$, given by
\begin{align}
D_J &= \sum_i \frac{(J^\text{obs}_i-J^\text{mod}_i)^2}{\sigma_i^2};
\label{DJ} \\
D_{X_\text{max}} &= 2 \sum_{ij} k_{i,j}^\text{obs} \, \ln\left(\frac{k_{i,j}^\text{obs}}{n^\text{obs}_i\, G^\text{mod}_{i,j}}\right).
\label{DXmax}	
\end{align}

$D_J$ is related to the energy spectrum, whose likelihood is treated as the product of Gaussian distributions, where in each $i$-th energy bin $J_i^\text{obs}$ is the observed flux, $\sigma_i$ is its statistical uncertainty, and $J_i^\text{mod}$ is the model prediction as described in Section~\ref{propa}. 
$D_{X_\text{max}}$ is a product of multinomial distributions describing the likelihood for the \xmax distributions\footnote{It is equivalent to considering a Poissonian deviance when it is summed over all bins and the model is normalised to the data.}, where $k_{i,j}^\text{obs}$ is the number of observed events in the $i$-th energy bin and in the $j$-th \xmax bin, $n_i^\text{obs} = \sum_j k_{i,j}^\text{obs}$ is the total number of events in the $i$-th energy bin, and $G_{i,j}^\text{mod}$ are the model predictions following the generalised Gumbel functions described in Section~\ref{gumbel}, normalised so that $\sum_j G_{i,j}^\text{mod}=1$ for each $i$. 

The best-fit parameter values for each scenario are then those with which the total deviance $D = D_J + D_{X_\text{max}}$ attains its minimum value $D_\text{min}$, which we locate using the Minuit package~\cite{James:1975dr}; the statistical uncertainties on the spectral parameters correspond to the half extent of the 1D profile in the parameter space where $D \le D_\text{min} + 1$, as computed using the \texttt{MINOS} routine of Minuit; the uncertainties on the emissivity and on the mass fractions are computed with Monte Carlo simulations, as explained in the next section.

\section{Results in the reference scenarios}
\label{sec:refresults}

The fit results depend on the choice of the distribution of sources, the propagation and the HIM. In this section, all the results are obtained by using \TALYS\ for the photodisintegration cross sections, the Gilmore model for the EBL spectrum and evolution, and the \EPOS\ HIM.  Other combinations of models will be discussed in Section~\ref{sysmodel}. In order to focus on the simplest case, in this section we assume a flat cosmological evolution for the extragalactic sources, whereas the effect of other choices of source evolution are investigated in Section~\ref{SourceEvolution}.

We reported the statistical uncertainties on all the estimated parameters, which are evaluated as follows: we fitted $n_\text{mock}=1000$ simulated data sets, generated from the best-fit solution with statistics equal to the real data set, and we calculated the one standard deviation uncertainties from the 16th and 84th percentiles of the corresponding distribution of each parameter. Since the uncertainties on the spectral parameters $\gamma$ and $R_\text{cut}$ are directly estimated by the minimiser and then can be easily obtained from Minuit, we verified that the two approaches provide compatible results. For all the other results illustrated in this work, we chose to only report the uncertainties on $\gamma$ and $R_\text{cut}$ from Minuit to make the results display clearer. Note also that in the cases where the rigidity cutoff is unconstrained we report only the lower bound above which the fit is not sensitive to the exact parameter value.  

\newcommand{\LEa}{LE}
\newcommand{\HEa}{HE}
\newcommand{\LEb}{LE}
\newcommand{\HEb}{HE}

\begin{table}
    \small
    \renewcommand\arraystretch{1.4}
    \centering
     \begin{threeparttable}
    \begin{tabular}{|l|cc|cc|}
        \multicolumn{1}{c}{} & \multicolumn{2}{c}{\textsc{Scenario 1}} & \multicolumn{2}{c}{\textsc{Scenario 2}}  \\
        \hline
        Galactic contribution (at Earth) & \multicolumn{2}{c |}{pure N} & \multicolumn{2}{c|}{---} \\
        \hline
        $J_0^\text{Gal}/(\text{eV}^{-1}\,\text{km}^{-2}\,\text{sr}^{-1}\,\text{yr}^{-1})$ & \multicolumn{2}{c|}{$(1.06 \pm 0.04){\times}10^{-13}$} & \multicolumn{2}{c|}{---} \\
        $\log_{10}(R_\text{cut}^\text{Gal}/\text{V})$ & \multicolumn{2}{c|}{$17.48 \pm 0.02$} & \multicolumn{2}{c|}{---} \\
        \hline
        \hline
        EG components (at the escape)  & \LEa & \HEa & \LEb & \HEb \\
        \hline
        $\mathcal{L}_0/(10^{44}\,\text{erg}\,\text{Mpc}^{-3}\,\text{yr}^{-1})$~\tnote{*} & $6.54 \pm 0.36$ & $5.00 \pm 0.35$ & $11.35\pm0.15$ & $5.07 \pm 0.06$ \\
        $\gamma$ & $3.34 \pm 0.07$ & $-1.47 \pm 0.13$ & $3.52 \pm 0.03$ & $-1.99 \pm 0.11$ \\
        $\log_{10}(R_\text{cut}/\text{V})$ & ${>}19.3$ & $18.19 \pm 0.02$ & ${>}19.4$ & $18.15 \pm 0.01$ \\
        $I_\text{H}$  (\%) & 100 (fixed) & $\phantom{0}0.0 \pm 0.0$ & $48.7 \pm 0.3$ & $\phantom{0}0.0\pm 0.0$ \\
        $I_\text{He}$ (\%) & --- & $24.5\pm 3.0$ &  $\phantom{0}7.3 \pm 0.4$& $23.6\pm 1.6$ \\
        $I_\text{N}$  (\%) & --- & $68.1\pm 5.0$ & $44.0 \pm 0.4$ & $72.1\pm 3.3$ \\
        $I_\text{Si}$ (\%) & --- &  $\phantom{0}4.9\pm 3.9$ &  $\phantom{0}0.0 \pm 0.0$ &  $\phantom{0}1.3 \pm 1.3$ \\
        $I_\text{Fe}$ (\%) & --- &  $\phantom{0}2.5 \pm 0.2$ &  $\phantom{0}0.0 \pm 0.0$ &  $\phantom{0}3.1 \pm 1.3$ \\
        \hline
        $D_J$ ($N_J$) & \multicolumn{2}{c|}{48.6 (24)} & \multicolumn{2}{c|}{56.6 (24)} \\
        $D_{X_\text{max}}$ ($N_{X_\text{max}}$) & \multicolumn{2}{c|}{537.4 (329)} & \multicolumn{2}{c|}{516.5 (329)} \\
        $D$ ($N$) & \multicolumn{2}{c|}{586.0 (353)} & \multicolumn{2}{c|}{573.1 (353)} \\
        \hline
    \end{tabular}
    \begin{tablenotes}
      \item[*] from $E_\text{min}=10^{17.8}$\,eV.
      \end{tablenotes}
       \end{threeparttable}
    \caption{Best-fit parameters obtained in the two reference scenarios. \textsc{Scenario 1} (Section~\ref{scenarioA}): a Galactic contribution of pure nitrogen, a low-energy extragalactic component of pure protons (\LEa), and a high-energy extragalactic component with a mixed mass composition (\HEa). \textsc{Scenario 2} (Section~\ref{scenarioB}): two mixed extragalactic components (\LEb\ and \HEb) overlapping in the ankle energy region.}
    \label{tab:mainresults}
\end{table}

\newcommand{\lastXmax}{The shaded grey area indicates the energy region where energy-by-energy estimates of the mass composition are not available (i.e.\ above the median of the highest energy bin used for \xmax data) and mass predictions are mainly based on the shape of the all-particle spectrum.}

\subsection{Scenario 1: extragalactic and Galactic populations}
\label{scenarioA}

In the first of the two scenarios we are considering, we assume an extragalactic population with a mixed mass composition dominating at high energies (``\HEa''), plus an additional extragalactic component dominating at low energies (``\LEa'') which in this scenario is of pure protons, similar to~\cite{protonspectrum}. The two extragalactic components are not necessarily produced in two different types of astrophysical environments. A \LEa\ population could e.g.\ arise from the photodisintegration of \HEa\ cosmic rays by the
photon fields in the environment of their sources, and the subsequent escape and beta decay
of the secondary neutrons thereby produced~\cite{UFA2015}. In this case, the LE proton component would not be independent of the HE one, because the processes originating the LE component impose relations between the features of the two components. 
The heavier nuclei at energies below the ankle are instead assumed to originate from a Galactic population.

We found that a Galactic component at Earth of pure nitrogen, extending up to a relatively high energy $Z\,R_\text{cut}^\text{Gal} \approx 2{\times}10^{18}$\,eV, provides the best fit to the data. In fact, heavier compositions with no nitrogen result in deviances $D \gtrsim 1000$, and in the (Si+N) and (N+He) cases the best fits are obtained with $f_\text{Si} = 0$ and $f_\text{He} = 0$, respectively.  Hence, in the following figures and tables we only show the results obtained in the case of pure nitrogen.~\footnote{A discussion about the possible explanations for such a Galactic contribution can be found in Sec.~\ref{disc_scenarios}.}

\begin{figure}
	\centering
	\def\w{0.49}
	\subfigure{\includegraphics[width=\w\linewidth]{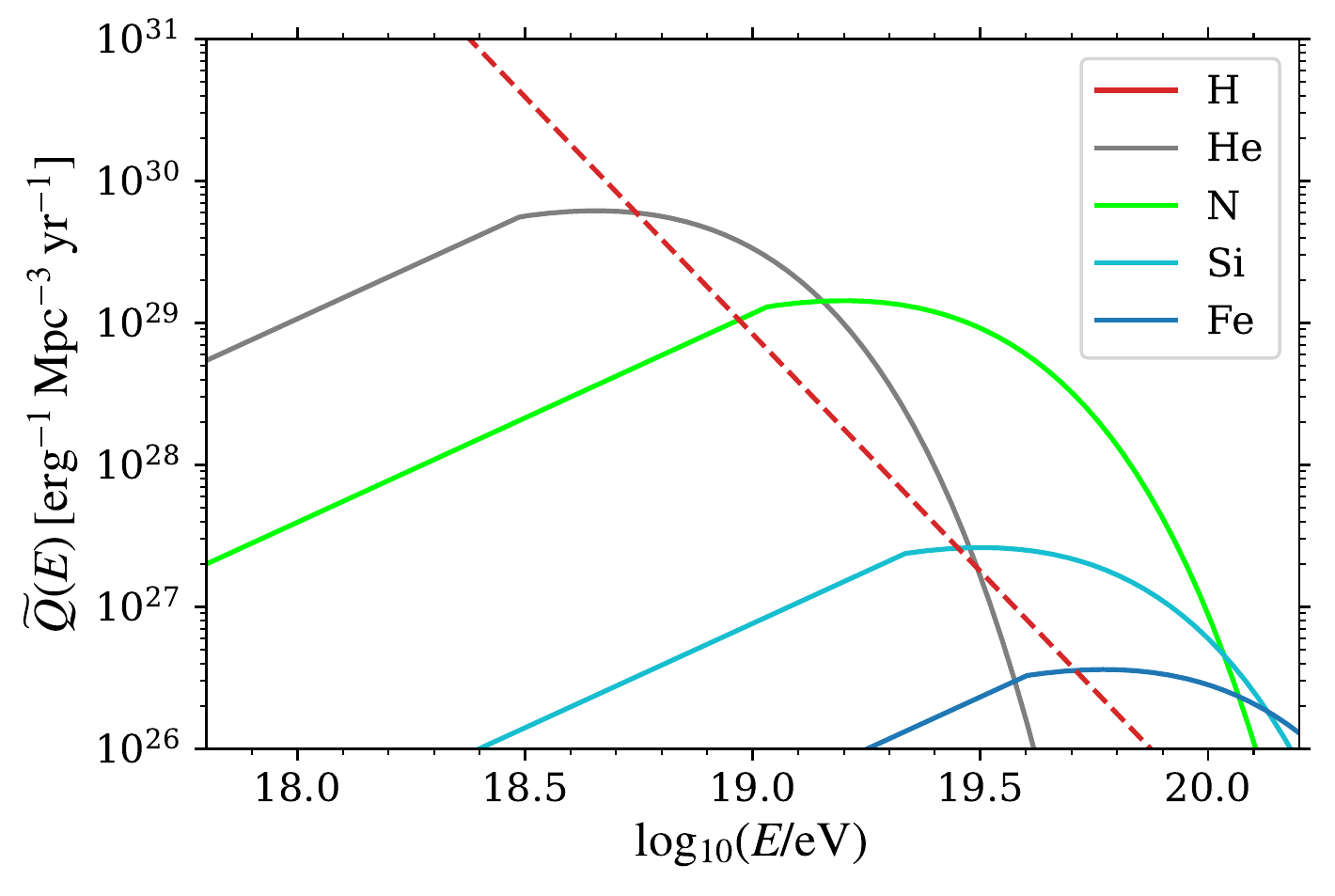}\label{gal1}}\hfill
	\subfigure{\includegraphics[width=\w\linewidth]{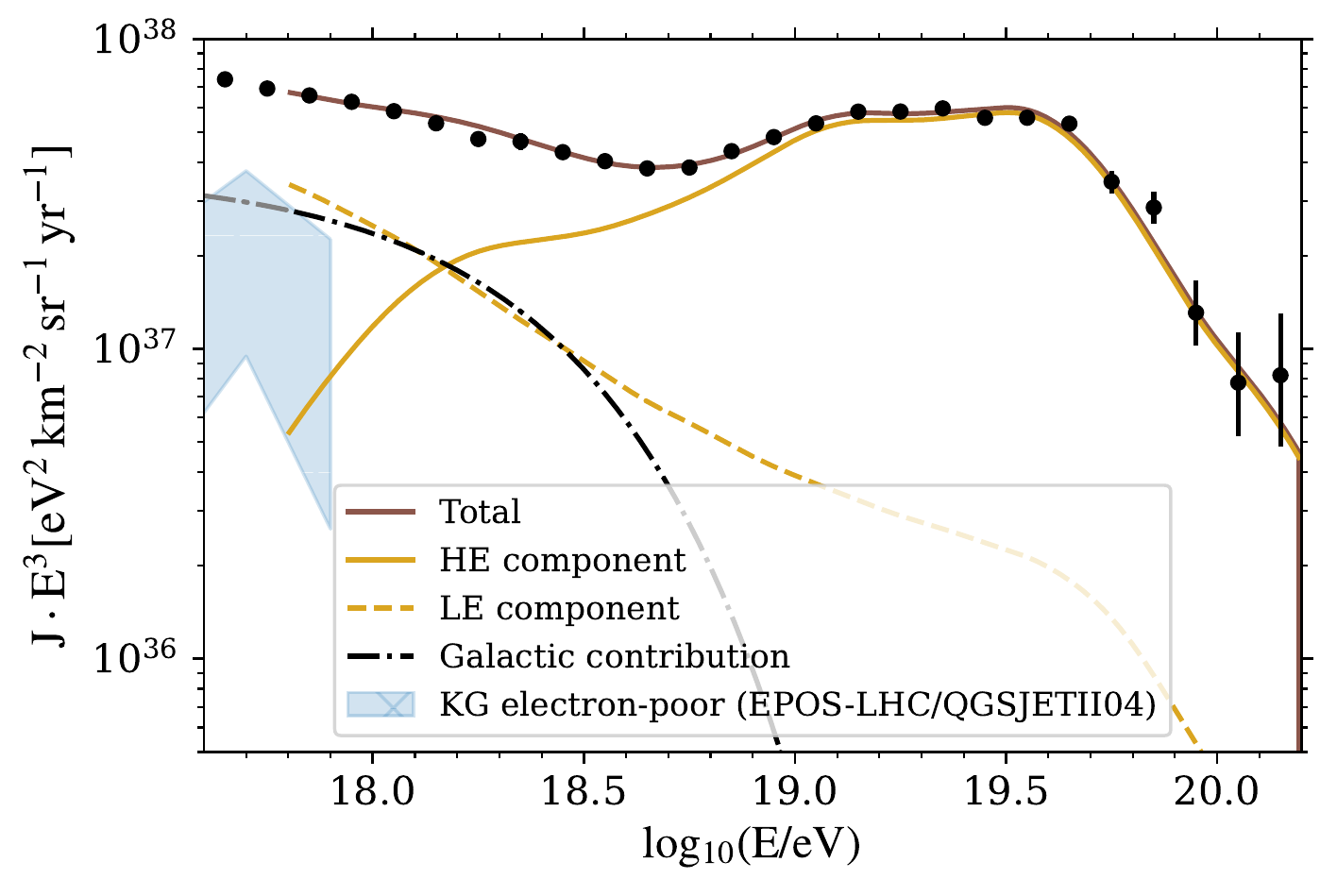}\label{gal2}}
	\caption{\textsc{Scenario 1}. \emph{Left:} The generation rate at the extragalactic sources for each representative mass; the \LEa\ and \HEa\ contributions are shown as dashed and solid lines, respectively. 
	\emph{Right:} The corresponding best-fit results for the all-particle energy spectrum at Earth, given by the superposition of three components. For comparison, also the electron-poor spectrum measured by KASCADE-Grande~\cite{KASCADE} is shown (see the text for details).}
	\label{fig:galcomp}
\end{figure}

\begin{figure}
	\centering
	\def\w{0.49}
	\subfigure{\includegraphics[width=\w\linewidth]{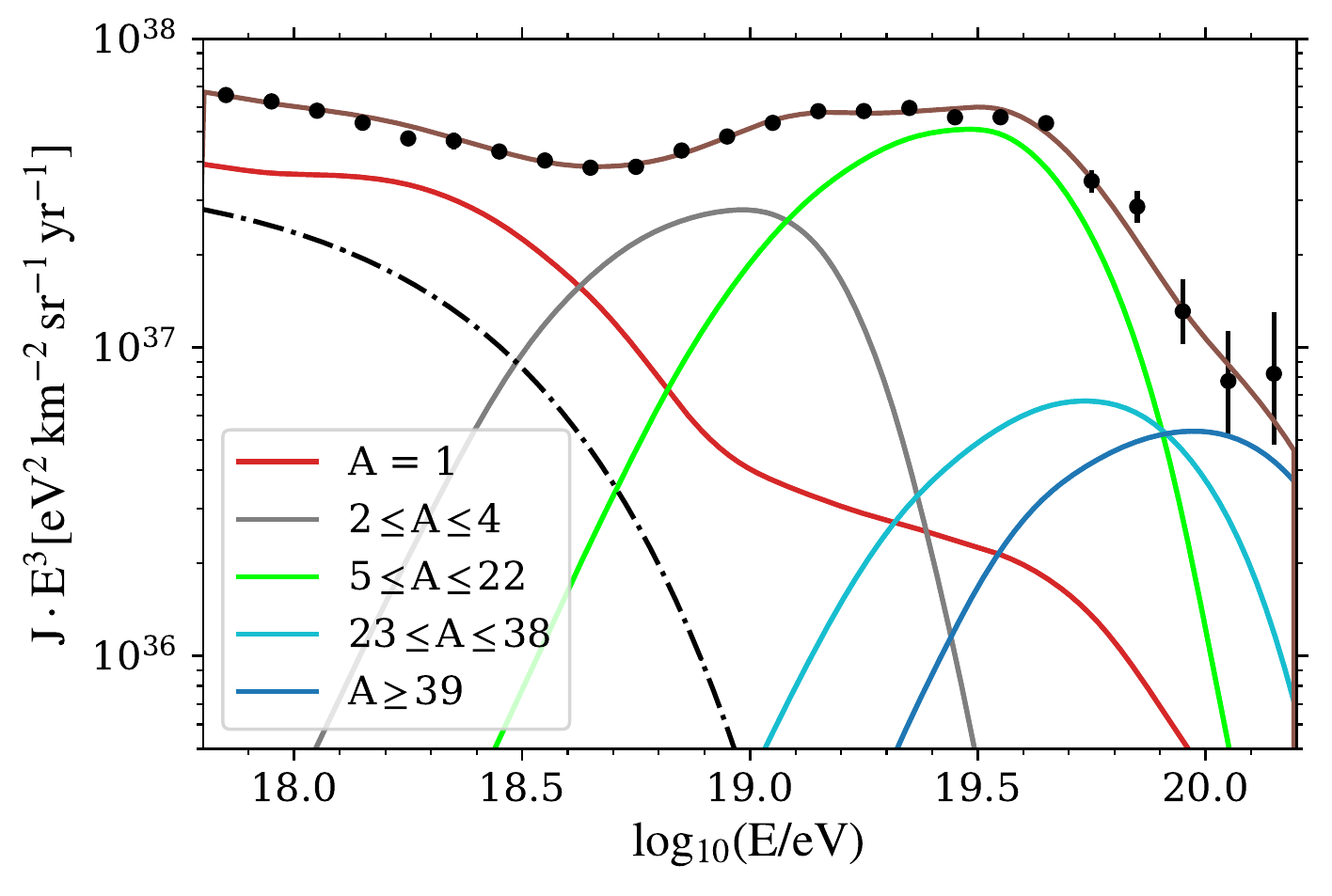}\label{galcomp1}}\hfill
	\subfigure{\includegraphics[width=\w\linewidth]{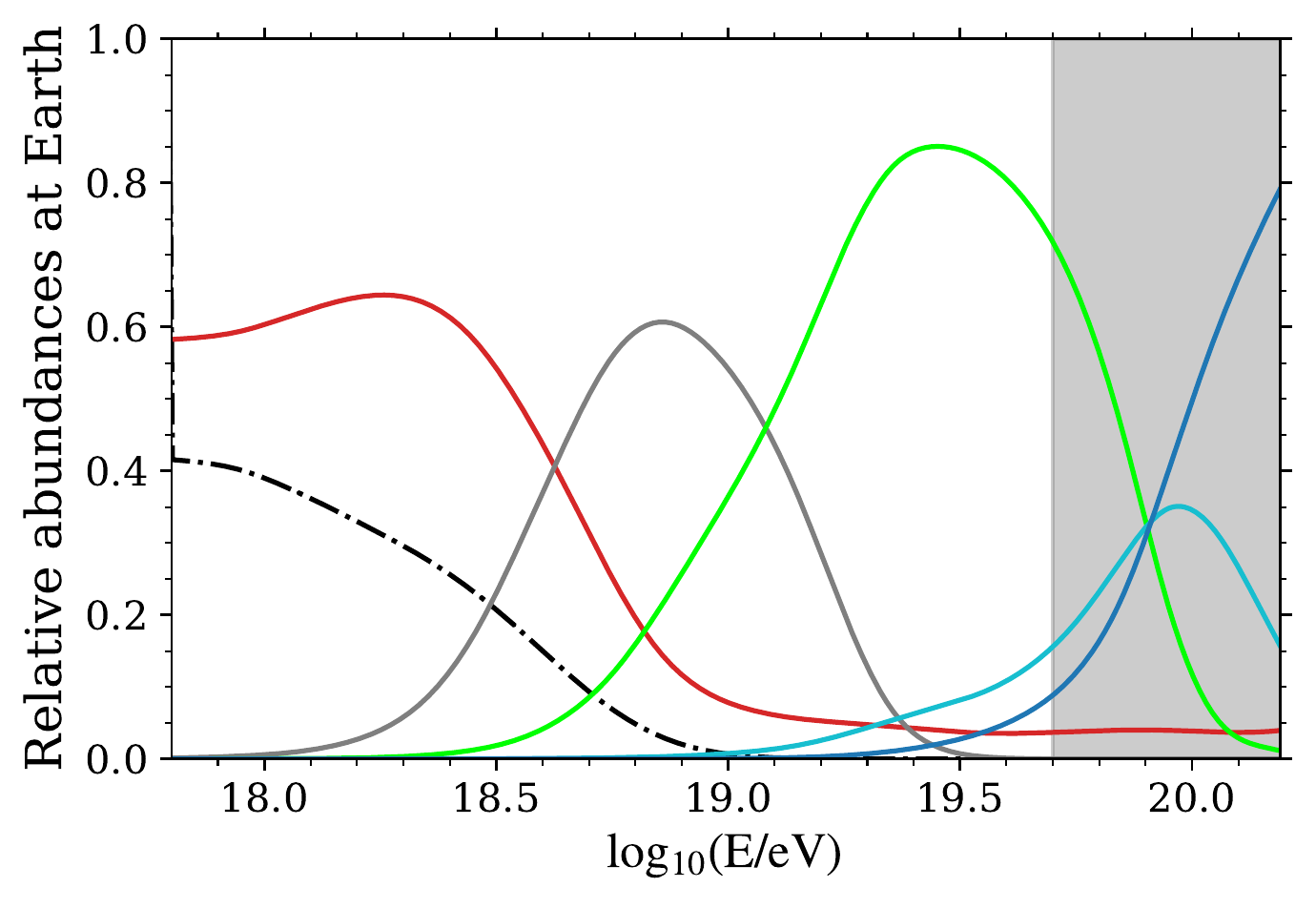}\label{galcomp2}}
	\caption{\textsc{Scenario 1}. \emph{Left:} the Galactic contribution (dot-dashed line) and the extragalactic contributions (grouped according to mass number) to the energy spectrum at the top of atmosphere. 
	\emph{Right:} the corresponding relative abundances as a function of the energy.
	}
	\label{fig:galcomp2}
\end{figure} 

\begin{figure}
	\centering
	\includegraphics[width=0.9\linewidth]{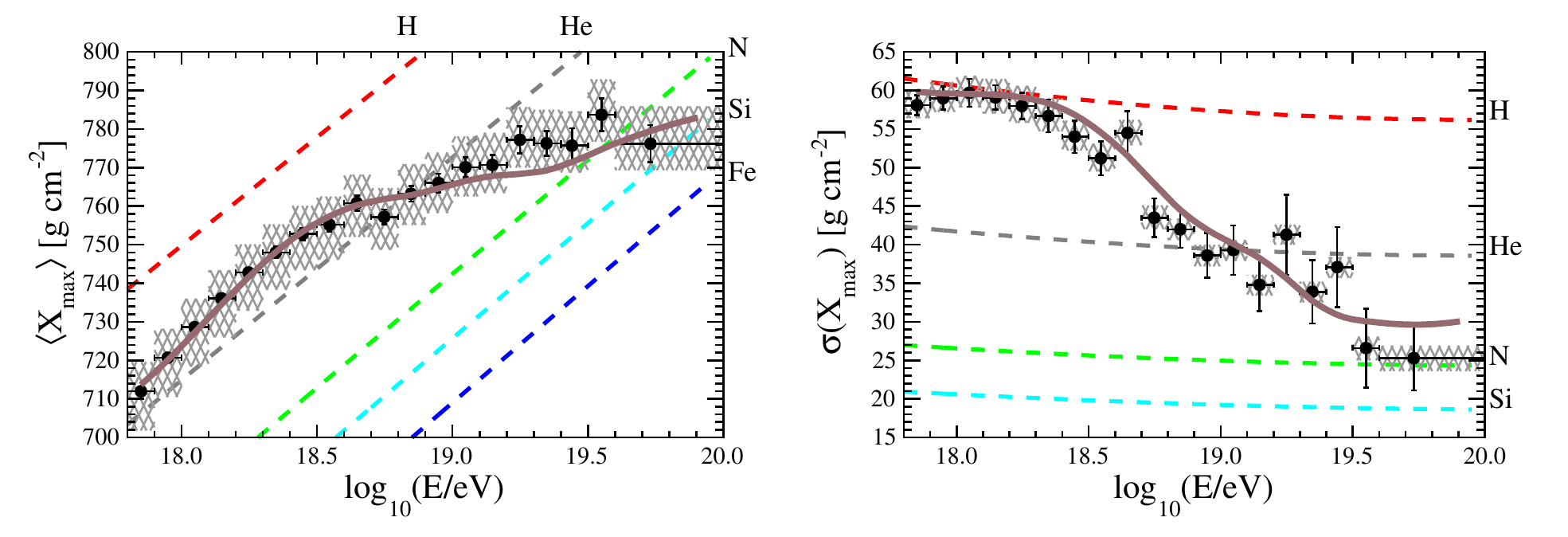}
	\caption{\textsc{Scenario 1}. First two moments of the \xmax distributions as predicted by the best-fit results, along with the measured values and the predictions for pure compositions of various nuclear species according to \EPOS\ (dashed lines).}
	\label{fig:moments_Galcomp}
\end{figure}

The best-fit results are shown in the central column (``\textsc{Scenario 1}'') of Table~\ref{tab:mainresults}. The \HEa\ component has a very hard energy spectrum ($\gamma<0$), a rather low rigidity cutoff and a mass composition dominated by medium-mass elements. The \LEa\ component exhibits a very soft energy spectrum, requiring a larger estimated source emissivity than that of the \HEa\ one and a rigidity cutoff which is much higher than that of the \HEa\ component. The estimated generation rate at the sources and the corresponding best-fit energy spectra at Earth together with the measured data are shown in Fig.~\ref{fig:galcomp}. Fig.~\ref{fig:galcomp} (right) also shows the end of the electron-poor spectrum measured by KASCADE-Grande~\cite{KASCADE}, as a blue band including all the systematic uncertainties and the dependence on the HIMs. This shows that the Galactic spectrum resulting from our best fit is in reasonable agreement with these measurements. Besides, one should consider that the electron-poor subsample given by KASCADE-Grande is obtained by using a selection criterion which depends on the hadronic interaction model and lies between the CNO group and silicon, hence in any case it provides only a lower bound to a Galactic contribution like the one preferred by our data. 

In Fig.~\ref{fig:galcomp2}, the Galactic contribution and the partial extragalactic ones are grouped according to the mass number. In Fig.~\ref{fig:moments_Galcomp} the predicted first two moments of the \xmax distributions are shown as a function of the energy and compared with the measured ones. \lastXmax

We notice that in our \textsc{Scenario 1} the proton component is included through a free parameter in the HE mixed component, while  in~\cite{protonspectrum} protons, being supposed to be generated from in-source interactions, are included only in the LE one; however, a much softer LE spectrum with respect to the HE component is found in both analyses. In our \textsc{Scenario 1}, the proton fraction of the HE component is found to be negligible and therefore the scenario is consistent with~\cite{protonspectrum}.

The rigidity cutoffs of the two extragalactic populations were fitted independently of each other;  the best-fit value of the \HEa\ component is estimated to be much lower than that of the \LEa\ one. Imposing a smaller rigidity cutoff for the \LEa\ component would worsen the fit. For example, requiring the two components to have the same rigidity cutoff, as hypothesized in~\cite{protonspectrum}, would increase the deviance by $\Delta D = +28$ (from 586 to 614), mainly due to a worsening of the energy spectrum fit.  However, note that such a difference is smaller than the one caused by the systematic uncertainties, which is illustrated in Section~\ref{sysunc}, so neither configuration can be strongly preferred over the other. Further details will be discussed in Section~\ref{disc_scenarios}.

\subsection{Scenario 2: two mixed extragalactic populations}
\label{scenarioB}

An alternative way to describe the data in the energy region of interest is assuming that the ankle around $10^{18.7}$\,eV is due to the superposition of two extragalactic components, one dominating at LE and the other at HE. We assume that the two components are both ejected according to energy spectra described by Eq.~\eqref{eq1bis} but with different parameter values, since they are reasonably associated to two different populations of sources. We are here implicitly assuming that a possible Galactic contribution is subdominant in the considered energy range.

The best-fit parameter values are listed in the column ``\textsc{Scenario 2}'' of Table~\ref{tab:mainresults}. The spectral parameters in both energy ranges as well as the composition of the \HEb\ one are similar to those found in the previous scenario. The composition of the \LEb\ component is a mix of mostly protons and nitrogen, similar to the sum of the Galactic and \LEa\ extragalactic components in the previous scenario.

\begin{figure}
	\centering
	\def\w{0.49}
	\subfigure{\includegraphics[width=\w\linewidth]{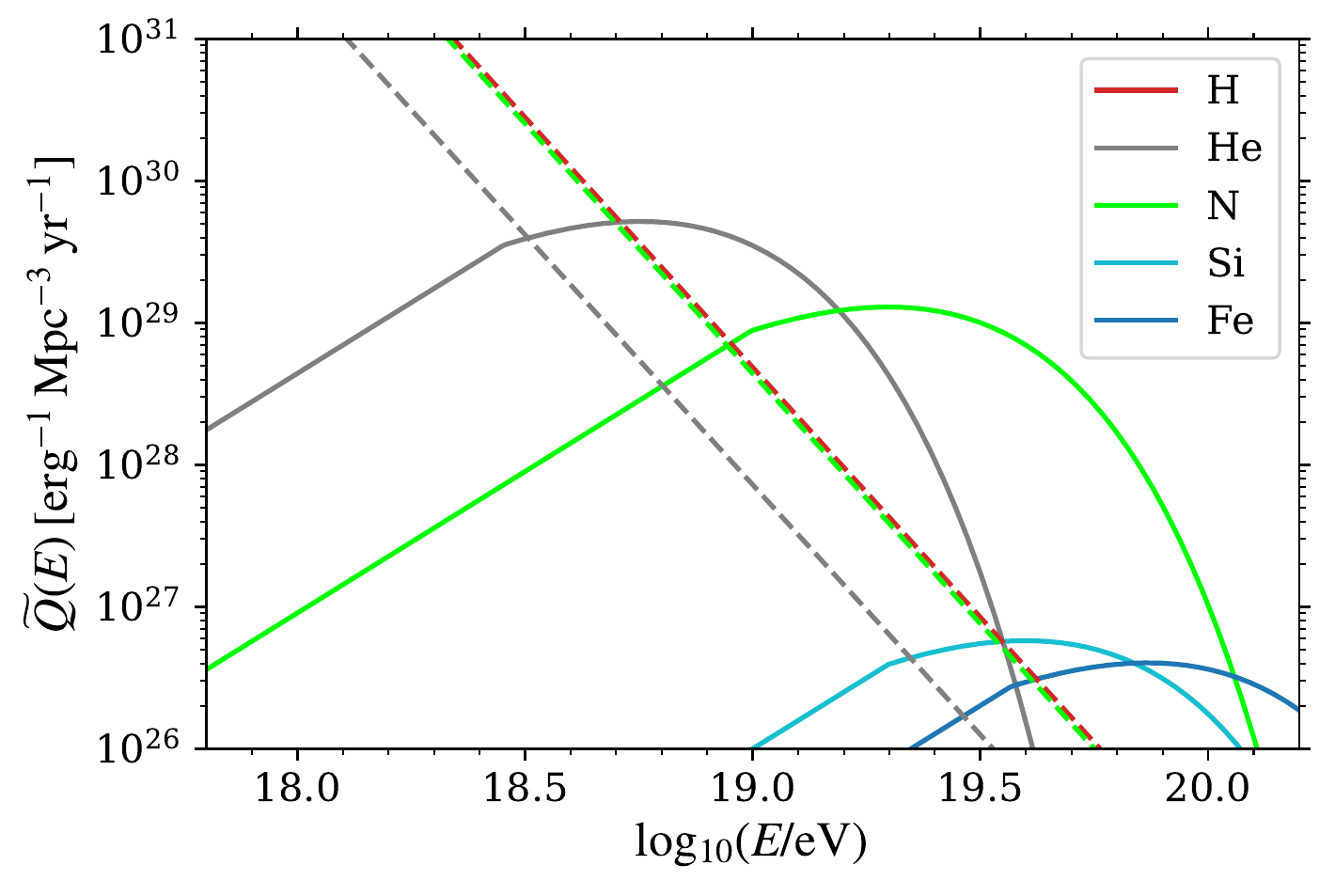}\label{mixed1}}\hfill
	\subfigure{\includegraphics[width=\w\linewidth]{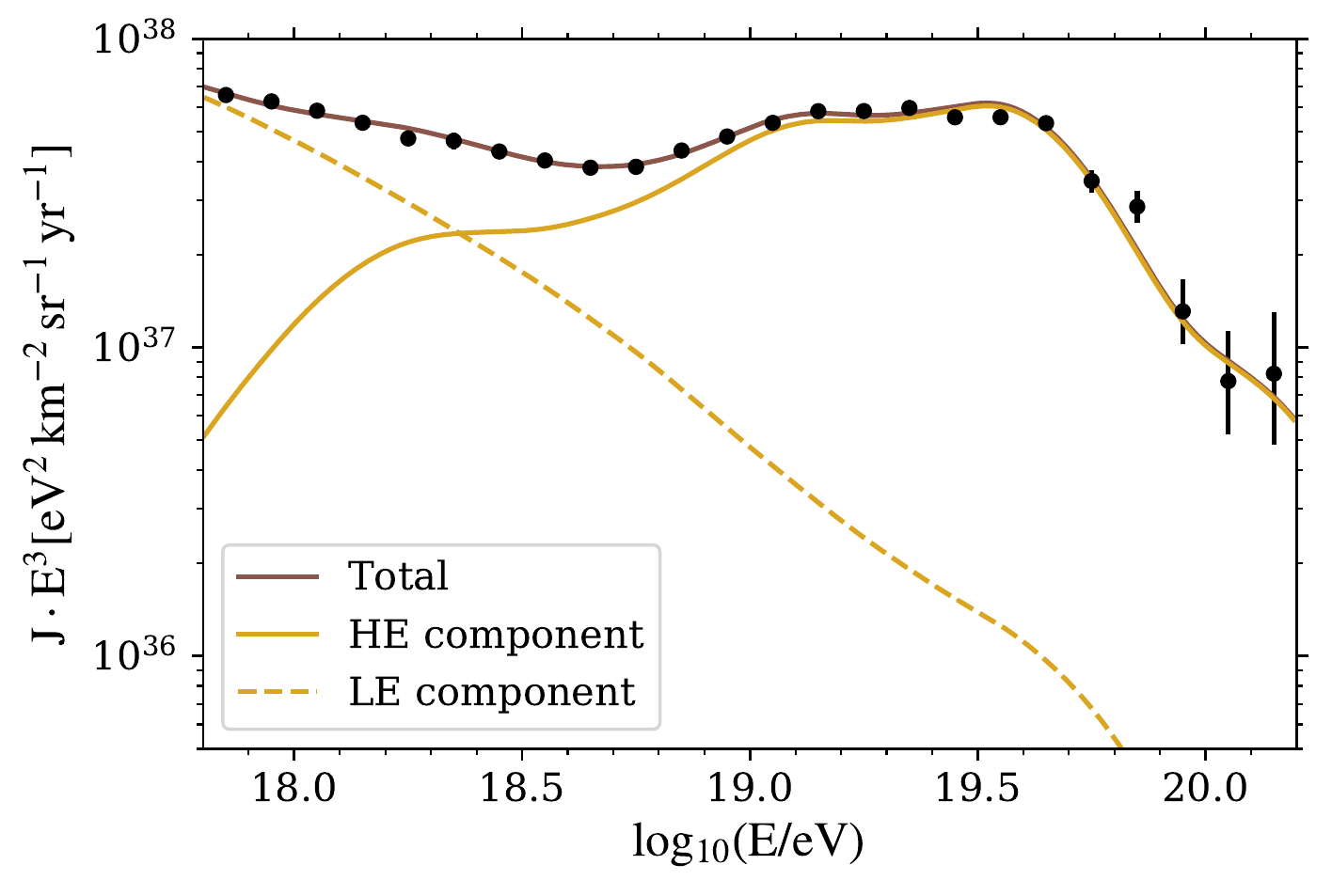}\label{mixed2}}
	\caption{\textsc{Scenario 2}. \emph{Left:} The generation rate at the sources for each representative mass; 
	the \LEb\ and \HEb\ contributions are shown as dashed and solid lines, respectively.
	 \emph{Right:} The corresponding best fit results for the all-particle energy spectrum at the Earth, given by the superposition of the \LEb\ and \HEb\ extragalactic components.}
	\label{fig:mixedcomp}
\end{figure}

The estimated generation rate at the sources is shown in the left panel of Fig.~\ref{fig:mixedcomp} for each component and each ejected nuclear species. After the propagation through the intergalactic medium, the partial contributions of the two components overlap in the ankle region and provide a total flux which describes the measured spectrum in the whole considered energy region, as shown in the right panel of Fig.~\ref{fig:mixedcomp}. 

\begin{figure}
	\centering
	\def\w{0.49}
	\subfigure{\includegraphics[width=\w\linewidth]{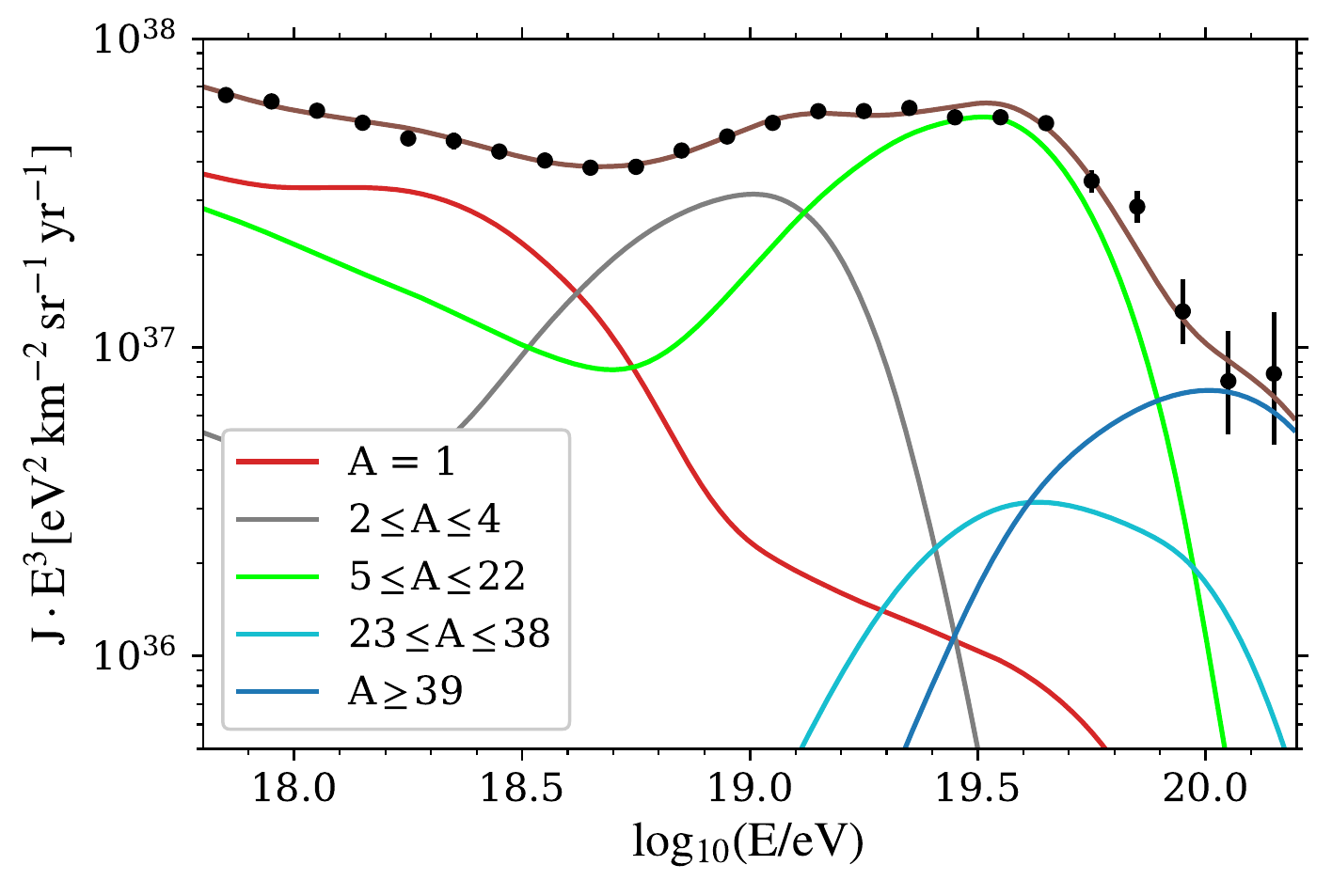}\label{alt1}}\hfill
	\subfigure{\includegraphics[width=\w\linewidth]{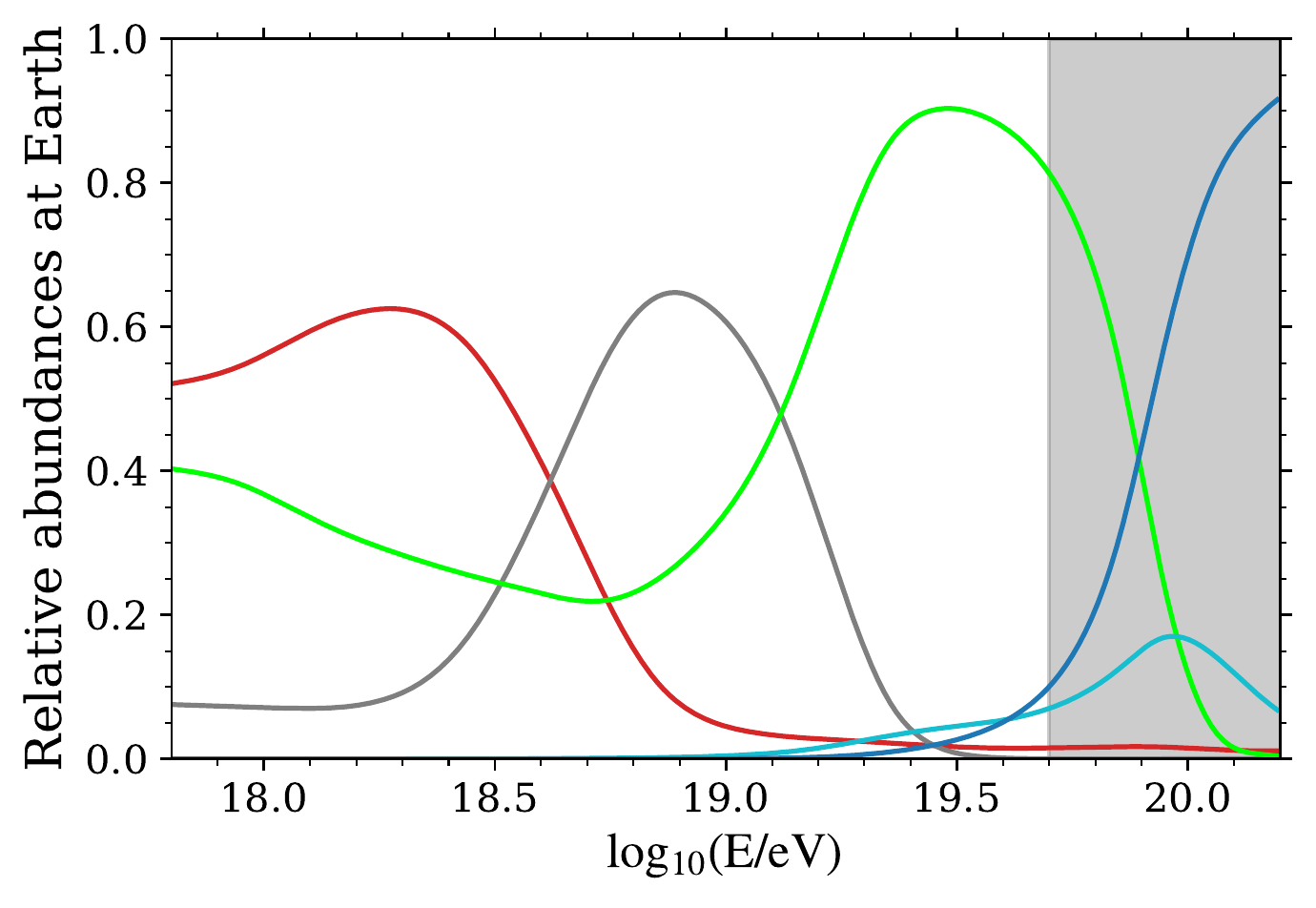}\label{alt2}}
	\caption{\textsc{Scenario 2}. \emph{Left:} Partial contributions to the energy spectrum at the top of the atmosphere grouped according to mass number. 
	\emph{Right:} the corresponding relative abundances as a function of the energy. 
	}
	\label{alt_scenario}
\end{figure}

\begin{figure}
	\centering
	\includegraphics[width=0.90\linewidth]{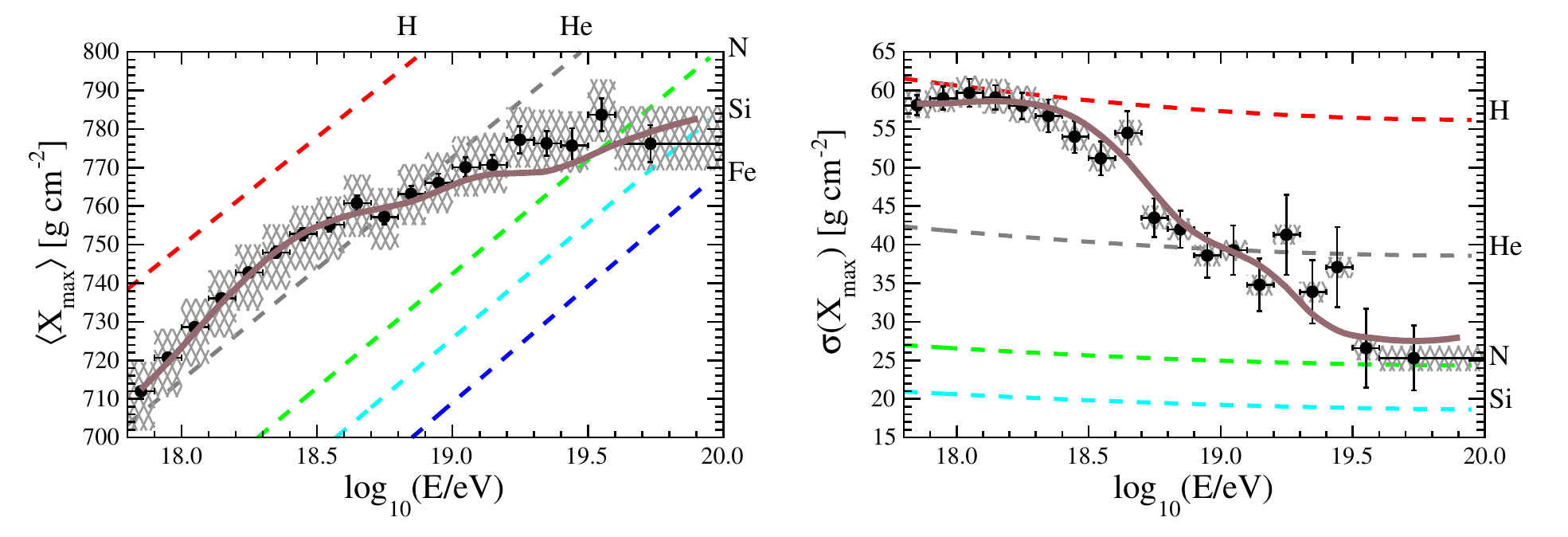}
	\caption{\textsc{Scenario 2}. First two moments of the \xmax distributions as predicted by the best-fit results, along with the measured values and the predictions for pure compositions of various nuclear species according to \EPOS.}
	\label{fig:moments}
\end{figure}

We report also the contributions at the top of the atmosphere grouped according to mass number (Fig.~\ref{alt_scenario}) and the first two moments of the \xmax distributions (Fig.~\ref{fig:moments}).

\begin{figure}
	\centering
	\def\w{0.49}
	\subfigure{\includegraphics[width=\w\linewidth]{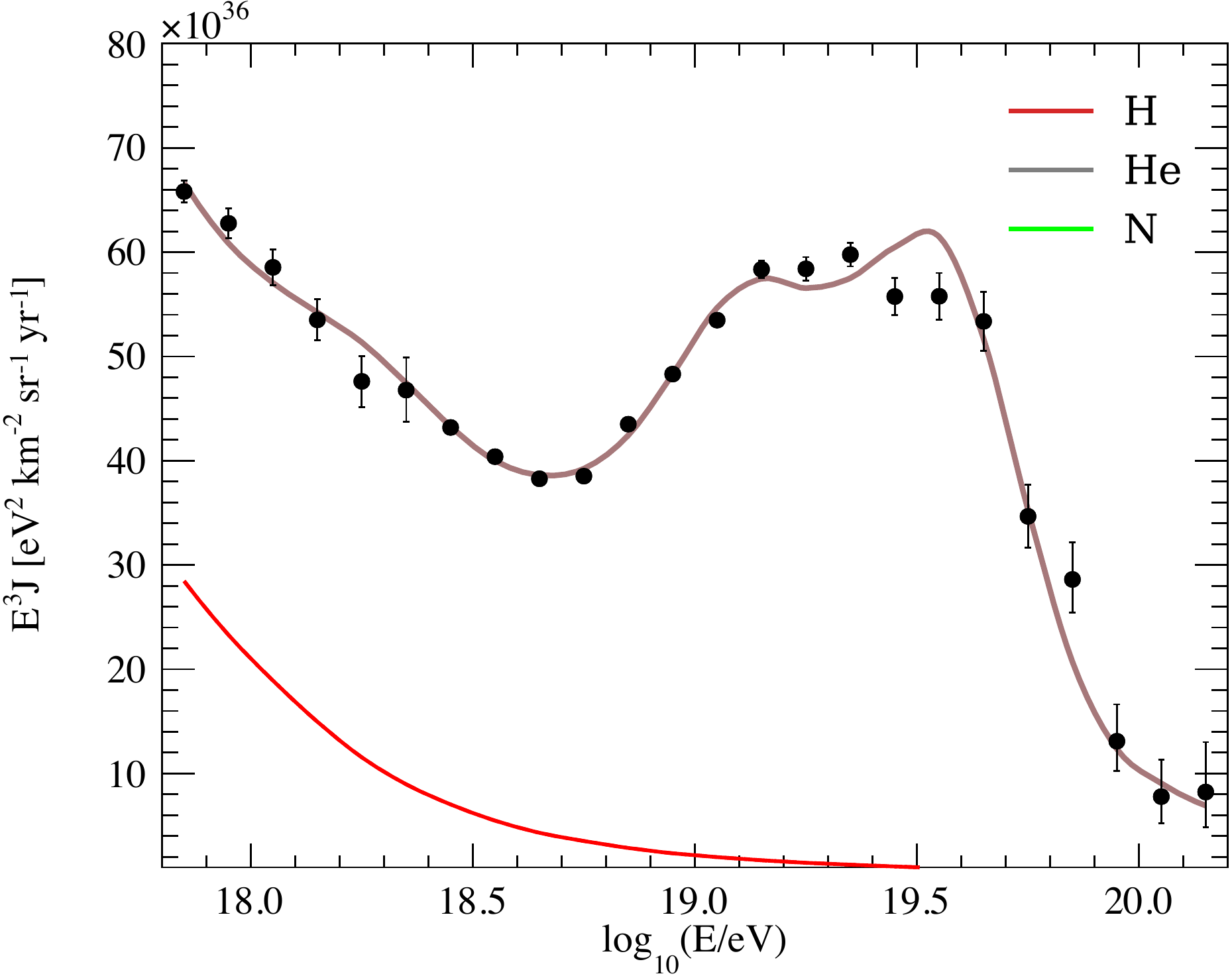}\label{spectrum_injA_fromH_LE}}\hfill
	\subfigure{\includegraphics[width=\w\linewidth]{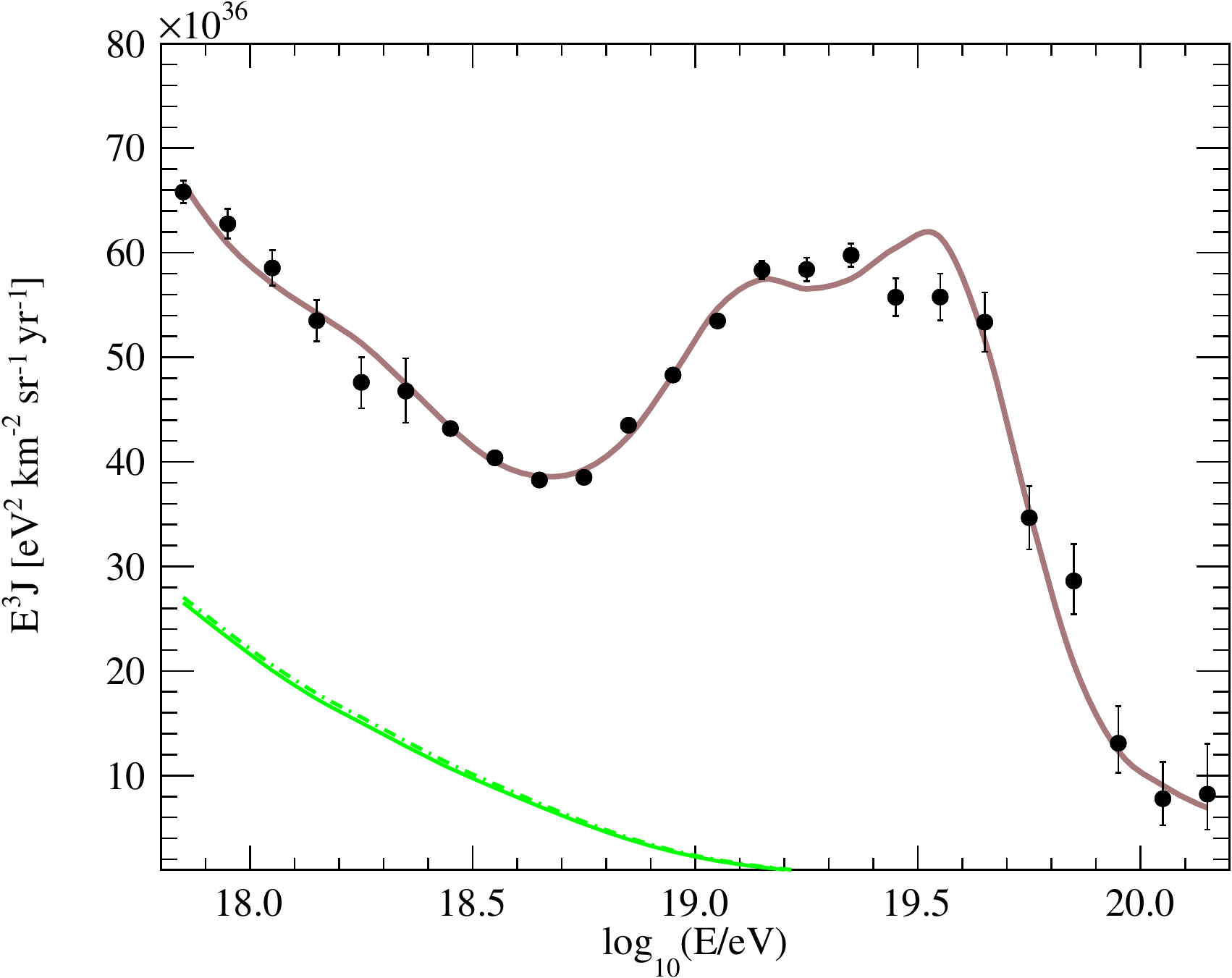}\label{spectrum_injA_fromN_LE}}
	\\
	\subfigure{\includegraphics[width=\w\linewidth]{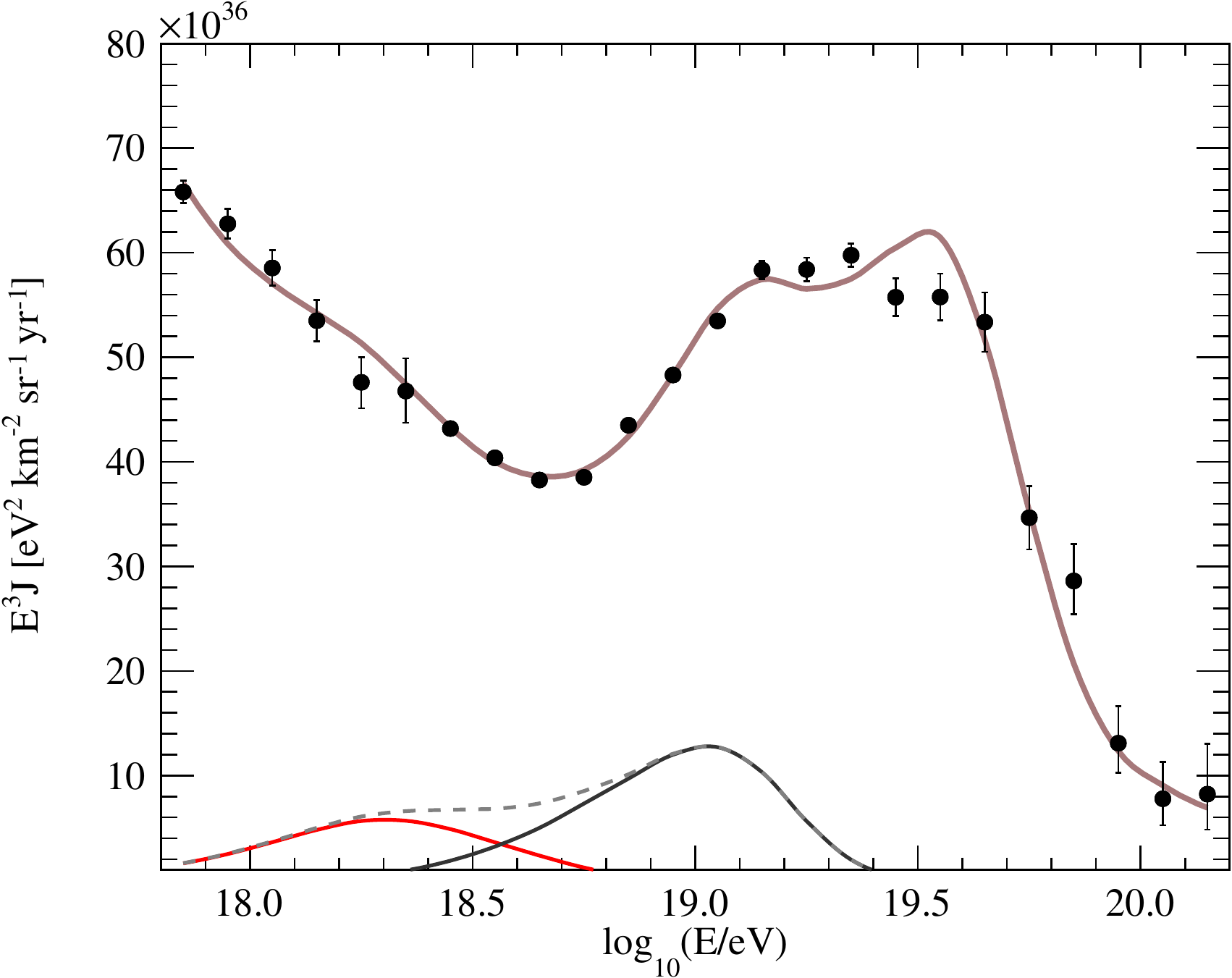}\label{spectrum_injA_fromHe_HE}}\hfill
	\subfigure{\includegraphics[width=\w\linewidth]{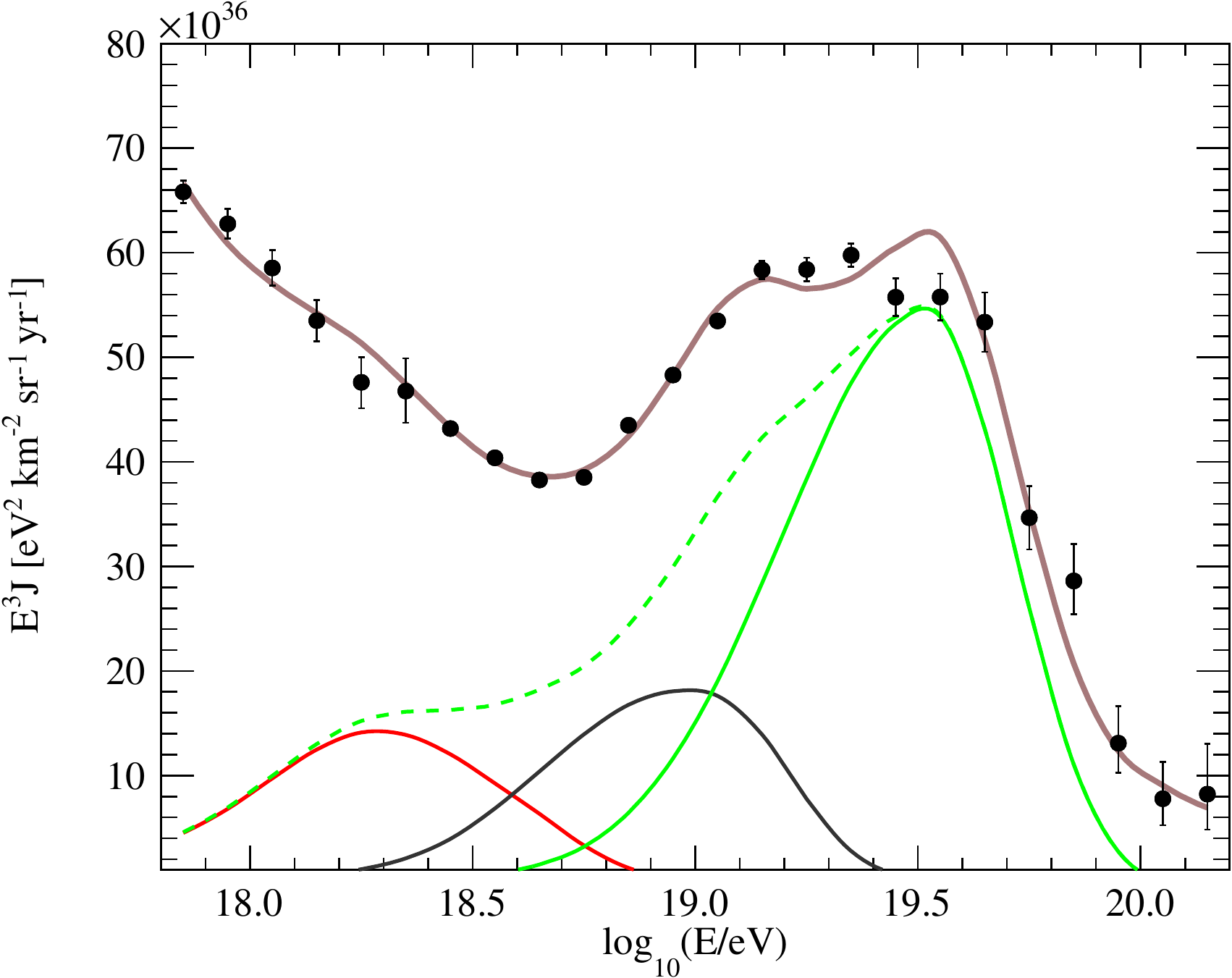}\label{spectrum_injA_fromN_HE}}
	\caption{\textsc{Scenario 2}. The flux at Earth produced by the dominant nuclear species at the sources for each component (dashed lines) and the partial contributions to them grouped according to the mass number $A$ of their secondary particles at Earth (solid lines). The main contributions from the \LEb\ component are shown on the top row (left:~H, right:~N) and the main ones from the \HEb\ component on the bottom row (left:~He, right:~N). The curves are colour-coded as in the previous plots.}
	\label{partial_contributions}
\end{figure}

In Fig.~\ref{partial_contributions}, the propagated fluxes produced by each ejected nucleus heavier than hydrogen are shown (dashed lines) along with their partial contributions from different mass groups of secondary particles at the Earth (solid lines). Note that so far only the statistical uncertainties have been taken into account and the visible minor features in the energy spectrum that are not described by our model are actually encompassed within the systematic uncertainties discussed in Section~\ref{sysunc}. Besides, it is worth stressing that further extending the fit to lower energies will require to include the effect of intergalactic magnetic fields, here neglected (see Section~\ref{disc_scenarios}), to avoid the overestimation of measured fluxes below the current fit threshold.

The plots on the top of Fig.~\ref{partial_contributions} show the contributions from the \LEb\ component, whereas the ones on the bottom refer to the \HEb\ one. From the comparison of the primary and secondary contributions, it is clear that the photodisintegration plays no significant role in the propagation of the \LEb\ component, whose observed composition is essentially the same as the one ejected at the sources. Within the \HEb\ component the intersection of the helium and nitrogen groups at Earth might be responsible of the change of the slope at the instep, as already pointed out in Ref.~\cite{paoICRC21}.

Although the values of the source rigidity cutoff $R_\text{cut}^\text{\HEb}$ are lower than
approximately $10^{18.5}$\,V, 
the shape of the cutoff is such that the ejected nuclei (especially medium-mass ones) can still undergo a substantial amount of photodisintegration during their propagation, with a major impact on the all-particle spectrum. In particular, as shown in the fourth panel of Fig.~\ref{partial_contributions}, the secondary nucleons and helium nuclei from such interactions contribute to around half of the all-particle spectrum at the ankle energy.

\subsection{Discussion of astrophysical scenarios}
\label{disc_scenarios}

Using two different populations of extragalactic sources dominating at high and low energy (HE and LE respectively) allows to easily reproduce the ankle feature. In both proposed scenarios, the HE extragalactic population has a mixed mass composition, in agreement with what was found in \CF\ for the fit above the ankle. Conversely, the two scenarios differ in the mass composition of the LE population, which in one case is mixed, while in the other case it is composed of pure protons, requiring an additional medium-mass Galactic component to match the observations~\cite{Aloisio:2013hya}.

A common finding between the two proposed scenarios is that the HE component requires very hard spectra with low rigidity cutoffs and intermediate mass compositions, while the LE component requires much steeper spectra. 

The negative spectral index of the \HEa\ component produces very hard elemental fluxes at Earth, with little overlap between different masses; this is required to obtain a good description of the very pronounced spectral features of the measured energy spectrum and the rather narrow \xmax distributions. We stress here that the spectral index found as outcome of the fit in this study, that includes the extragalactic propagation only, is related to the UHECR spectrum escaping the source environment. This can differ from the accelerated one due to energy-dependent effects concerning interactions and diffusion in the source environments, justifying our finding in the HE component in both scenarios.

Alternative explanations to the interplay between the interaction rates and the diffusion one can be provided to justify the steepness of the LE spectrum, especially regarding the \textsc{Scenario 2}.
For instance, if the assumption of identical sources is relaxed and different maximal energies are taken into account, the effective energy spectrum obtained by integrating over them would be steeper than the one of each individual source, as demonstrated in Ref.~\cite{Semikov2006}. Besides, it is also important to remember that in this work we are considering an effective energy spectrum which encompasses also the effects of intergalactic magnetic fields, here neglected. Due to the so-called magnetic horizon effect, if the closest sources are far enough (${>}10$\,Mpc), i.e.\ if the source density is small enough, the time needed for the particles to reach the Earth may become larger than the lifetime of the sources. This would cause a suppression of the flux at low energies~\cite{MR2020}, which makes the observed spectrum harder than the actual one escaping from the sources. For example, in a preliminary study~\cite{wittkowski} a softer energy spectrum was estimated in presence of a relatively strong IGMF in the case of the above-ankle fit. 

In terms of mass composition at LE, we find that the data can be described by a mix of nitrogen and hydrogen in both scenarios, their relative contributions respectively decreasing and increasing with energy. The need for a medium-mass contribution in this energy range was already known from the independent fits to the \xmax distributions~\cite{PAOcomp,bellido2017}; however, with this analysis, it is possible to discuss the origin of the inferred composition at the escape from the sources.
Galactic supernova remnants are expected to accelerate iron nuclei up to ${\sim}10^{17}$\,eV, but lighter particles such as nitrogen nuclei can reach only energies of the order of $3{\times}10^{16}$\,eV according to the rigidity dependent scenario~\cite{Cristofari:2020mdf}. However, a secondary Galactic component able to reach much higher energies has been considered by different authors. If non-linear amplifications of magnetic fields can happen upstream of supernova shocks~\cite{BellLucek}, then cosmic particles could be accelerated to energies of the order of $Z{\times}10^{17}$\,eV.  Based on this model, Hillas~\cite{Hillas2005} proposed acceleration of particles through Type~II explosions into dense stellar winds (where very strong magnetic fields should exist). GCRs accelerated in supernova remnants and diffusing out of the disk could be captured in termination shocks produced by strong Galactic winds, and be re-accelerated back into the disk~\cite{Blasi:2017caw}.\footnote{at energies which depend on the balance between advection and diffusion, as higher energy particles can diffuse faster and reach the disk with higher efficiency.} Explosion of supernovae in the winds of Wolf--Rayet stars~\cite{Crowther:2006dd, Rosslowe} are expected to happen, although for a quite small fraction (${\sim}1/7$) of cases~\cite{gal-yam} and reach energies up to more than $10^{18}$\,eV if the magnetic field in the wind is as high as 100\,G or higher~\cite{bier93}.  This mechanism would provide a higher contribution to the total flux of cosmic rays at lower energies (below the knee) and a higher cutoff energy when compared to the previous one~\cite{wr}. In particular, depending on the compositions of the Wolf--Rayet winds, such explosions may accelerate nitrogen nuclei up to an energy cutoff of ${\sim}10^{18}$\,eV and helium up to a few $10^{17}$\,eV, which would make plausible to observe the tail of this Galactic component in the energy range included in our fit~\cite{gal-yam,wr}. 
In the context of the \textsc{Scenario 1}, the obtained results suggest to rule out models foreseeing a dominance of Galactic iron in the region below the ankle, like the one originally proposed by Hillas, or those assuming a contribution from re-acceleration in Galactic strong winds. Models proposing a contribution from explosions in the winds of Wolf-Rayet-like stars would describe our data better, as for reasonable choices of parameters they provide compositions dominated by the CNO group. In addition, being independent of the scenario, the result on mass composition at LE strongly confirms what found in Ref.~\cite{PierreAuger:2016qzj} about the needed mixture at the ankle. The possibility of a mixing with heavier nuclear species such as iron is therefore excluded around the ankle region. On the contrary, the small percentage of iron found by the fit at HE seems to be only required by the energy spectrum at the highest energies, being the composition data absent in that energy range, and in particular also depends on the shape of the cutoff function. In fact, as noted in Ref.~\cite{Heinze2019}, a low rigidity cutoff will require the presence of an elemental group at $Z\,R_\text{cut}$ to populate the spectrum at UHE. Our updated composition fraction fits presented in Ref.~\cite{bellido2017} are indeed compatible with the onset of a heavy component at UHE above $10^{19.4}$\,eV.

The HE rigidity cutoff found as a result of the fit suggests that the maximum energy emitted at the sources is not high enough to entirely attribute the spectrum features, in particular the suppression at the highest energies, to propagation effects. However, due to the fact that we are evaluating the spectrum at the escape, this result cannot fully be used to constrain the maximum energy at the acceleration, being the interactions in source potentially also responsible for reducing the maximum energy, as for instance studied in Refs.~\cite{Biehl:2017zlw,Rodrigues:2017fmu}.
As concerns the \LEa\ component, the fit
is degenerate with respect to $R_\text{cut}^\text{\LEa}$ for values ${\gg}10^{19.5}$\,V, thus fixing this parameter to
any arbitrarily higher value provides the same best-fit results. Such a degeneracy is visible in the figures in Appendix~\ref{dev_profiles}, where the values of the total deviance obtained by scanning over $R_\text{cut}^\text{\LEa}$ (re-optimizing all other parameters for each $R_\text{cut}^\text{\LEa}$ value) are shown. This can be explained by the fact that the estimated energy spectrum of this component is very steep, and hence it is rapidly suppressed even in the absence of an exponential cutoff, making the energy range where this component is the dominant one rather narrow (as shown in the right panel of Fig.~\ref{fig:galcomp}-right) and the fit is insensitive to the details of its shape. 
Furthermore, in this energy region the propagation effects on the spectrum and composition are minimal, the only non-negligible process being the adiabatic energy loss due to the expansion of the Universe. For these reasons, both the two possible scenarios we used provide a description of the data set with very similar deviance values; firm conclusions about a favoured scenario cannot be reached without further
investigating the Galactic-to-extragalactic transition region.
Even so, it is worth noting that the case with two extragalactic mixed components provides a better fit of the \xmax measurements but a worse description of the very pronounced features in the energy spectrum.
One way in which a Galactic and an extragalactic below-ankle medium-mass composition would differ is in their distribution of arrival directions, which are not considered in this work.
As shown in Refs.~\cite{PAOlsa,PAOcomp}, a large fraction of GCRs below the ankle can be excluded by the low level of anisotropy and the measurements of composition. This conclusion was also drawn in Ref.~\cite{Gia2012} by considering possible variations of the parameters of the Galactic magnetic field and by including intermediate nuclei.
However, in our \textsc{Scenario 1} the anisotropy of the Galactic component could be diluted by the large isotropic extragalactic contribution present, which is of the order of $60\%$~of the all-particle flux around 1\,EeV and increases at higher energies.

\subsection{Comparisons to the combined fit above the ankle}
\label{comparisonCF}
The main qualitative features of the \HEb~component at injection in our best fit are the same as in \CF, namely a mixed mass composition dominated by the nitrogen group, a much harder spectrum than predicted in the case of Fermi acceleration, and a rigidity cutoff well below the threshold for pion production on CMB photons.
On the other hand, there are a few noticeable quantitative differences.

In Ref.~\cite{combinedfit}, in the scenarios with no source evolution and with systematic uncertainties on energies and \xmax neglected, the best-fit spectral index sometimes also assumed positive values, while here it is always found to be negative. Likewise, the cutoff rigidity $\log_{10}(R_\text{cut}/\text{V})$, which is strongly correlated with $\gamma$, shows a narrower range of variation here with respect to our previous findings.  Part of this change is due to the \LEb~component contributing to a non-negligible fraction of the total flux even at energies within the fitting range of our previous work (namely $E\ge10^{18.7}$\,eV), as shown in Fig.~\ref{fig:mixedcomp}\subref{mixed2}, hence the addition of such a contribution requires the low-energy tail of the \HEb~component to be lowered, i.e.\ its spectrum hardened. 

The hardening of the spectral index also causes a lowering of the cutoff rigidity due to the correlation between these two parameters. A smaller part of the effect is due to the treatment of the finite energy resolution of the detector via the forward-folding technique, which may bias the fit against very hard spectra in the case that the total flux at energies below the start of the fitting range is underestimated, as it was in Ref.~\cite{combinedfit} due to the absence of a \LEb~component. On the contrary, the current work reasonably reproduces the total flux below the ankle and does not use a forward-folding technique, hence it is not affected by such a bias. A counter-effect, although of considerably smaller magnitude (see Appendix~\ref{app:overdensity}), is obtained when including a local overdensity in the otherwise homogenous and isotropic distribution of the sources, as done here but not in Ref.~\cite{combinedfit}.

Another difference is the predicted mass composition in the highest-energy part of the spectrum: in Ref.~\cite{combinedfit} the best-fit fraction of iron was 0 and the end of the spectrum was dominated by silicon, whereas here we infer a best-fit fraction of iron of about 3\%.
This is because the number of events above $10^{20}$\,eV has increased from 5 to 15 thanks to an improved determination of the energy scale, and in our model the observed cutoff is due to the photodisintegration of nuclei, whose threshold is roughly proportional to the mass number.

The extension of the combined fit to the data below the ankle energy, which have much smaller statistical uncertainties than at higher energies, causes a substantially worse goodness of the fit than in our previous work. Indeed, in Ref.~\cite{combinedfit} only the first two bins had statistical uncertainties less than 1\%, whereas in the data used here this applies to all the first five bins after the SD-1500 threshold ($\log_{10}(E/\text{eV}) \in [18.4,18.9)$). Besides,  the widths of the \xmax distributions used in this work are narrower by a few g/cm$^2$. This is due to new constraints used in the shower profile fit in order to improve the resolution at low energies, which typically result in deeper \xmax estimates for shallow events and vice versa with respect to the old constraints. Since the \xmax distributions are already as narrow as predicted by the model with a nearly pure mass composition at each energy (right panel of Fig.~\ref{fig:moments}), further narrowing them results in a worse fit.

In the same paper~\cite{combinedfit}, the extension of the fit to lower energies was also explored, following an approximate procedure instead of a proper fit. The possible presence of a Galactic component was also considered therein, using an extrapolation of KASCADE-Grande data and assuming that it was Fe-dominated.  In the current analysis this dominance is excluded. We notice that the new result about a preference of a lighter  mass composition has been made possible thanks to a proper evaluation of the fit deviance and the increased statistical accuracy of the data.

\section{Effect of the systematic uncertainties}
\label{sysunc}
Since the scenarios described in Sections~\ref{scenarioA} and \ref{scenarioB} were found to be nearly equivalent in practice, in this and the following sections we will only study variations on \textsc{Scenario 2}, with no Galactic component and two mixed extragalactic populations. Such scenario is the one on which the effects of different assumptions about the distribution and evolution of extragalactic sources and the propagation in intergalactic space is expected to be more noticeable.

\subsection{Experimental uncertainties}
\label{nuisance}
\newcommand{\Dsyst}{D_\text{syst}(\xmax)}

The energy scale and the \xmax scale are the most important sources of experimental systematic uncertainties. For the energy scale, an energy independent uncertainty $\Delta E/E = 14\%$ is adopted
in the whole considered energy region~\cite{PierreAuger:2021hun}. As concerns the systematic uncertainties on the measured \xmax 
values, they are asymmetric and slightly energy-dependent, 
ranging from 6 to 9\,g/cm$^2$~\cite{xmax_dist}.

Regarding the energy scale uncertainty, we followed the same approach used in \CF, which consists of shifting all the measured energies by one systematic standard deviation in each direction. On the other hand, as concerns the \xmax scale uncertainty, it is worth noticing that, while the correlations are nearly perfect (${\sim}0.998$) in the case of first-neighbour energy bins, they can go down to~$\sim 0.6$ between the lowest and the highest energy bins, hence we chose to use a more complete approach than the one used in Ref.~\cite{combinedfit}, which we describe in Appendix~\ref{nuisance-expl}. Two nuisance parameters are added to the fit, corresponding to the principal components of the covariance, allowing different shifts at different energies. However, for a direct comparison with the approach used in Ref.~\cite{combinedfit}, the results obtained by considering all the possible combinations of shifting the measured energies and \xmax values by one systematic standard deviation in each direction are shown in Appendix~\ref{sysexp}.

\begin{figure}
    \centering
    \def\w{0.49}
    \includegraphics[page=1,width=\w\columnwidth]{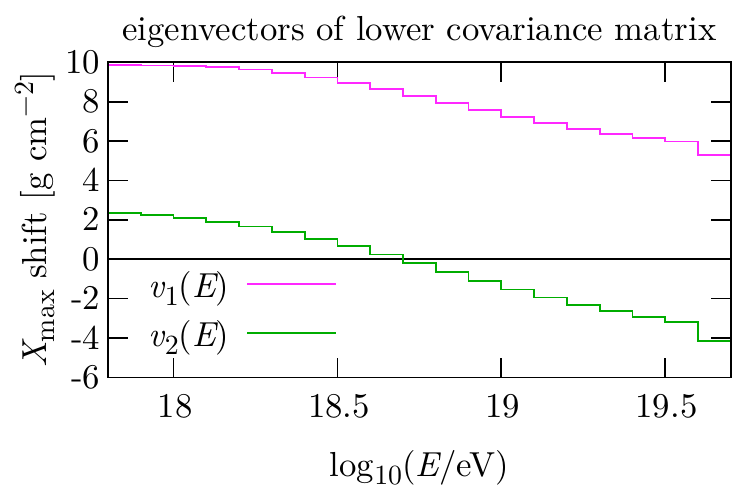}
    \includegraphics[page=2,width=\w\columnwidth]{eigenvectors}
    \caption{The first two eigenvectors of the covariance matrix of lower (left) and upper (right) systematic uncertainties in \xmax (see the text for details).}
    \label{fig:eigenvectors}
\end{figure}

In the approach based on two nuisance parameters $a$ and $b$, the first two eigenvectors of the covariance matrix define two functions of energy, $v_1(E_i)$ and $v_2(E_i)$, plotted in Fig.~\ref{fig:eigenvectors}; all the \xmax distributions are shifted by a quantity $a\,v_{1}(E)+b\,v_{2}(E)$, and an additional term $\Dsyst=a^2+b^2$ is added to the deviance.
The parameter $a$ shifts all the \xmax distributions in the same direction by an energy-dependent amount, whereas $b$ has an opposite effect on the high-energy and the low-energy distributions.\footnote{Since the systematic uncertainties are asymmetrical, we actually have two different covariance matrices, one for lower and one for upper uncertainties.  We use the former when $a<0$ and the latter when $a>0$.}

\begin{table}
\renewcommand\arraystretch{1.4}
	\centering\footnotesize
	\centering	
	\makebox[\textwidth][c]{%
	\begin{threeparttable}
	\begin{tabular}{ | l | c  c | c  c | c  c|}
	\hline
	 $\Delta E / \sigma_\text{syst}$ & \multicolumn{2}{c|}{$-1$} & \multicolumn{2}{c|}{0} & \multicolumn{2}{c|}{+1} \\ \hline \hline
	         & \LEb & \HEb & \LEb & \HEb & \LEb & \HEb \\
			\cline{2-7}
			$\mathcal{L}_0/(\text{erg}\,\text{Mpc}^{-3}\,\text{yr}^{-1})$~\tnote{*} & $7.9{\times}10^{44}$ & $3.7{\times}10^{44}$ & $11.5{\times}10^{44}$ & $4.9{\times}10^{44}$  & $16.1{\times}10^{44}$ & $6.1{\times}10^{44}$ \\
			$\gamma$ & $3.47 \pm 0.03$ & 	$-1.82 \pm 0.11$ & $3.47 \pm 0.03$ &  $-1.92 \pm 0.13$ & $3.45 \pm 0.03$ & $-1.79 \pm 0.14$\\  
			$\log_{10}(R_\text{cut}/\text{V})$ & ${>}19.2$ & $18.12 \pm 0.01$  &  ${>}19.3$ &  $18.15 \pm 0.01$ & $>19.3$ & $18.19 \pm 0.02$ \\ 
			$I_\text{H}$   (\%) & 48.2 & 0.0 & 49.6 & 0.0 &51.6 & 0.0 \\ 
			$I_\text{He}$ (\%) & 14.2 & 25.7   & 10.3 & 21.3 &  7.2 & 16.4 \\ 
			$I_\text{N}$ (\%) & 37.6 & 71.2  & 40.1 & 74.3 & 41.3 & 75.4 \\ 
			$I_\text{Si}$ (\%)  & 0.0 & 0.0 & 0.0 & 0.3 & 0.0 & 4.0 \\ 
			$I_\text{Fe}$ (\%) & 0.0 & 3.1    & 0.0 & 4.1 & 0.0 & 4.2 \\ 
			$a$ & \multicolumn{2}{c|}{$-0.59\pm0.09$} & \multicolumn{2}{c|}{$-0.20\pm0.09$}& \multicolumn{2}{c|}{$\phantom{+}0.08\pm0.09$}\\
			$b$ & \multicolumn{2}{c|}{$\phantom{+}0.9\pm0.3$} & \multicolumn{2}{c|}{$\phantom{+}0.9\pm0.3$} & \multicolumn{2}{c|}{$\phantom{+}1.2\pm0.3$}\\\hline
			$D_J$ ($N_J$)  & \multicolumn{2}{c|}{47.0 (24)}   & \multicolumn{2}{c|}{38.7 (24)} & \multicolumn{2}{c|}{70.5 (24)}  \\ 
			$D_{\xmax}$ ($N_{\xmax}$)  & \multicolumn{2}{c|}{507.2 (329)}    & \multicolumn{2}{c|}{499.8 (329)} & \multicolumn{2}{c|}{493.4 (329)} \\
			$D$ ($N$)  &  \multicolumn{2}{c|}{554.1 (353)}  & \multicolumn{2}{c|}{558.6 (353)}  & \multicolumn{2}{c|}{563.9 (353)}\\   \hline
		\end{tabular}
		\begin{tablenotes}
      \item[*] from $E_\text{min}=10^{17.8}$\,eV.
      \end{tablenotes}
       \end{threeparttable}
		}
	\smallskip
	\caption{The estimated best fit parameters obtained when introducing the nuisance parameters $a$ and $b$ and considering the energy scale uncertainty effect with shifts of one standard deviation in each direction. }
	\label{table:sysXmaxnui}		
\end{table}
The results so obtained are shown in Table~\ref{table:sysXmaxnui}, where the additional $\Dsyst$ is always ${\sim}1$ and included in $D_{\xmax}$. The three cases with no shift and a shift in the energy scale of one standard deviation in each direction are considered. 

\begin{figure}
	\footnotesize
	\centering
    \def\w{0.49}
	\subfigure{\includegraphics[width=\w\linewidth]{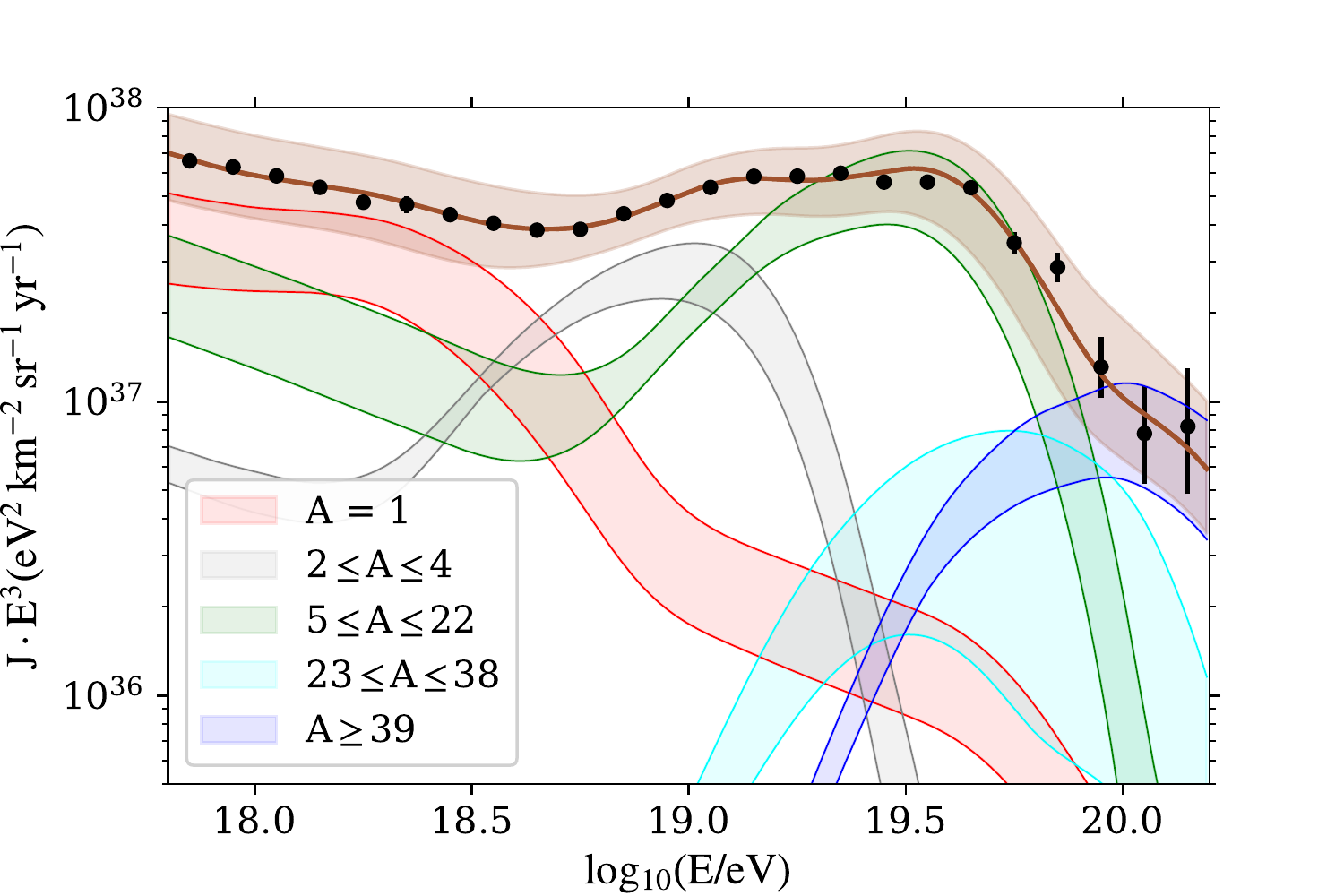}}\hfill
	\subfigure{\includegraphics[width=\w\linewidth]{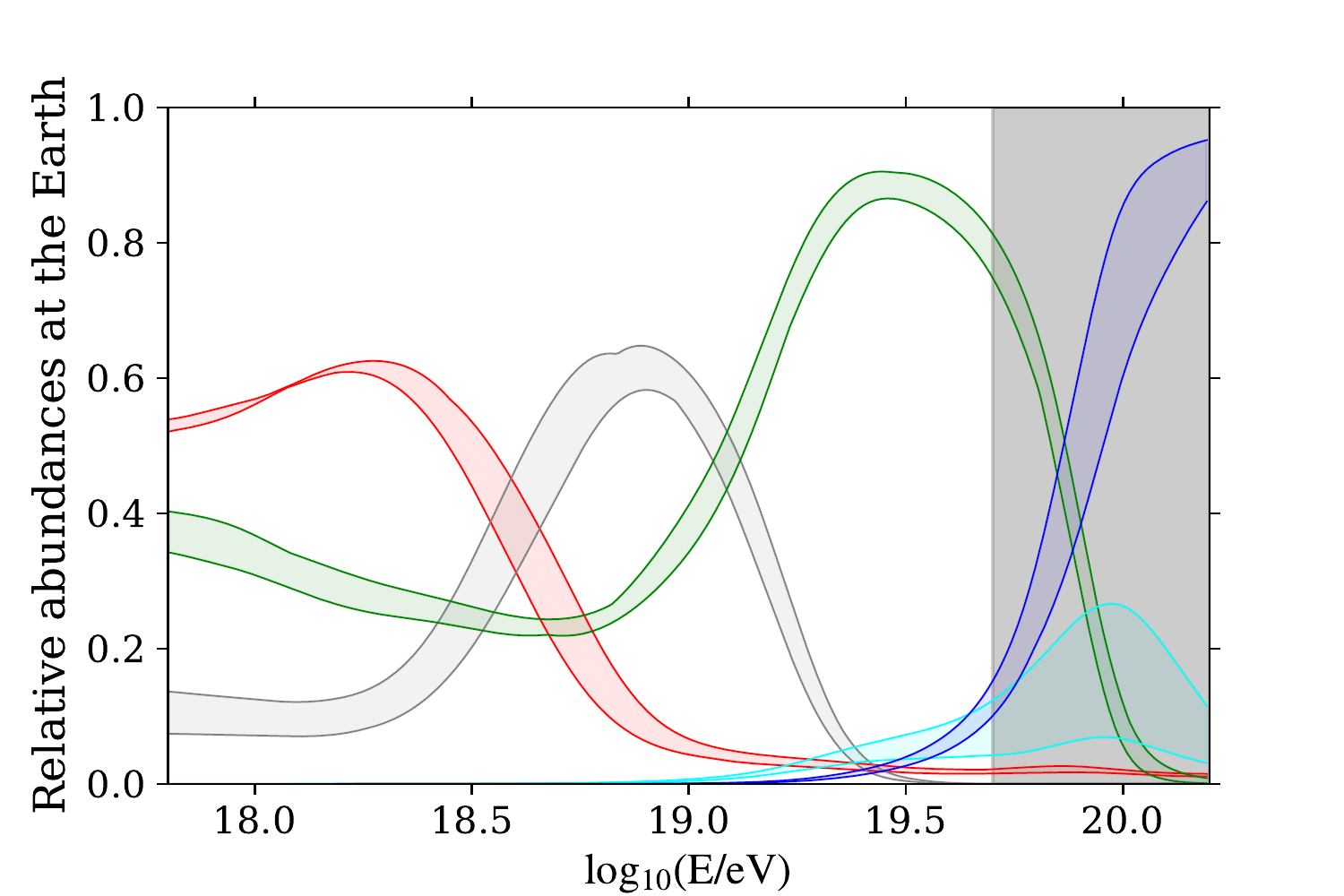}}
	\caption{\emph{Left:} the combined effect of the experimental uncertainties on the energy spectrum. \emph{Right:} the effect on the relative abundances at the top of the atmosphere. The bands represent the variations induced by considering the configurations in Table~\ref{table:sysXmaxnui}. \lastXmax}
	\label{xmaxnui_comparison_2comp}
\end{figure}

\begin{figure}
	\centering
	\includegraphics[width=0.49\linewidth]{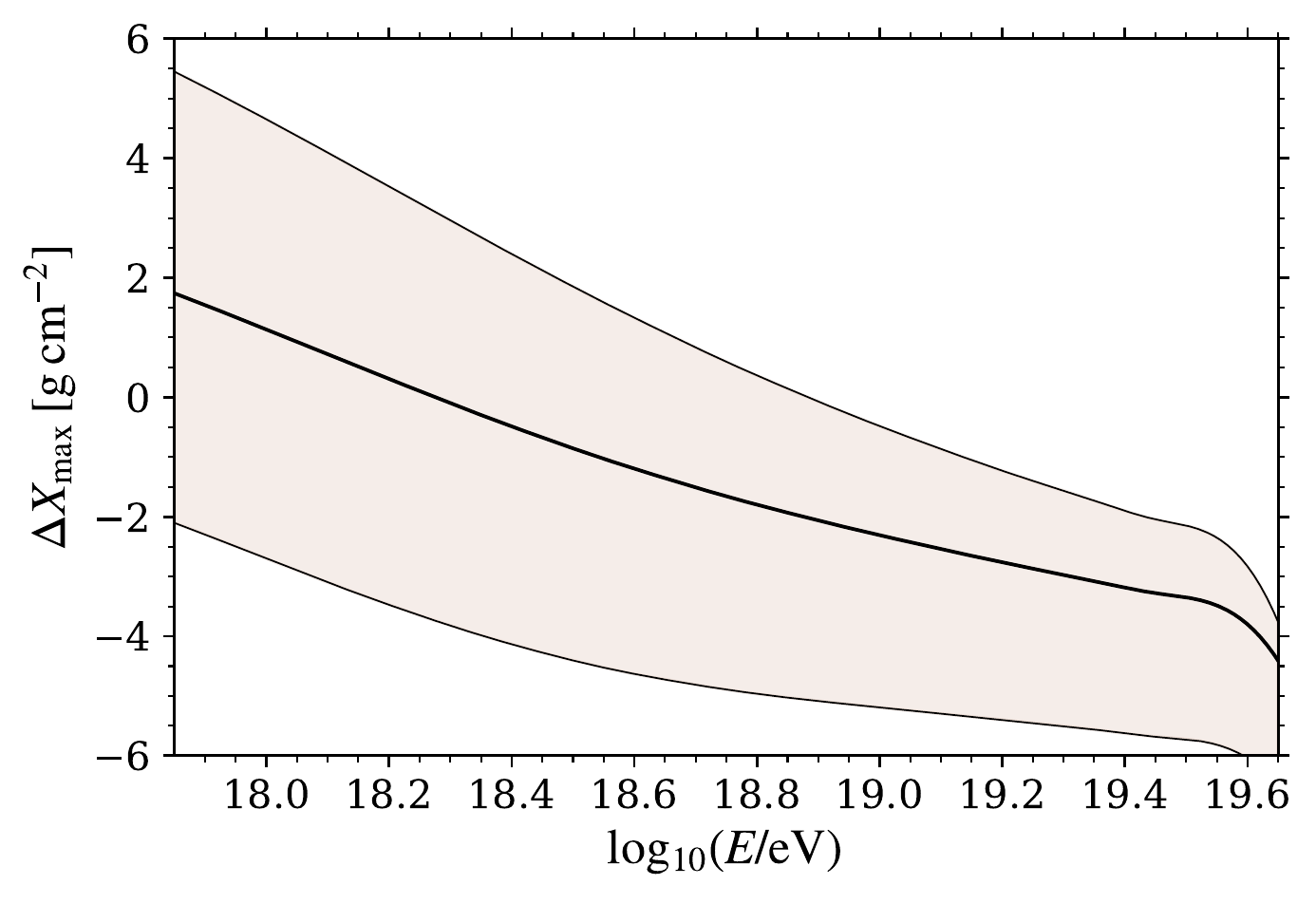}
	\caption{The shifts in the \xmax scale induced by the best-fit parameters $a$ and $b$ listed in Table~\ref{table:sysXmaxnui}. The central black line refers to the case with no shift in the energy scale, and the band represents the effect of shifting the energies by one standard deviation in either direction.}
	\label{fig:sysXmaxnui}
\end{figure} 

The variations on the predicted fluxes at Earth, obtained by considering the configurations of Table~\ref{table:sysXmaxnui}, are shown in Fig.~\ref{xmaxnui_comparison_2comp}.  The rather large uncertainty on the predicted total fluxes (brown band) is mainly due to the $\pm14\%$ shifts in the energy scale, which significantly affects only the estimated source emissivities. On the other hand, the nuisance parameters allow the \xmax distributions to shift to find a better agreement between the predicted and the observed fluxes. Thus the total deviance decreases, but the other estimated best fit parameters are almost unchanged and the modifications on the predicted fluxes and abundances at Earth are rather small. 

Despite some differences in the estimated nominal values, in general the nuisance parameters $a$ and $b$ induce positive (negative) shifts in the \xmax scale at low (high) energies. Alternatively, when the energy scale uncertainty is also considered, they can induce a negative but smaller shift also at low energy. The shifts in the \xmax scale corresponding to the best fit nuisance parameters obtained in the three energy scale configurations of Table~\ref{table:sysXmaxnui} are shown in Fig.~\ref{fig:sysXmaxnui}.

Note that in principle the same approach could be extended also to the treatment of the energy scale uncertainty by introducing an additional nuisance parameter. However, considering that the energy scale systematic uncertainties have a subdominant effect on the goodness-of-fit, as shown in Appendix~\ref{sysexp}, we chose to explore this more complete approach only for the \xmax scale uncertainty.

Besides, we also verified that the effects of uncertainties in the acceptance and resolution~\cite{xmax_dist} of the \xmax data set 
are negligible: very small differences on the deviance and almost no changes in the fit parameters are observed when such uncertainties are included as nuisance parameters. Hence, these effects are not shown here and will not be considered further in this work.

\subsection{Uncertainties from propagation and shower models}
\label{sysmodel}
The propagation models and the HIM are other sources of systematic uncertainties; we explored their effects by repeating the fit considering
different combinations of them with respect to those used in the reference configuration. As regards the photodisintegration, we tested the PSB model, that neglects
photodisintegration channels in which alpha particles rather than single nucleons are
ejected. The cross sections for such channels are difficult to measure, and the few available data~\cite{afanasev} appear to be overestimated in \TALYS\ by around an order of magnitude, so neglecting such channels altogether as done in PSB is not necessarily less accurate~\cite{AlvesBatista:2015jem}. Besides, as concerns the EBL spectrum and evolution, we tested also the Dom\'inguez model, which has a higher spectral energy density in the far infrared with respect to the Gilmore one. 
Regarding the HIM, we verified that \QGSJET\ cannot properly describe our data ($D \gtrsim 1000$ in all cases), and is thus excluded from this analysis. Instead of fixing a single HIM, we allow for the possibility to describe our data with an
intermediate model between \EPOS\ and \SIBYLL\ by introducing an additional
nuisance parameter $\delta_\text{HIM}$, limited between 0 and 1. In this way each HIM-dependent Gumbel
parameter is interpolated as alpha as $\alpha_\text{HIM} = \delta_\text{HIM}\,\alpha_\text{\EPOS}+ (1-\delta_\text{HIM})\,\alpha_\text{\SIBYLL}$,\footnote{For a given primary mass and energy, the Gumbel distribution parameters $\mu,\sigma,\lambda$ are linear functions of the HIM-dependent parameters $a_i,b_i,c_i$, so it makes no difference whether we interpolate the former or the latter.} so that $\delta_\text{HIM}=0$ corresponds to ``pure'' \SIBYLL\ and $\delta_\text{HIM}=1$ to ``pure'' \EPOS. \footnote{This is just an approximation, as the ``true'' model is not necessarily a linear interpolation between \EPOS\ and \SIBYLL.}

\begin{table}
	\renewcommand\arraystretch{1.4}
	\centering\small
	\begin{threeparttable}
	\begin{tabular}{| r | c  c | c  c |}
	\hline
		\multicolumn{1}{|c}{} & \multicolumn{2}{|c|}{\TALYS} & \multicolumn{2}{|c|}{PSB} \\
		\hline
		\multicolumn{1}{|l|}{Gilmore EBL}& \LEb & \HEb & \LEb & \HEb \\
		\hline
		$\mathcal{L}_0/(10^{44}\,\text{erg}\,\text{Mpc}^{-3}\,\text{yr}^{-1})$~\tnote{*} & $11.4$ & $5.1$ &$11.1$ & $4.9$ \\
		$\gamma$ & $3.52 \pm 0.03$ & $-1.99 \pm 0.11$ & $3.51 \pm 0.03$ &  $-1.89 \pm 0.18$ \\
		$\log_{10}(R_\text{cut}/\text{V})$ & ${>}19.4$ & $18.15 \pm 0.01$  &  $>19.5$ & $18.16 \pm 0.02$ \\ 
		$I_\text{H}$   (\%) & 48.7 & 
		0.0 & 49.1 & 0.2 \\ 
		$I_\text{He}$ (\%) &  7.3 & 23.6 & 11.1 & 48.3 \\ 
		$I_\text{N}$ (\%) & 44.0 & 72.1 & 39.8 & 41.5 \\ 
		$I_\text{Si}$ (\%) & 
		0.0 & 1.3  &  
		0.0 & 8.5 \\ 
		$I_\text{Fe}$ (\%) & 0.0 & 3.1  & 0.0 & 1.5 \\
		$\delta_\text{HIM}$ & \multicolumn{2}{c|}{$1.0$ (limit)} & \multicolumn{2}{c|}{$0.96  ^{+0.04}_{-0.12}$}
		\\
		\hline
		$D_J$ ($N_J$) & \multicolumn{2}{c|}{56.6 (24)}  & \multicolumn{2}{|c|}{50.7 (24)} \\ 
		$D_{\xmax}$ ($N_{\xmax}$) & \multicolumn{2}{c|}{516.5 (329)}  & \multicolumn{2}{c|}{529.0 (329)} \\
		$D$ ($N$) & \multicolumn{2}{c|}{573.1 (353)}  & \multicolumn{2}{c|}{579.7 (353)} \\
		\hline
		\hline
		\multicolumn{1}{|l|}{Dom\'inguez EBL}& \LEb & \HEb & \LEb & \HEb \\
		\hline
		$\mathcal{L}_0/(10^{44}\,\text{erg}\,\text{Mpc}^{-3}\,\text{yr}^{-1})$~\tnote{*} & $9.2$ & $7.3$ & $8.7$ & $7.3$ \\
		$\gamma$ & $3.67 \pm 0.06$ & $-0.87 \pm 0.08$ & $3.71 \pm 0.06$ & $-0.85 \pm 0.08$ \\
		$\log_{10}(R_\text{cut}/\text{V})$ &  $18.01 \pm 0.06$ &  $18.23 \pm 0.01$ &  $18.00 \pm 0.07$ &  $18.22 \pm 0.01$ \\
		$I_\text{H}$   (\%) & 41.4 & 0.0 & 42.4 & 0.0 \\
		$I_\text{He}$  (\%) & 7.4 & 17.2 & 8.6 & 48.2\\
		$I_\text{N}$   (\%) & 51.6 & 78.0 & 49.0 & 42.1 \\
		$I_\text{Si}$  (\%) & 
		0.0& 2.1 & 
		0.0 & 8.2 \\
		$I_\text{Fe}$  (\%) & 0.0 & 2.7 & 0.0 & 1.6 \\
		$\delta_\text{HIM}$  & \multicolumn{2}{c|}{$0.88  \pm 0.11$} & \multicolumn{2}{c|}{$0.88  \pm 0.11$} \\
		\hline
		$D_J$ ($N_J$) & \multicolumn{2}{c|}{42.5 (24)} & \multicolumn{2}{c|}{39.9 (24)} \\
		$D_{\xmax}$ ($N_{\xmax}$) & \multicolumn{2}{c|}{561.6 (329)} & \multicolumn{2}{c|}{568.6 (329)} \\
		$D$ ($N$)  & \multicolumn{2}{c|}{604.2 (353)}  & \multicolumn{2}{c|}{608.5 (353)}  \\
		\hline
	\end{tabular}
		\begin{tablenotes}
      \item[*] from $E_\text{min}=10^{17.8}$\,eV.
      \end{tablenotes}
       \end{threeparttable}
	\smallskip
	\caption{Best fit results obtained by using different combinations of propagation models. The uncertainty due to the HIM choice is considered by fitting the nuisance parameter $\delta_\text{HIM}$. 
	}
	\label{table:propmodels}	
\end{table}

\begin{figure}
	\footnotesize
	\centering
	\def\w{0.49}
	\subfigure{\includegraphics[width=\w\linewidth]{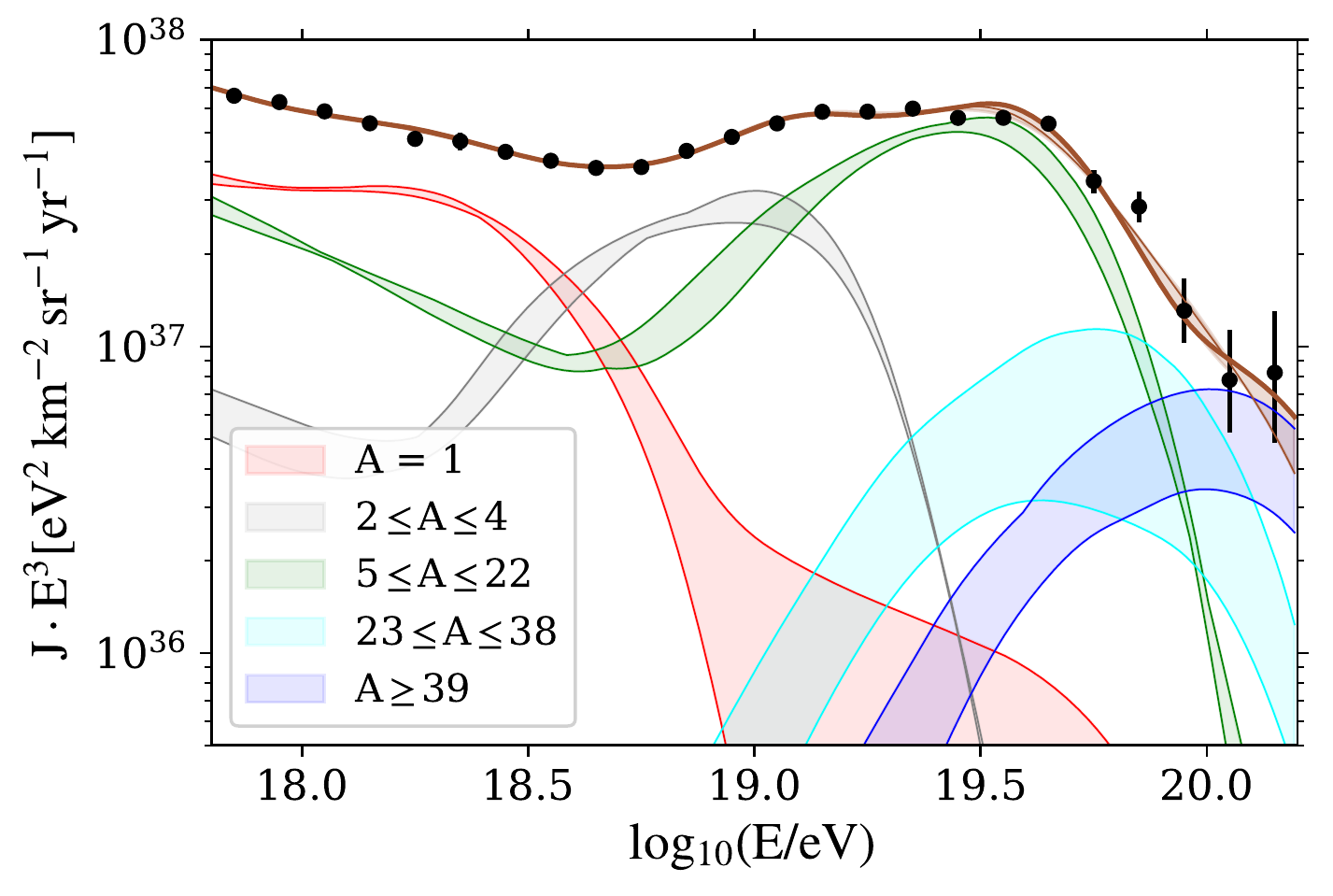}}\hfill
	\subfigure{\includegraphics[width=\w\linewidth]{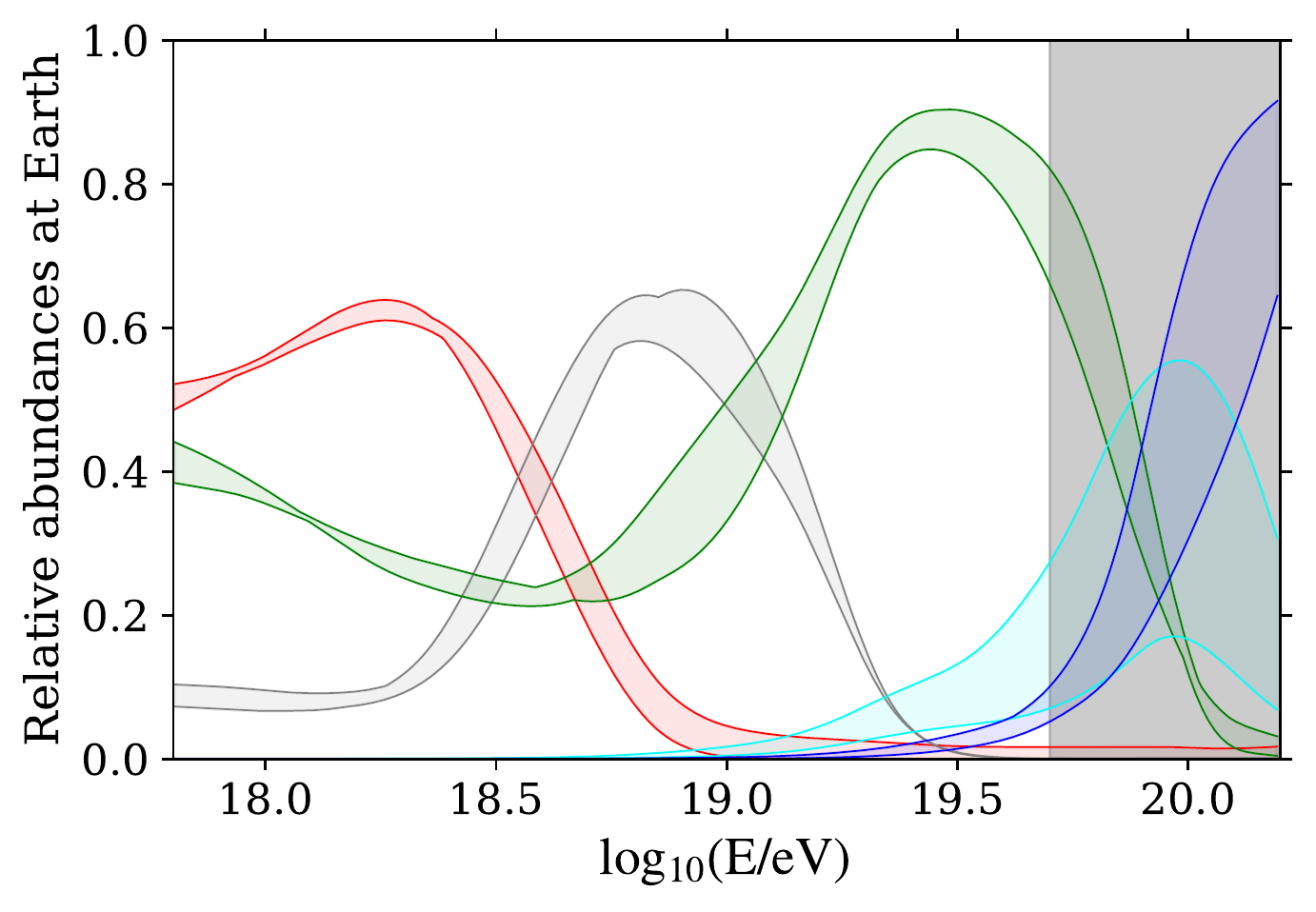}}
	\caption{\emph{Left:} the effect of the uncertainties from models on the energy spectrum. \emph{Right:} the effect on the relative abundances at the top of the atmosphere. The bands represent the maximal variations given by the results in Table~\ref{table:propmodels}. \lastXmax}
	\label{models_comparison_2comp}
\end{figure}

The results thus obtained are summarised in Table~\ref{table:propmodels} and their effect on the predicted fluxes at Earth is shown in Fig.~\ref{models_comparison_2comp}.

\begin{table}
	\renewcommand\arraystretch{1.4}
	\centering\small
	\begin{threeparttable}
	 
	\begin{tabular}{| r | c  c | c  c |}
	\hline
		\multicolumn{1}{|l}{\TALYS} & \multicolumn{2}{|c|}{\EPOS} & \multicolumn{2}{|c|}{\SIBYLL} \\
		\cline{2-5} 
		\multicolumn{1}{|l|}{Gilmore EBL}& \LEb & \HEb & \LEb & \HEb \\
		\hline
		$\mathcal{L}_0/(10^{44}\,\text{erg}\,\text{Mpc}^{-3}\,\text{yr}^{-1})$~\tnote{*} & $11.4$ & $5.1$ & $10.8$ & $4.9$ \\
		$\gamma$ & $3.52 \pm 0.03$ & $-1.99 \pm 0.11$ & $3.40 \pm 0.02$ &  $-1.30 \pm 0.19$ \\
		$\log_{10}(R_\text{cut}/\text{V})$ & ${>}19.4$ & $18.15 \pm 0.01$  &  $18.26 \pm 0.05$ & $18.19 \pm 0.02$ \\ 
		$I_\text{H}$   (\%) & 48.7 & 
		0.0 & 15.6 & 0.0 \\ 
		$I_\text{He}$ (\%) &  7.3 & 23.6 & 46.2 & 20.9 \\ 
		$I_\text{N}$ (\%) & 44.0 & 72.1 & 38.2 & 70.7 \\ 
		$I_\text{Si}$ (\%) & 
		0.0 & 1.3  &  
		0.0 & 5.4 \\ 
		$I_\text{Fe}$ (\%) & 0.0 & 3.1  & 0.0 & 3.0 \\
		\hline
		$D_J$ ($N_J$) & \multicolumn{2}{c|}{56.6 (24)}  & \multicolumn{2}{|c|}{42.7 (24)} \\ 
		$D_{\xmax}$ ($N_{\xmax}$) & \multicolumn{2}{c|}{516.5 (329)}  & \multicolumn{2}{c|}{592.2 (329)} \\
		$D$ ($N$) & \multicolumn{2}{c|}{573.1 (353)}  & \multicolumn{2}{c|}{634.9 (353)} \\
		\hline
	\end{tabular}
	\begin{tablenotes}
      \item[*] from $E_\text{min}=10^{17.8}$\,eV.
      \end{tablenotes}
       \end{threeparttable}
	\smallskip
	\caption{Comparison between the best fit results obtained by using \EPOS\ and \SIBYLL\ as the HIM in the \TALYS+Gilmore configuration.}
	\label{table:hadronicmodels}	
\end{table}

Regardless of the propagation models configuration, our data appear to be better described by pure \EPOS\ or by intermediate models much closer to \EPOS\ than to \SIBYLL, making the HIM choice the dominant uncertainty among the ones from models in terms of predictions at Earth. For example, from Table~\ref{table:hadronicmodels} it is clear that a significant worsening of the deviance is obtained when \SIBYLL\ is assumed as the HIM and the reference propagation models configuration is used. 
As concerns the propagation models effects, even if the impact on the deviance and on the predicted fluxes at the Earth is smaller, some changes in the best fit parameters at the sources are observed, which are in agreement with what is expected to compensate the differences in the propagation to produce similar fluxes at the Earth. 
When the photodisintegration cross sections are modelled with PSB instead of \TALYS, the absence of secondary alpha-particle production during propagation must be compensated by a larger amount of helium ejected at the sources.
When the EBL spectrum is based on the Dom\'inguez model, the \LEb\ component is suppressed at lower energy with an upper-constrained value of $R_\text{cut}$ to compensate the larger amount of secondary particles below the ankle provided by the \HEb\ component. 
The lowest deviance is obtained in the \TALYS+Gilmore
configuration. However, the impact of changing the propagation models on the deviance and on the predicted fluxes at Earth is encompassed by the effect of the experimental systematic uncertainties. 

\section{Cosmological evolution of sources}
\label{sec:variations}
\label{SourceEvolution}

\subsection{Impact on UHECR parameters}

We repeated the fit considering, for each population of sources, three different models for the cosmological evolution of the source emissivity, parameterised as $\propto (1+z)^m$, namely $m = +5$, $+3$ and $-3$, in addition to the no-evolution ($m=0$) case considered so far. 
As in the previous section, the study of variations is restricted to \textsc{Scenario 2}, which is the most general one and does not imply possible mutual dependencies between the two extragalactic components that could constrain our assumptions on the source evolution.
UHECRs are simulated up to $z_\text{max}=10$; however, due to the energy losses in the propagation, practically all nuclei reaching us with energies in the range we are fitting ($E \ge 10^{17.8}$\,eV) originate from $z \lesssim 3$, and in particular those with $E \gtrsim 10^{18.4}$\,eV from $z \le 1$.

\begin{figure}
	\centering
	\includegraphics[width=0.6\linewidth]{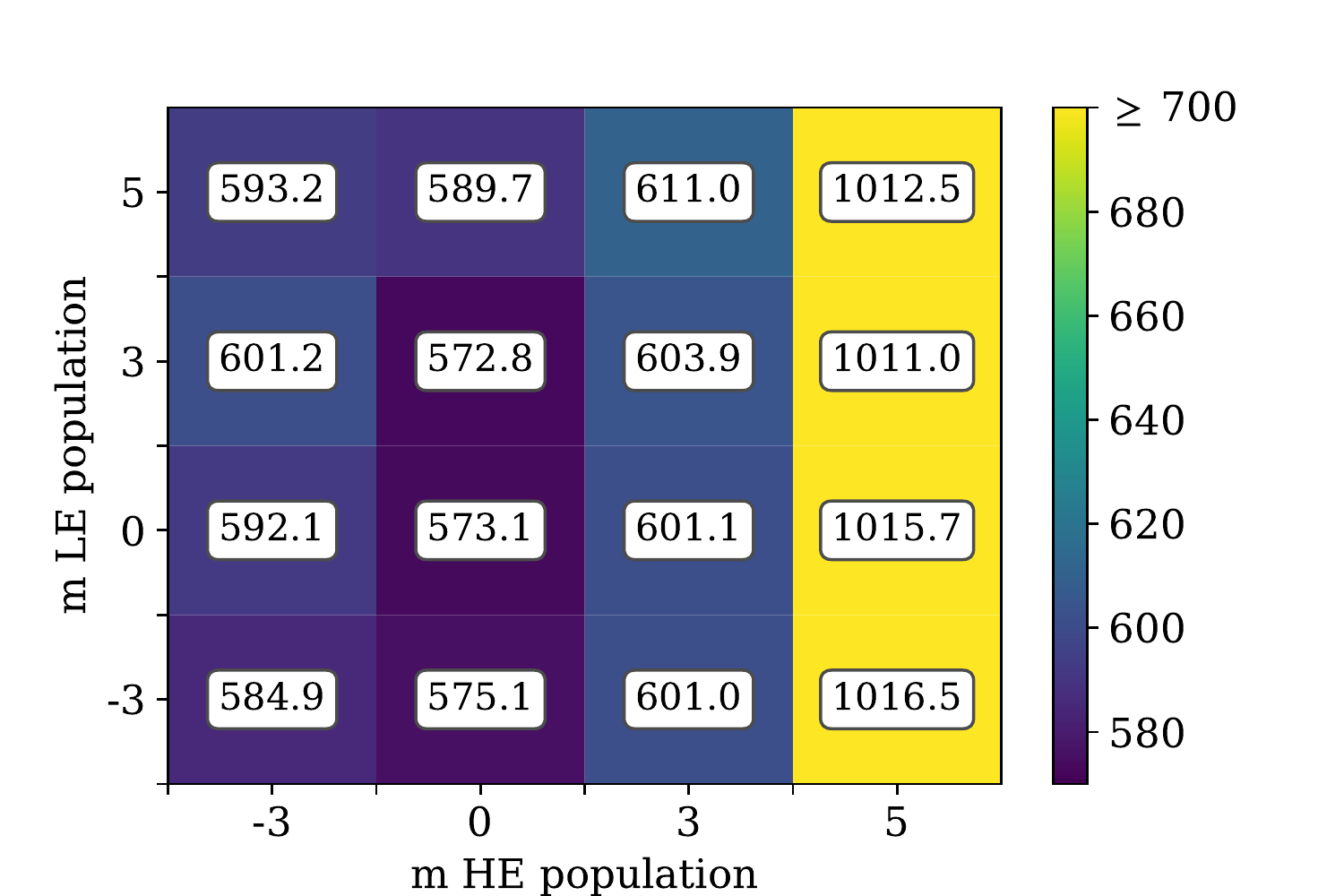}
	\caption{The total deviance is colour-coded as in the right bar and shown for all the possible combinations of source evolution of the two populations. The values of $m$ are shown on the axes.} 
	\label{fig_sourceevol1}
\end{figure}

\begin{figure}
	\footnotesize
	\centering
	\def\h{0.37}
	\subfigure[]{\includegraphics[height=\h\linewidth]{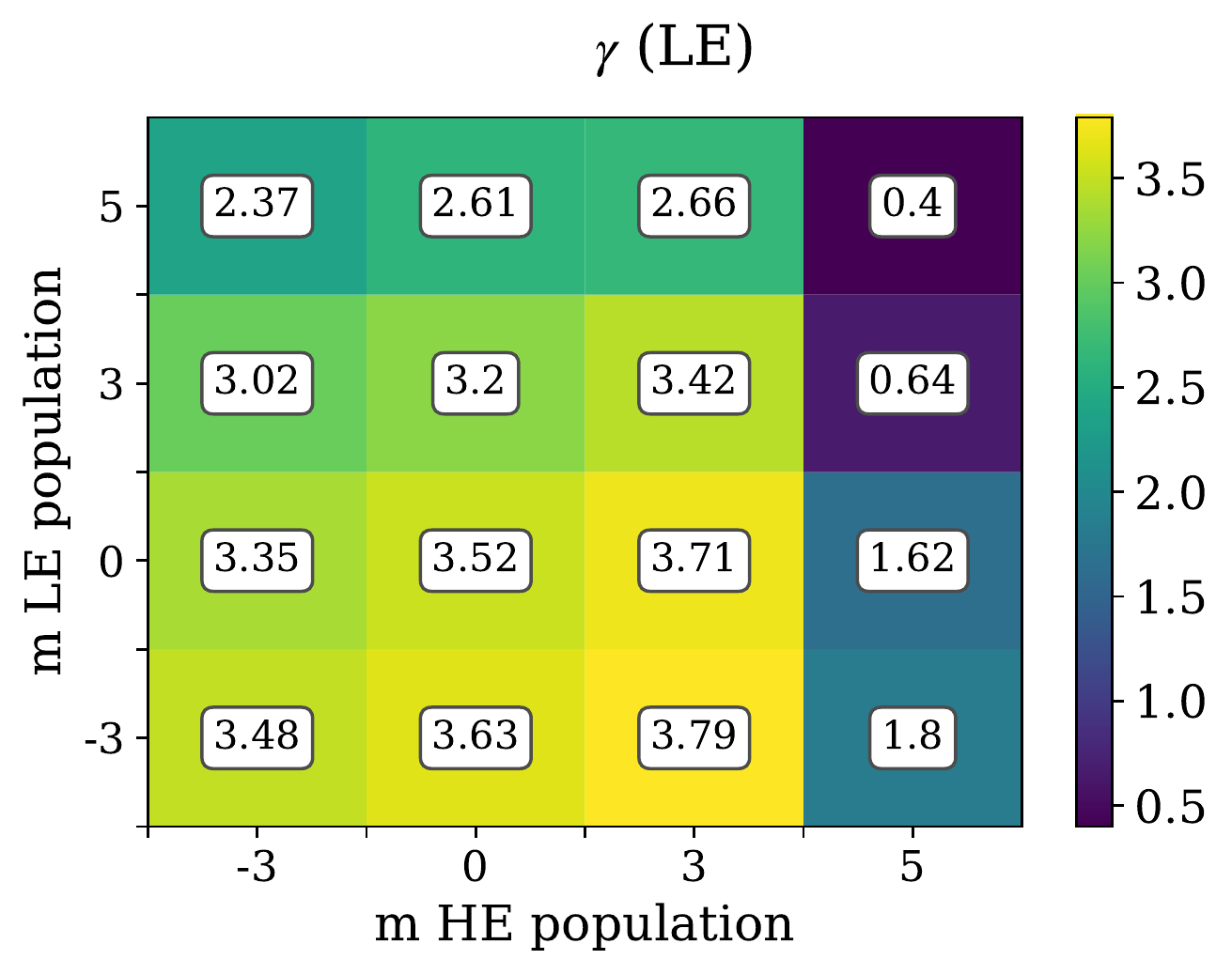}\label{source1}}\hfill
	\subfigure[]{\includegraphics[height=\h\linewidth]{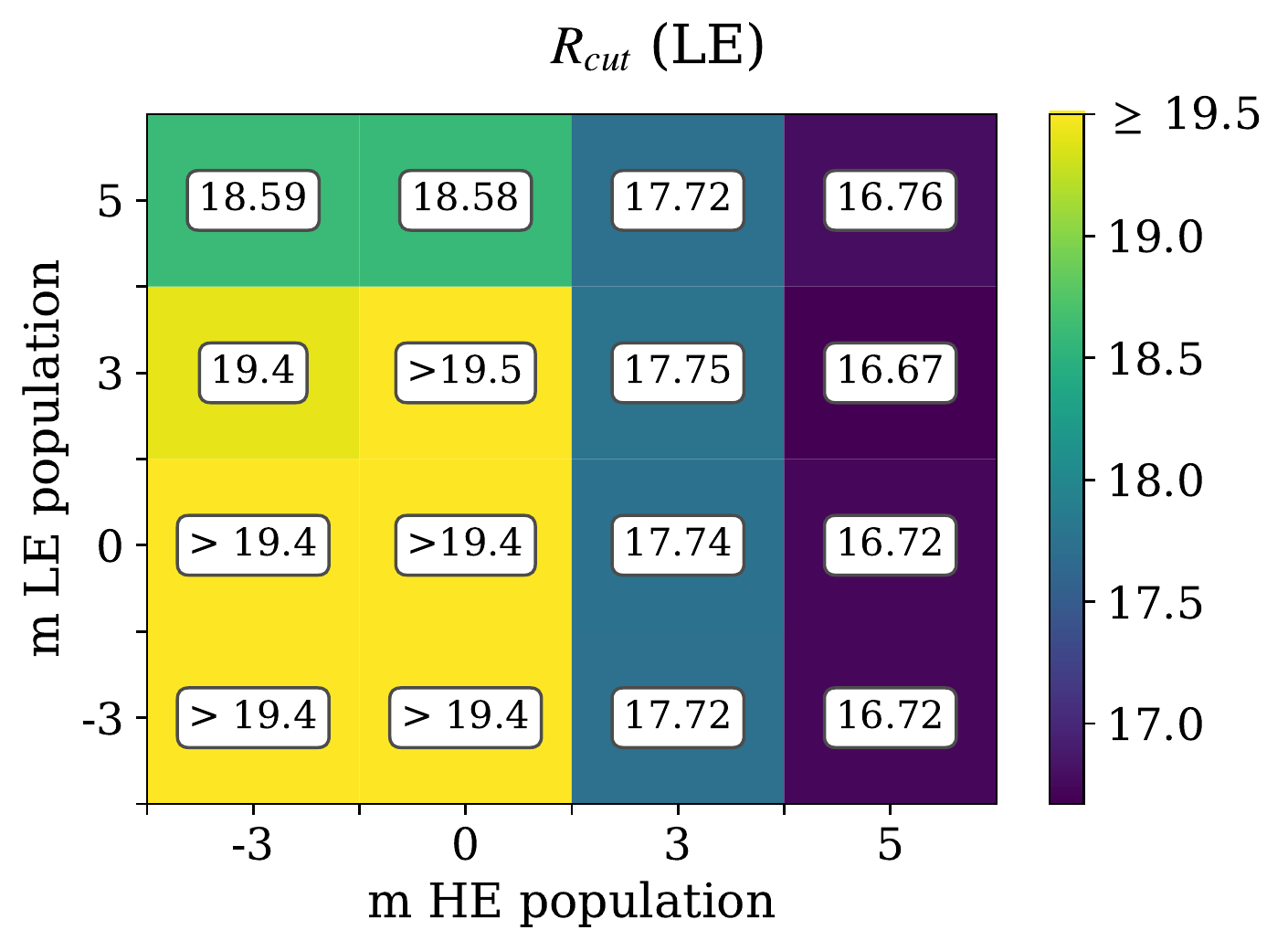}\label{source2}}
	\\
	\subfigure[]{\includegraphics[height=\h\linewidth]{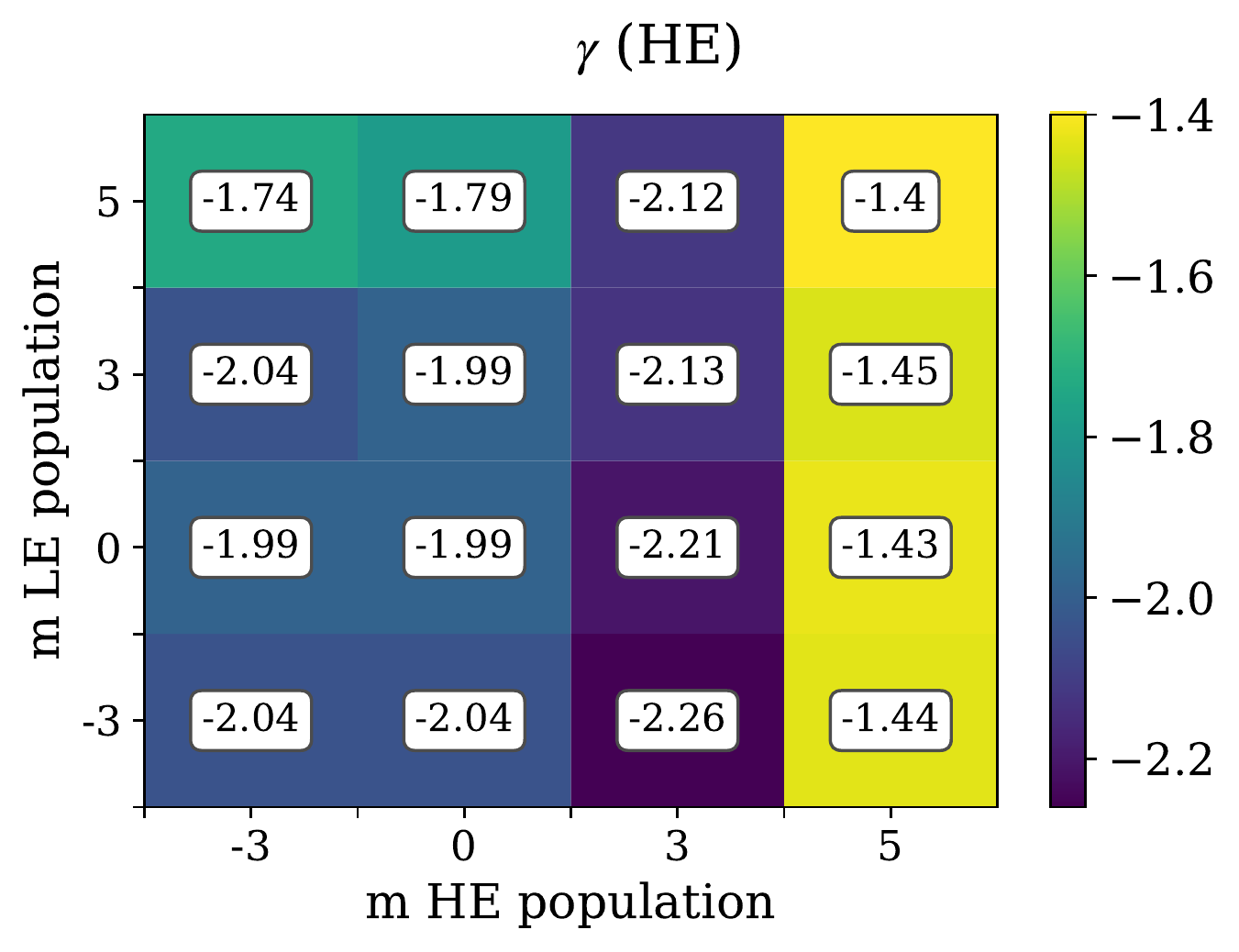}\label{source3}}\hfill
	\subfigure[]{\includegraphics[height=\h\linewidth]{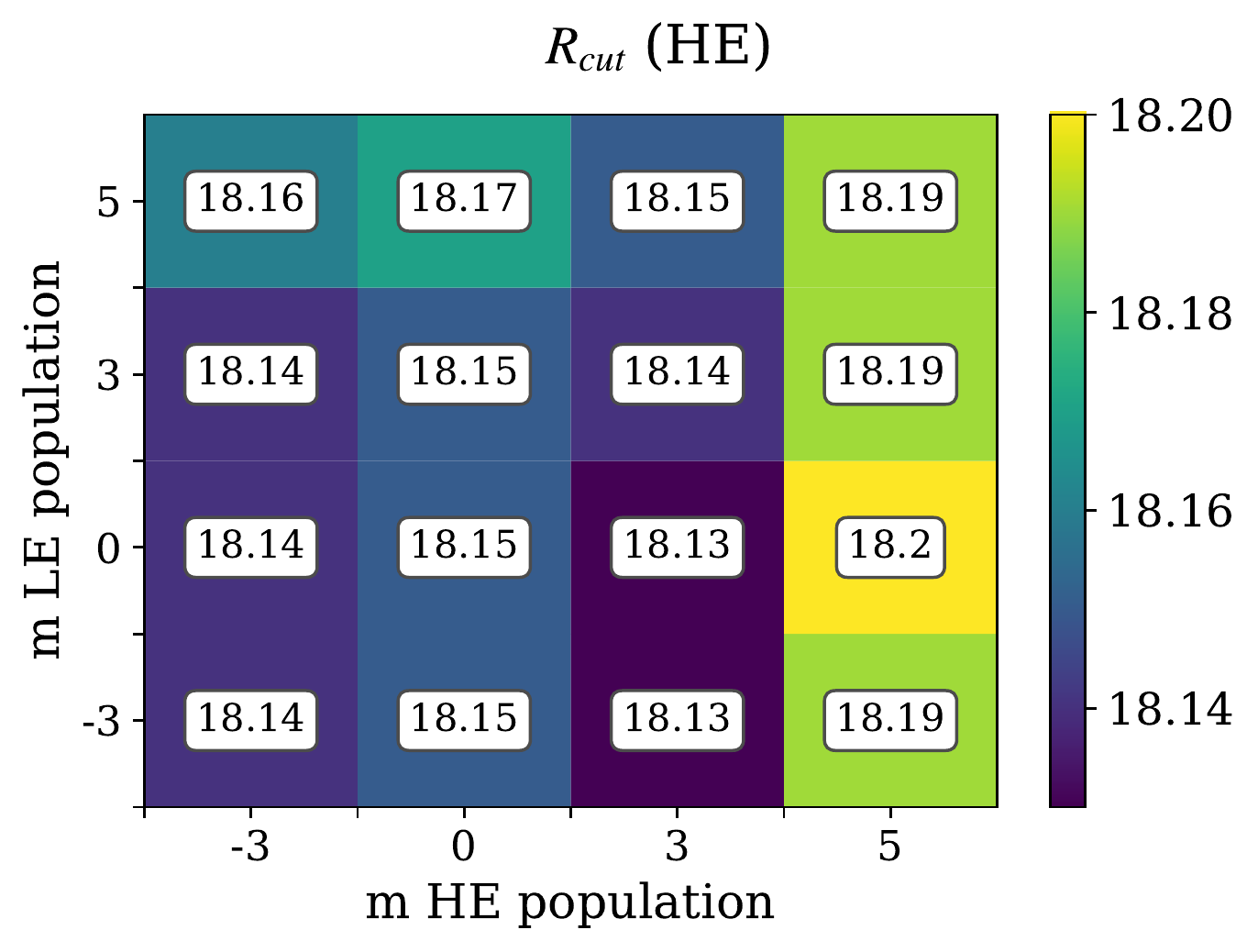}\label{source4}}
	\caption{The estimated spectral parameters of the \LEb~(top row) and \HEb~(bottom row) component are colour-coded as in the corresponding right bars and are shown as a function of the source evolution of the two populations.}
	\label{sourceevol_spectralpars}
\end{figure}

At low redshifts ($z \lesssim 1$), a strong positive ($m=5$) evolution could be associated to jetted AGN (high-luminosity BL Lacs and FSRQs) observed in gamma rays~\cite{Ajello:2013lka} or to non-jetted AGN such as high-luminosity Seyfert galaxies~\cite{Ueda:2014tma}. A weaker positive evolution ($m=3$) can be connected to the SFR evolution~\cite{Madau:2014bja}. The case of no-evolution ($m=0$) can be instead associated to the stellar-mass density~\cite{Madau:2014bja}, non-jetted AGN (low-luminosity Seyfert galaxies observed in X-rays \cite{Ueda:2014tma})
and jetted AGN (intermediate-luminosity BL Lacs and FSRQs~\cite{Ajello:2013lka}). Negative evolutions ($m=-3$) can trace jetted AGN (low-luminosity BL Lacs observed in gamma rays~\cite{Ajello:2013lka}) or non-jetted AGN (radio-galaxies observed in gamma rays~\cite{Fukazawa:2022gwm}), as well as the evolution with redshift of tidal disruption events (TDEs)~\cite{Kochanek:2016zzg}.  
At higher redshifts ($z \gtrsim 1$), the evolution of some of these classes of sources is uncertain. In this work we show results using $(1+z)^m$ with constant $m$ in the entire redshift range, but we have verified that other possibilities for the behaviour of the evolution at $z > 1$ have only a small impact on the \LEb~component, not affecting our main conclusions, and a completely negligible effect on the \HEb~component.  An exception to this is the flux of secondary neutrino and gamma rays, discussed in Section~\ref{sec:neutrinos}.

Since the \LEb\ and \HEb\ populations might be accelerated in different classes of sources, they could have different source evolutions. Hence, we consider all sixteen possible pairs of evolutions among $m \in \{-3, 0, +3, +5\}$. 
Our results are summarised in Fig.~\ref{fig_sourceevol1} for the total deviance and in Fig.~\ref{sourceevol_spectralpars} for the best-fit parameters.

\begin{figure}
	\centering
	\def\w{0.49}
	\subfigure[]{\includegraphics[width=\w\linewidth]{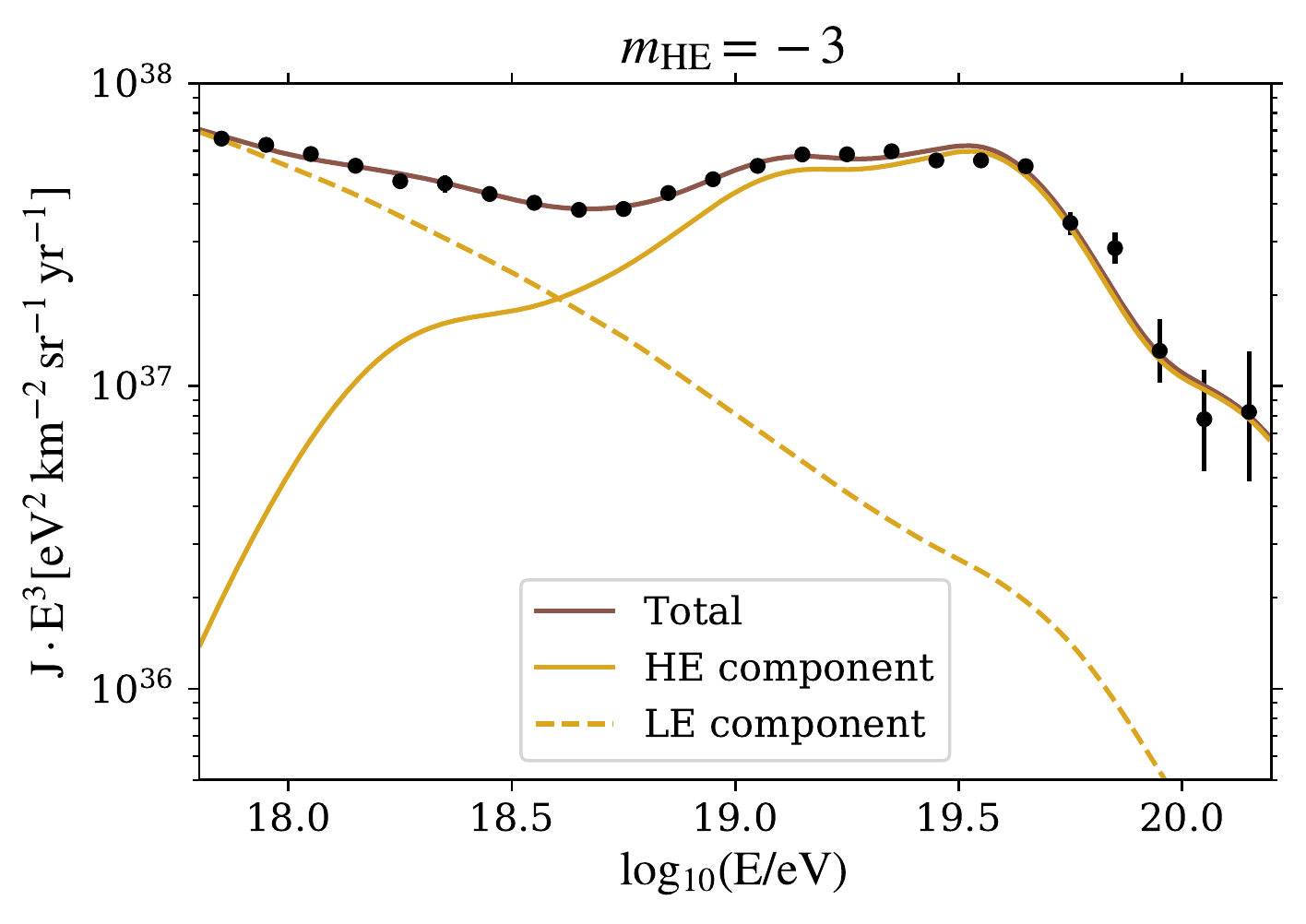}\label{2comp_sourceevol1}}\hfill
	\subfigure[]{\includegraphics[width=\w\linewidth]{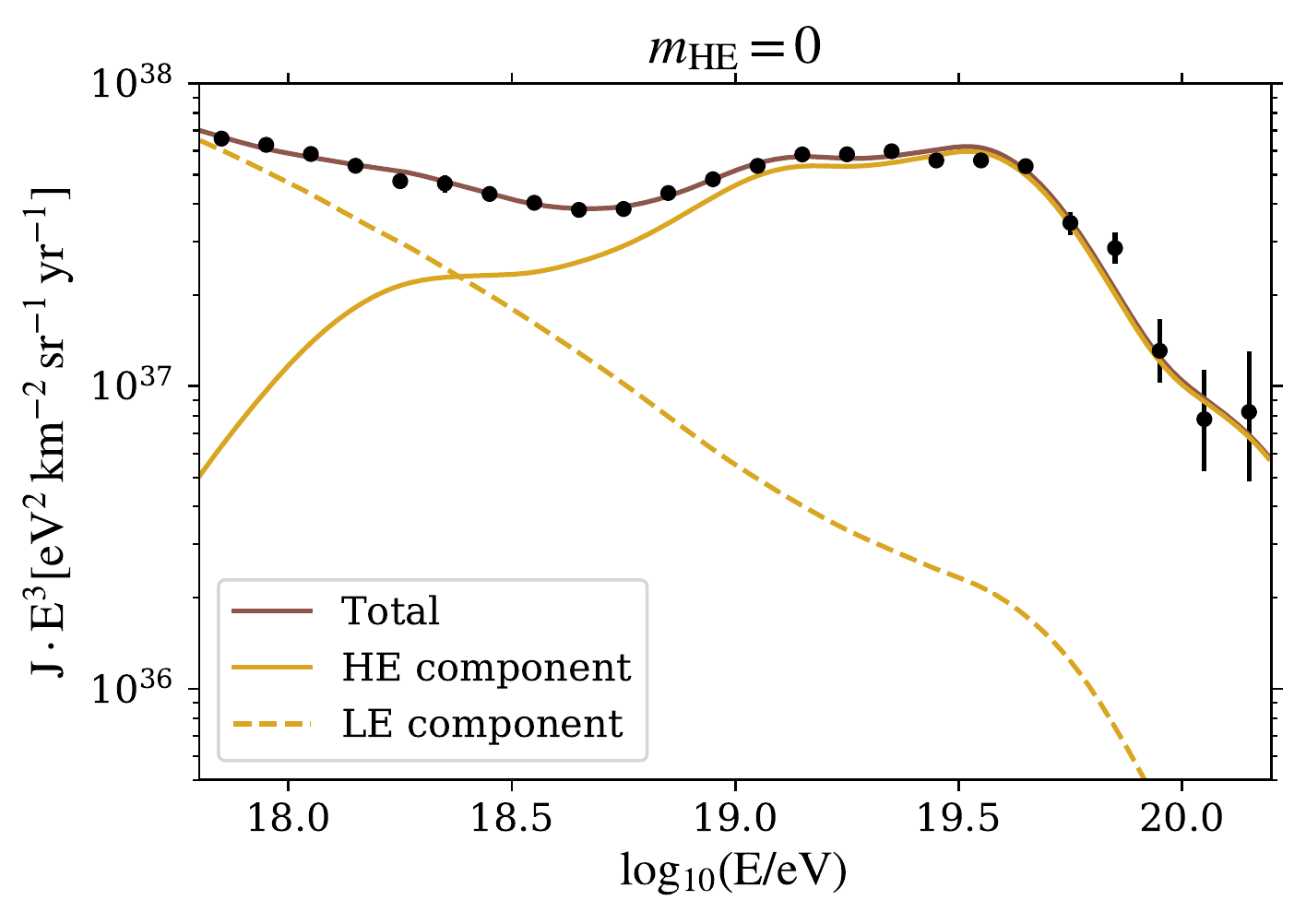}\label{2comp_sourceevol2}}
	\\
	\subfigure[]{\includegraphics[width=\w\linewidth]{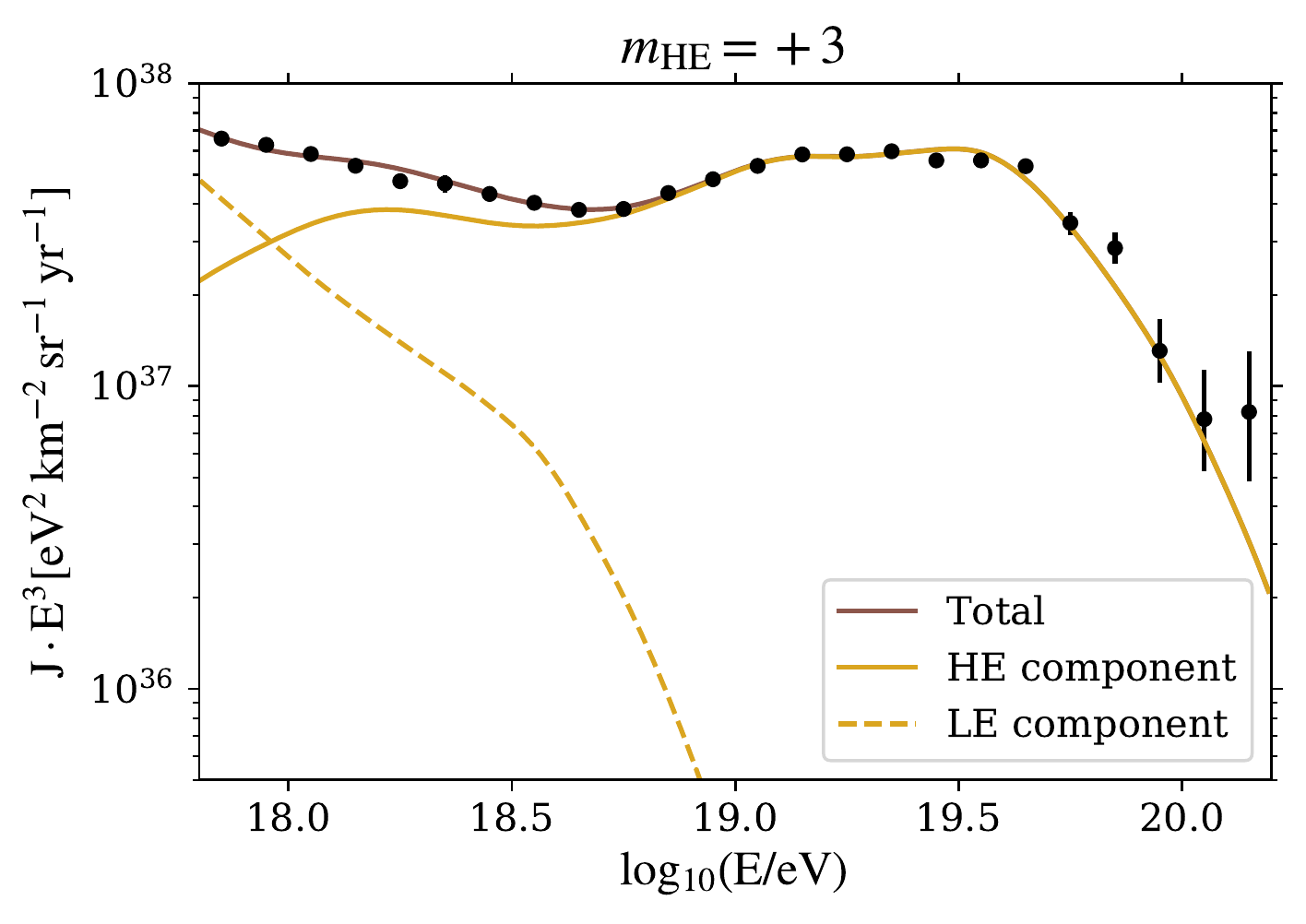}\label{2comp_sourceevol3}}\hfill
	\subfigure[]{\includegraphics[width=\w\linewidth]{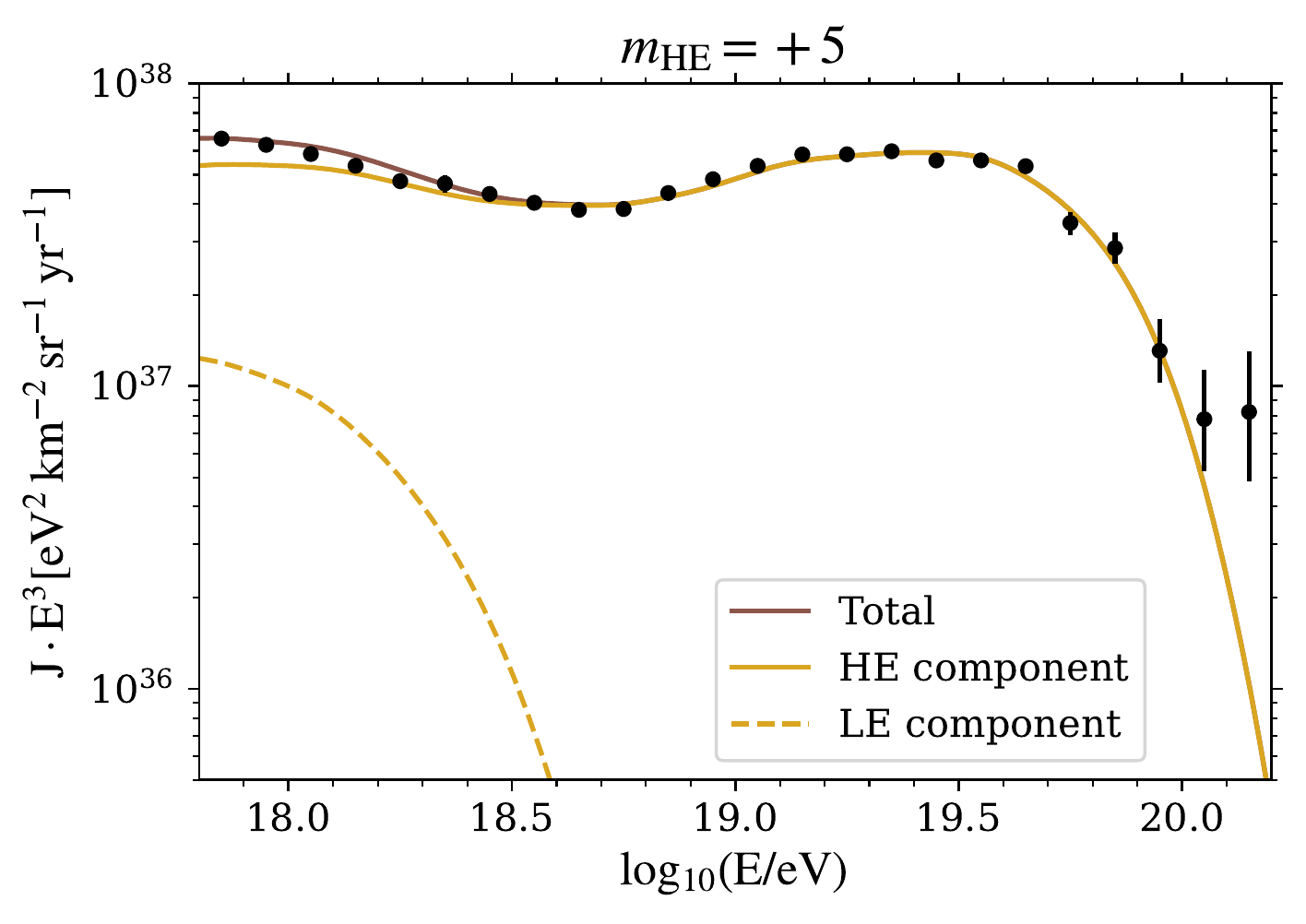}\label{2comp_sourceevol4}}
	\caption{The contributions at Earth of the two extragalactic components: for each considered source evolution of the \HEb\ component ($m=-3$~\subref{2comp_sourceevol1}, $m=0$~\subref{2comp_sourceevol2}, $m=3$~\subref{2comp_sourceevol3}, $m=5$~\subref{2comp_sourceevol4}), only the \LEb\ source evolution providing the lowest deviance is chosen as an example.}
	\label{2comp_sourceevol}
\end{figure}

A positive (negative) evolution means that particles were on average accelerated longer ago (more recently) than in the no-evolution scenario, and hence had the time to undergo more (fewer) interactions in intergalactic space.  The effects are more noticeable for the \HEb\ population, as interactions are more frequent at high energies.  This mostly affects the flux of secondary protons and helium produced at energies around the ankle, and it is at the origin of the observed anti-correlation between $m$ and the estimated spectral index, as found already in~\cite{Taylor:2015rla,combinedfit,Heinze:2015hhp,AlvesBatista:2018zui,Heinze2019}.  In Fig.~\ref{2comp_sourceevol}, one can appreciate the way the contribution of the \HEb\ component to the all-particle spectrum around the ankle increases with its evolution, and how the cutoff of the \LEb\ component consequently needs to be lowered (for this figure, the LE evolution providing the lowest deviance is shown).   In the case of a strong positive ($m=5$) evolution of the HE component, its secondary flux at ankle energies exceeds the observed all-particle spectrum, so that no good fit of the data is possible ($D \sim 1000$). Such scenarios (corresponding to the last column in the plots of Figs.~\ref{fig_sourceevol1} and~\ref{sourceevol_spectralpars}) will not be considered further. In the past they were mostly used for pure-proton composition if the energy range across the ankle was taken into account, as for instance in \cite{Gelmini:2011kg,Heinze:2015hhp}. 

In the case of a weak positive ($m=3$) evolution of the \HEb\ component, its secondaries around the ankle saturate the observed spectrum, so that a good fit is only possible if the \LEb\ component does not provide any more particles at these energies, requiring it to have an extremely soft ejection spectrum (Fig.~\ref{sourceevol_spectralpars}\subref{source1}) with a very low rigidity cutoff (Fig.~\ref{sourceevol_spectralpars}\subref{source2}).
In the case of no or negative evolution of the \HEb\ component, its secondaries are less than the observed all-particle spectrum, so that a contribution from the \LEb\ component is also needed, as was shown in Section~\ref{scenarioB}.  The scenarios with no evolution for the \HEb\ population appear to be favoured overall (Fig.~\ref{fig_sourceevol1}), though acceptable fits can also be found with a weak evolution ($m = \pm 3$).

The effects of the cosmological evolution are smaller in the case of the \LEb\ component.  A positive (negative) evolution requires a hardening (softening) of the ejection spectrum to compensate the larger (smaller) amount of low-energy particles (Fig.~\ref{sourceevol_spectralpars}\subref{source1}), and a strong positive evolution also requires a lower rigidity cutoff (Fig.~\ref{sourceevol_spectralpars}\subref{source2}). The deviance (Fig.~\ref{fig_sourceevol1}) appears to slightly favour scenarios with a weak or no evolution for this component, but is still acceptable with a strong one.
As for the ejection spectral parameters of the \HEb\ population, their best-fit values stay nearly unchanged among all scenarios with acceptable deviances, as shown in Figs.~\ref{sourceevol_spectralpars}\subref{source3} and~\ref{sourceevol_spectralpars}\subref{source4}.

\subsection{Expected neutrino and gamma-ray fluxes}\label{sec:neutrinos}
Cosmogenic neutrinos do not undergo any interactions during their propagation, except for adiabatic energy losses due to the expansion of the Universe and flavour oscillations, so they can reach us even from
very high redshifts,
from which we do not expect any high-energy nuclei to survive.
Hence, the comparison of the  
flux of the expected cosmogenic neutrinos associated with the best-fit results of each chosen scenario
with the measured fluxes (or, at higher energies, with the estimated upper limits) can possibly constrain the cosmological evolution of sources in ways complementary to those available from UHECR measurements.

The Pierre Auger Observatory is sensitive to neutrinos with energies above $10^{8}$\,GeV~\cite{Auger-neu2019}, which corresponds to the energy range for neutrinos coming from the pion photoproduction of UHECRs on the CMB and EBL photons. The energy of a cosmogenic neutrino is on average of the order of 5\% of the energy of the nucleon that produced it. No neutrinos were observed so far, hence 90\% C.L. upper limits have been set on $E^2J_\nu$, assuming an $E^{-2}$ spectral shape. 

\begin{figure}
	\centering
	\def\w{0.49}
	\subfigure{\includegraphics[width=\w\linewidth]{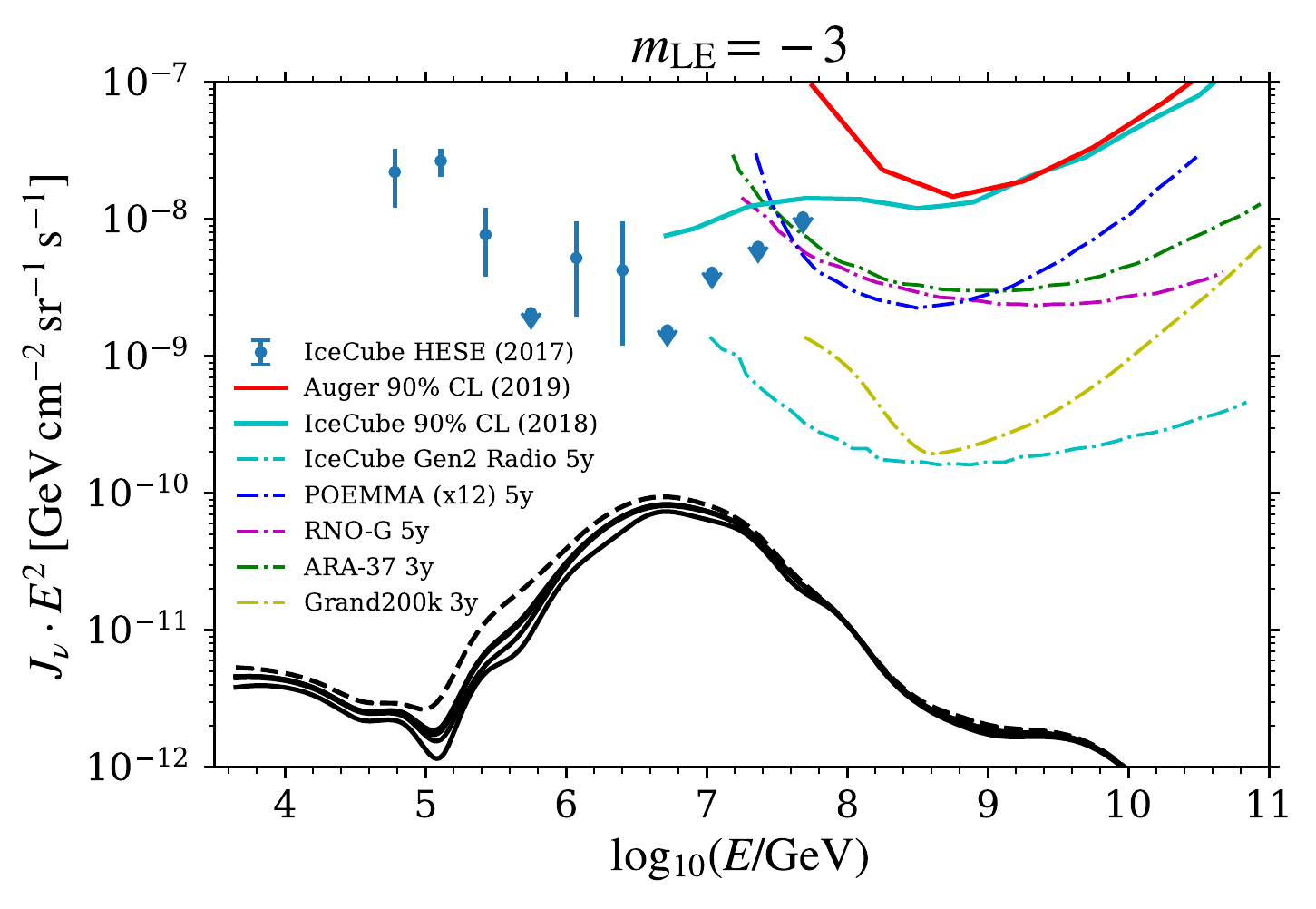}\label{fig:neutrinos11}}\hfill
	\subfigure{\includegraphics[width=\w\linewidth]{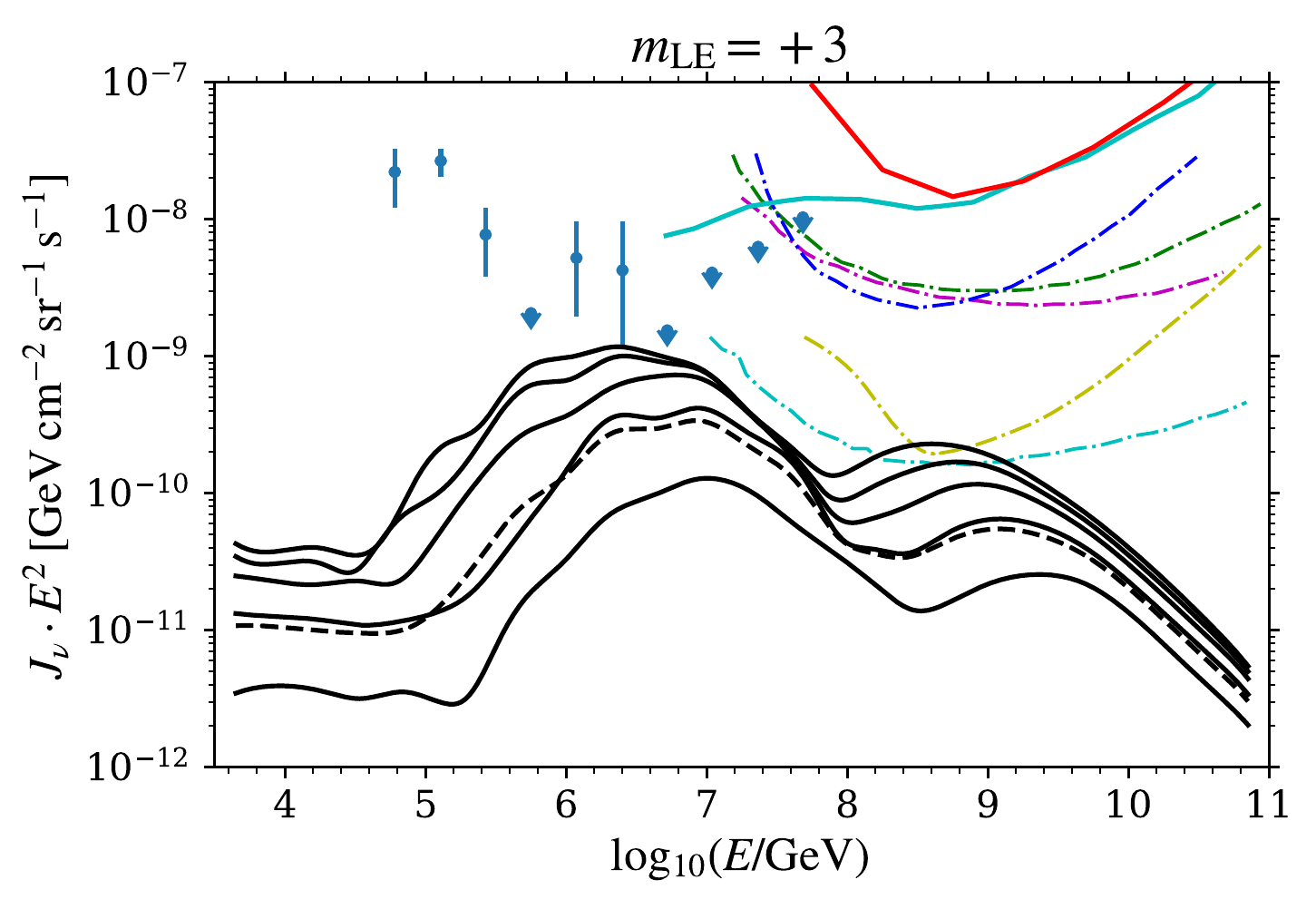}\label{fig:neutrinos12}}
	\\
	\subfigure{\includegraphics[width=\w\linewidth]{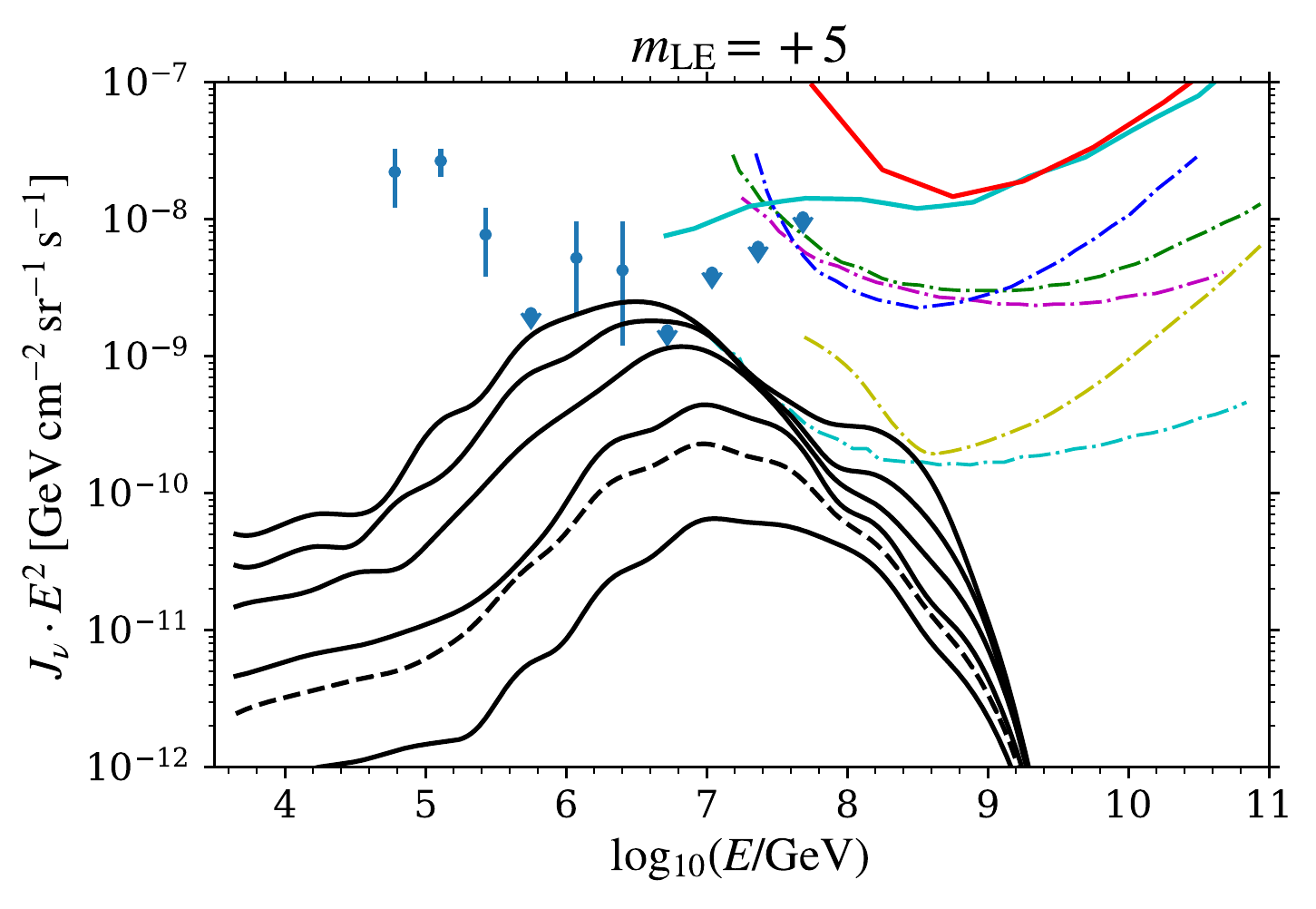}\label{fig:neutrinos13}}
	\caption{The predicted fluxes of neutrinos (single flavour) corresponding to the best fit results obtained by assuming a source evolution with $m=-3$ (top left), $m=3$ (top right) and $m=5$ (bottom)  for the \LEb\ component; in all the three cases the \HEb\ component has no source evolution ($m=0$). The black solid curves represent the fluxes corresponding (from the bottom to the top) to $z_\text{max} = 1$, 2, 3, 4 and 5, assuming a constant $m$ value in the whole redshift range.  The dashed black curve shows the expected fluxes corresponding to $z_\text{max}=3$ with a flat source evolution above $z=1$. 
	The observed IceCube HESE flux, the current upper limits from IceCube and Auger (solid lines), and the predicted sensitivities of future detectors (dot-dashed lines) are also shown for reference (see text).
	}
	\label{fig:neutrino1}
\end{figure}

They are currently among the most stringent ones in the UHE range and are shown in Fig.~\ref{fig:neutrino1}.
Since most of the predicted neutrinos have energies below the region where Auger could detect them, also the measurements up to $10^{8}$\,GeV~\cite{icecube2} and the upper limits~\cite{icecube} provided by IceCube are shown.  

Note however that neutrinos with $E \lesssim 10^{8}$\,GeV can be produced by nuclei injected with energies below the range of our fits, $E < 10^{17.8}$\,eV, where we extrapolate the injection spectrum as a power law with $\gamma \gtrsim 3$ down to indefinitely low energies; this is a rather extreme hypothesis, as it would require incredibly large integrated emissivities at low injection energies.  Hence, the predicted fluxes shown in Fig.~\ref{fig:neutrino1} below $10^{8}$\,GeV should be considered upper bounds to the predictions in more realistic scenarios, in which at $E \ll 10^{17.8}$\,eV the injection spectra are harder.

In general, the contribution of the \HEb\ population to the flux of expected neutrinos is negligible, regardless of its cosmological evolution: due to its rather low rigidity cutoff, even when the estimated fraction of protons is not negligible, the pion photoproduction interactions cannot occur on CMB photons, but only on the EBL ones. 
The latter, despite having a lower energy threshold,  contributes to the neutrino flux to a lesser extent because of the much greater interaction length.  As a consequence, the neutrino fluxes shown in Fig.~\ref{fig:neutrino1} are entirely provided by the \LEb\ population of sources, and are thus sensitive to the assumptions on the source evolution of such component. 

In the case of a flat or negative source evolution for the \LEb\ component, the expected neutrino fluxes are well below the current observations and the future detectors sensitivity; the case with~$m=-3$ is shown on the top left panel of Fig.~\ref{fig:neutrino1}. 
The predicted flux increases in the case of a positive source evolution for the \LEb\ population, e.g.\ as shown in the top right panel of Fig.~\ref{fig:neutrino1} ($m=3$), and the largest increase is obtained with a strong ($m=5$) source evolution, shown in the bottom panel of Fig.~\ref{fig:neutrino1}, corresponding to the best fit for the LE component with this evolution. A peak is predicted at ${\sim}10^{7}$\,GeV, corresponding to pion production on EBL photons; this is visible in the lower curve of Fig.~\ref{fig:neutrino1} (top right and bottom panels), corresponding to $z_\text{max}=1$, and is shifted towards lower energies for increasing values of $z_\text{max}$. The evolution of the source distribution with redshift is however uncertain above $z=1$, and this can influence the expected neutrino flux. As an example, in the bottom panel the intermediate solid black line, corresponding to an evolution $(1+z)^5$ up to $z_\text{max}=3$ can be compared to the dashed black line, corresponding to $(1+z)^5$ up to $z_\text{max}=1$ and to a flat evolution in the redshift range $1<z<z_\text{max}=3$, which is more than one order of magnitude lower than the former. The maximum rigidity of the LE component has also a strong impact in the neutrino flux;  
for example, in the case of a source evolution with $m=3$ for the LE component (top right panel), the rigidity found from the fit is $R_\text{cut}\sim10^{21}$\,V, hence the peak corresponding to the UHECR interactions with CMB photons is visible at the highest energies. 

It is worth noting that future neutrino detectors will provide an improved sensitivity to cosmogenic neutrinos at energies above $10^{8}$\,GeV. As shown in the top right and bottom plots of Fig.~\ref{fig:neutrino1},  our predictions in the cases of positive source evolutions would be constrained by the most stringent future limits, provided by the next-generation detector upgrade of IceCube~\cite{icecube_nextgen} and by planned detectors~\cite{grand,ara,poemma}. We can conclude that, if the sensitivity of the next-generation neutrino detectors are exploited, the neutrino fluxes predicted for the simple two-component scenario proposed here may put some additional constraints on the source properties, for example excluding some source evolutions for the \LEb\ component and/or limiting the possible values of its rigidity cutoff.  

\begin{figure}
    \centering
    \includegraphics[width=0.45\columnwidth,page=2]{photons}
    \hspace*{0.05\columnwidth}
    \includegraphics[width=0.45\columnwidth,page=1]{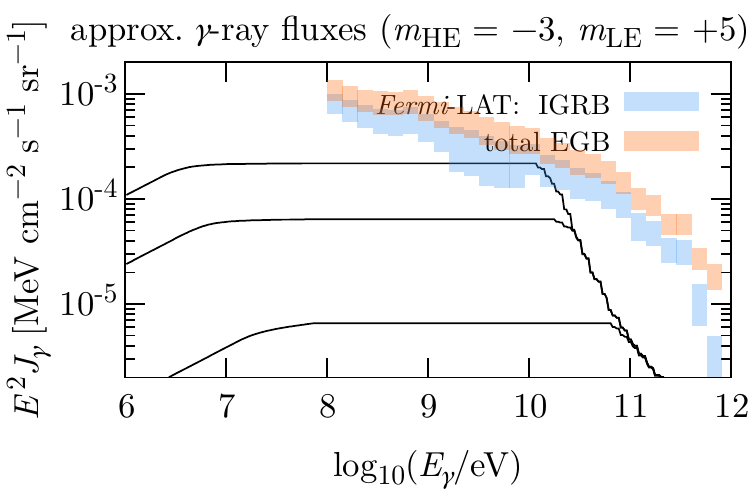}
    \caption{Fluxes at Earth of gamma-ray cascades produced in UHECR propagation in two example scenarios with (from bottom to top) $z_\text{max} = 1$, 3 and 5, estimated through the analytic approximation from Ref.~\cite{Berezinsky:2016feh}, and compared to \textit{Fermi}-LAT measurements from Ref.~\cite{Fermi-LAT:2014ryh} (with error bands showing uncertainties due to those on Galactic foreground models).  Note that this approximation is not very accurate for short source distances, and hence for high photon energies. All other scenarios in Section~\ref{SourceEvolution} with $D \lesssim 1000$ result in similar or lower fluxes.}
    \label{fig:photons}
\end{figure}

Another messenger of potential interest in the study of UHECRs is the flux of gamma-ray cascades produced in their propagation.  Photons and electrons produced in a photohadronic interaction in intergalactic space can initiate electrophotonic cascades via repeated pair production $\upgamma + \upgamma \to \text{e}^+ + \text{e}^-$ and synchrotron emission or inverse Compton scattering $\text{e}^\pm + \upgamma \to \text{e}^\pm + \upgamma$, until all the secondaries have $E \lesssim 100$\,GeV.  Provided the first interaction happens far enough that the cascade has the room to fully develop before reaching Earth, the shape of the final spectrum of gamma rays is nearly independent of the primary photon or electron energy, and only weakly dependent on the initial redshift~\cite{Berezinsky:2016feh}.  A detailed study of such cascades is outside the scope of this work, but a rough estimate of the resulting gamma-ray fluxes at $E \lesssim 100$\,GeV can be obtained by applying the analytical approximation from Ref.~\cite{Berezinsky:2016feh} to electrons and photons produced in SimProp simulations.  In this case, both the \LEb~and the \HEb~component have a non-negligible contribution, as the Lorentz factor threshold for electron--positron pair production by UHECRs on CMB photons is two orders of magnitude lower than for pion production.  The results are shown in Fig.~\ref{fig:photons}, and compared to \textit{Fermi}-LAT measurements~\cite{Fermi-LAT:2014ryh} of the isotropic diffuse gamma-ray background (IGRB) and the total extragalactic gamma-ray background (EGB), the former excluding and the latter including the emissions resolved into point sources. The strength of intergalactic magnetic fields, and hence
how the angular spread of the cascades compares with the angular resolution of the
telescope, is however not known. Even assuming that the magnetic fields are strong enough that the cascades resulting from UHECR propagation would have an angular spread much larger than the \textit{Fermi}-LAT resolution and hence be entirely comprised in the diffuse IGRB, and even considering the model with the highest Galactic foreground among those used in Ref.~\cite{Fermi-LAT:2014ryh}, the only scenarios in tension with the data would be the ones where the \HEb\ component has a strong positive evolution and $z_\text{max} \gtrsim 2$, or the \LEb\ component has a strong positive evolution and $z_\text{max} \gtrsim 4$.\footnote{In addition, the scenario where both components have a moderate positive evolution (not shown) is in tension with the data if $z_\text{max} \gtrsim 6$.}
As shown in Fig.~\ref{fig_sourceevol1} the former are already excluded by the deviance of our combined fit, and as shown in Fig.~\ref{fig:neutrino1} the latter may also result in amounts of cosmogenic neutrinos within the reach of future planned detectors.  Hence, it would appear that gamma-ray fluxes cannot provide any additional information
compared to that available from UHECR and neutrino data in most of the scenarios here considered. Note however that in some of the fits a very large fraction of the high energy IGRB is due to Bethe-Heitler production of extragalactic cosmic rays. Models of the contribution of sources to the IGRB attribute it almost entirely to unresolved point sources~\cite{2015ApJ, 2021Natur} and hence, once the accuracy of these models improves, the gamma-ray fluxes will provide very constraining boundary conditions to the cosmic-ray models.  

This result is comparable to those of earlier works assuming a mixed mass composition for UHECRs (e.g.~\cite{Globus:2017ehu, Muzio:2019leu}) and more pessimistic than those assuming a pure proton composition~(e.g.~\cite{Liu:2016brs}).

\section{Conclusions and outlook}
\label{conclusions}

In this paper we have shown that, using the energy spectrum and composition
data from the Pierre Auger Observatory, it is possible to constrain
astrophysical scenarios for the UHECR sources. 

We considered the hypothesis
of two extragalactic components, from two distinct populations, in presence
or not of a secondary Galactic contribution. The two components reasonably succeed to reproduce the ankle feature, whose sharpness, as observed in
Auger data, is hard to reproduce with other scenarios. 
Also the region above the ankle is reproduced including,
in particular, the newly observed feature at ${\sim}10^{19}$\,eV
(the `instep’), which originates from the interplay of light-to-intermediate
nuclei. Despite the fact that a definite conclusion on the presence of a
subdominant Galactic flux cannot be reached, our results show that its end is
compatible with the data only if it is composed by medium-mass nuclei. 

The possible systematic uncertainties from both
experimental and model sources, though large enough to affect the fit
parameters especially in the case of hadronic models describing interactions in atmosphere, do not spoil these conclusions. 

Based on this work, very strong source evolutions can be excluded, since they would cause a flux of secondary particles at the ankle exceeding the observed spectrum, even in presence of a negligible contribution from the LE component in that region. This conclusion could
not be reached with a fit limited to the energy region above the ankle. Finally, we show that for some of the considered scenarios the predictions of cosmogenic
neutrino fluxes might reach the sensitivity range of the next-generation detectors.

An extension of the combined fit to even lower energies will be more effective to investigate the
transition from Galactic to extragalactic cosmic rays, increasing the
lever-arm with the use of composition data from HEAT (High Elevation Auger
Telescopes).
Composition results below $10^{17.8}$\,eV have been already reported in a
preliminary analysis~\cite{bellido2017}. An update of the \xmax analysis in the
whole energy range is currently in progress and its results are expected
to push remarkably the sensitivity of the combined fit studies in the
transition region.

Further insight on the possible sources of UHECRs can be gained by extending the combined fit to include the arrival directions information to the spectrum and composition data. The results of a preliminary analysis were shown in Ref.~\cite{Bister-ICRC2021}.

In this analysis, the mass composition
data do not extend to energies where the suppression occurs, because of the
limited duty cycle of the FD. The interpretation of the suppression in the flux by differentiating between
a cutoff due to propagation effects and the maximum energy reached in the
sources can provide fundamental constraints on the sources of UHECRs and their properties.
In the near future, mass composition estimates will be obtained through \xmax and the muon content of showers by using machine learning techniques on SD data~\cite{PierreAuger:2021fkf,PierreAuger:2021nsq}. 

Furthermore, the Pierre Auger Observatory is currently
undergoing an upgrade, AugerPrime~\cite{PierreAuger:2016qzd,Castellina:2019irv}, that includes the deployment of
scintillators on top of the SD stations to help disentangle
the muonic and electromagnetic content of the showers. This will allow the
measurement of the mass composition beyond the present limit, help testing the presence of
 a possible sub-dominant light contribution at the highest energies
and cover the
suppression region to perform an analysis similar to the one presented here
with much larger statistics.

\appendix

\section{Parameterisation of the \xmax distributions}
\label{gumbel_par}
In this work the \xmax distributions are parameterised by fitting Gumbel distributions to CONEX~\cite{conex} simulations of H-, He-, N-, Si- and Fe-initiated showers with energies ranging from $10^{17}$\,eV to $10^{20}$\,eV. The parameters thus obtained are shown in Tab.~\ref{tab:gumbel}, corresponding to different hadronic interaction models. The coefficients $a_i$, $b_i$ and $c_i$ parameterise the expansion of the generalised Gumbel coefficients $\mu$, $\sigma$ and $\lambda$ in powers of $\lg E$ and~$\ln A$,  as described in Ref.~\cite{gumbel}.

In each energy bin, we use as $E$ the geometric mean of the energies of the observed FD events in the bin.  
From this, we computed the total \xmax distribution in each energy bin as $g_\text{tot}(\xmax|E) = \sum_A f_A(E) \, g(\xmax |E, A)$, where $f_A(E)$ is
the fraction of simulated events in the energy bin with mass number $A$.  Then we multiplied the distribution above by the acceptance function $\mathcal{A}(\xmax, E)$ and we convolved it by the detector resolution function $\mathcal{R}(\xmax^\text{rec}-\xmax|E)$, using for both the parametrisations from Ref.~\cite{xmax_dist} with the central values for the parameters.   
Hence, we can define the model prediction $G^\text{mod}_{ij}$ in the $i$-th energy bin and $j$-th \xmax\ bin, normalised so that $\sum_j G^\text{mod}_{ij} = 1$ for each $j$.

\begin{table}[h]
    \centering
    \small
\begin{tabular}{|c|ccc ccc ccc|}
\hline
\textbf{E} & $a_0$ & $a_1$ & $a_2$ & $b_0$ & $b_1$ & $b_2$ & $c_0$ & $c_1$ & $c_2$\\
\hline
$\mu$	& $775.457$ & $-10.399$ & $ -1.753$ & $\,58.529$ & $ -0.826$ & $\phantom{+}0.231$ & $ -1.408$ & $\phantom{+}0.226$ & $ -0.100$ \\
$\sigma$	& $\phantom{0}32.263$ & $\phantom{+0}3.943$ & $ -0.864$ & $\phantom{+}1.275$ & $ -1.812$ & $\phantom{+}0.232$ & --- & --- & --- \\
$\lambda$	& $\phantom{00}0.641$ & $\phantom{+0}0.220$ & $\phantom{+}0.171$ & $\phantom{+}0.073$ & $\phantom{+}0.035$ & $ -0.013$ & --- & --- & --- \\
\hline
\hline
\textbf{Q} & $a_0$ & $a_1$ & $a_2$ & $b_0$ & $b_1$ & $b_2$ & $c_0$ & $c_1$ & $c_2$\\
\hline
$\mu$	& $758.650$ & $-12.357$ & $ -1.245$ & $\,56.594$ & $ -1.012$ & $\phantom{+}0.229$ & $ -0.535$ & $ -0.173$ & $ -0.019$ \\
$\sigma$	& $\phantom{0}35.424$ & $\phantom{+0}6.759$ & $ -1.462$ & $ -0.796$ & $\phantom{+}0.202$ & $ -0.014$ & --- & --- & --- \\
$\lambda$	& $\phantom{00}0.672$ & $\phantom{+0}0.374$ & $\phantom{+}0.075$ & $\phantom{+}0.030$ & $\phantom{+}0.047$ & $ -0.001$ & --- & --- & --- \\
\hline
\hline
\textbf{S} & $a_0$ & $a_1$ & $a_2$ & $b_0$ & $b_1$ & $b_2$ & $c_0$ & $c_1$ & $c_2$\\
\hline
$\mu$	& $785.852$ & $-15.599$ & $ -1.069$ & $\,60.593$ & $ -0.786$ & $\phantom{+}0.201$ & $ -0.689$ & $ -0.295$ & $\phantom{+}0.040$ \\
$\sigma$	& $\phantom{0}41.035$ & $\phantom{0}{-}2.173$ & $ -0.306$ & $ -0.309$ & $ -1.165$ & $\phantom{+}0.225$ & --- & --- & --- \\
$\lambda$	& $\phantom{00}0.799$ & $\phantom{+0}0.235$ & $\phantom{+}0.009$ & $\phantom{+}0.063$ & $ -0.001$ & $\phantom{+}0.000$ & --- & --- & --- \\
\hline
\end{tabular}

    \caption{Parameters of the Gumbel distributions used in this work (\textbf{E}:~\EPOS, \textbf{Q}:~\QGSJET, \textbf{S}:~\SIBYLL; $\mu$ and~$\sigma$ in g/cm$^2$, $\lambda$ dimensionless).} 
    \label{tab:gumbel}
\end{table}

\section{Deviance profiles as a function of the LE rigidity cutoff}
\label{dev_profiles}

\begin{figure}[h]
	\centering
	\def\w{0.49}
	\subfigure{\includegraphics[width=\w\linewidth]{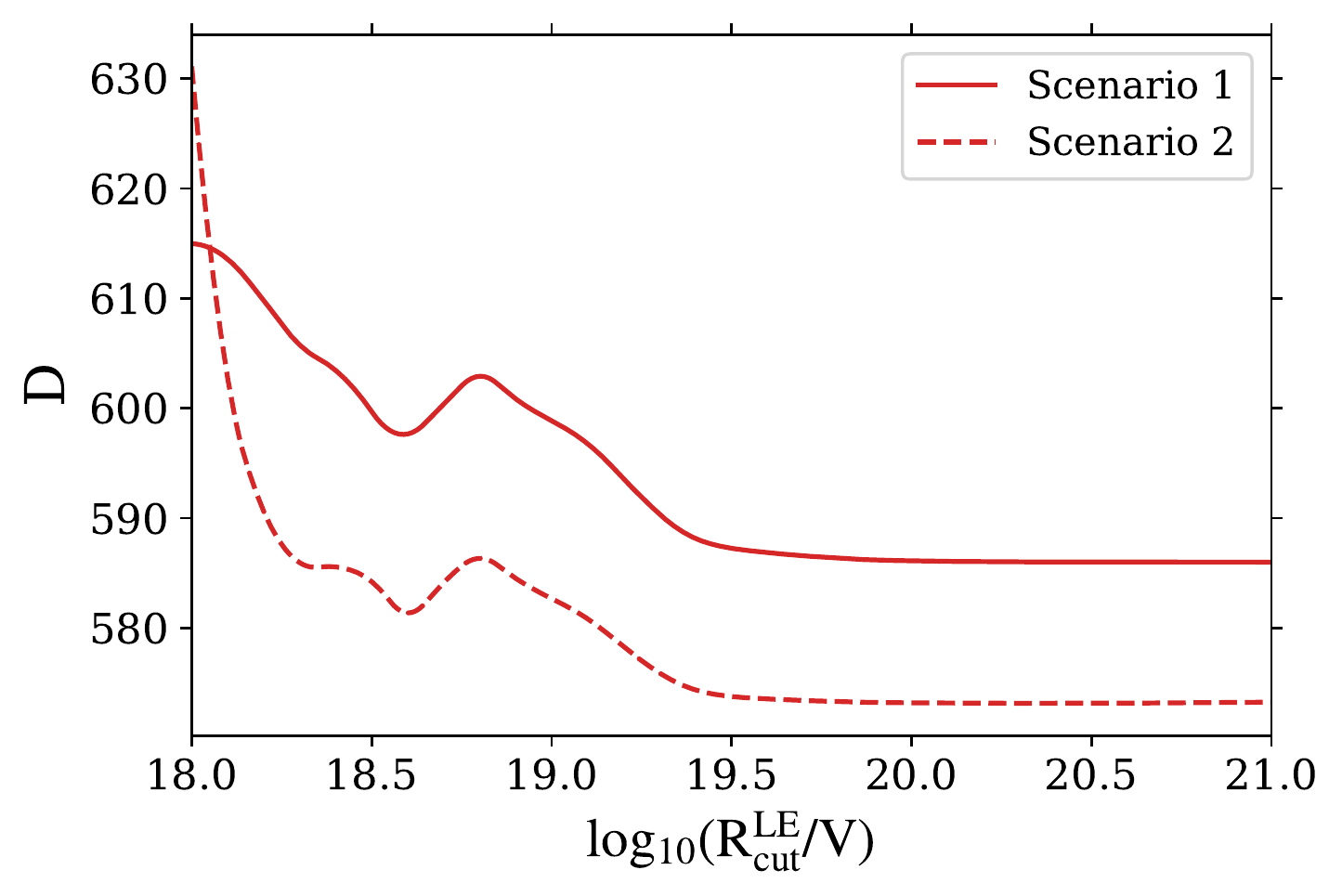}\label{scanRcut1}}
	\\
	\subfigure{\includegraphics[width=\w\linewidth]{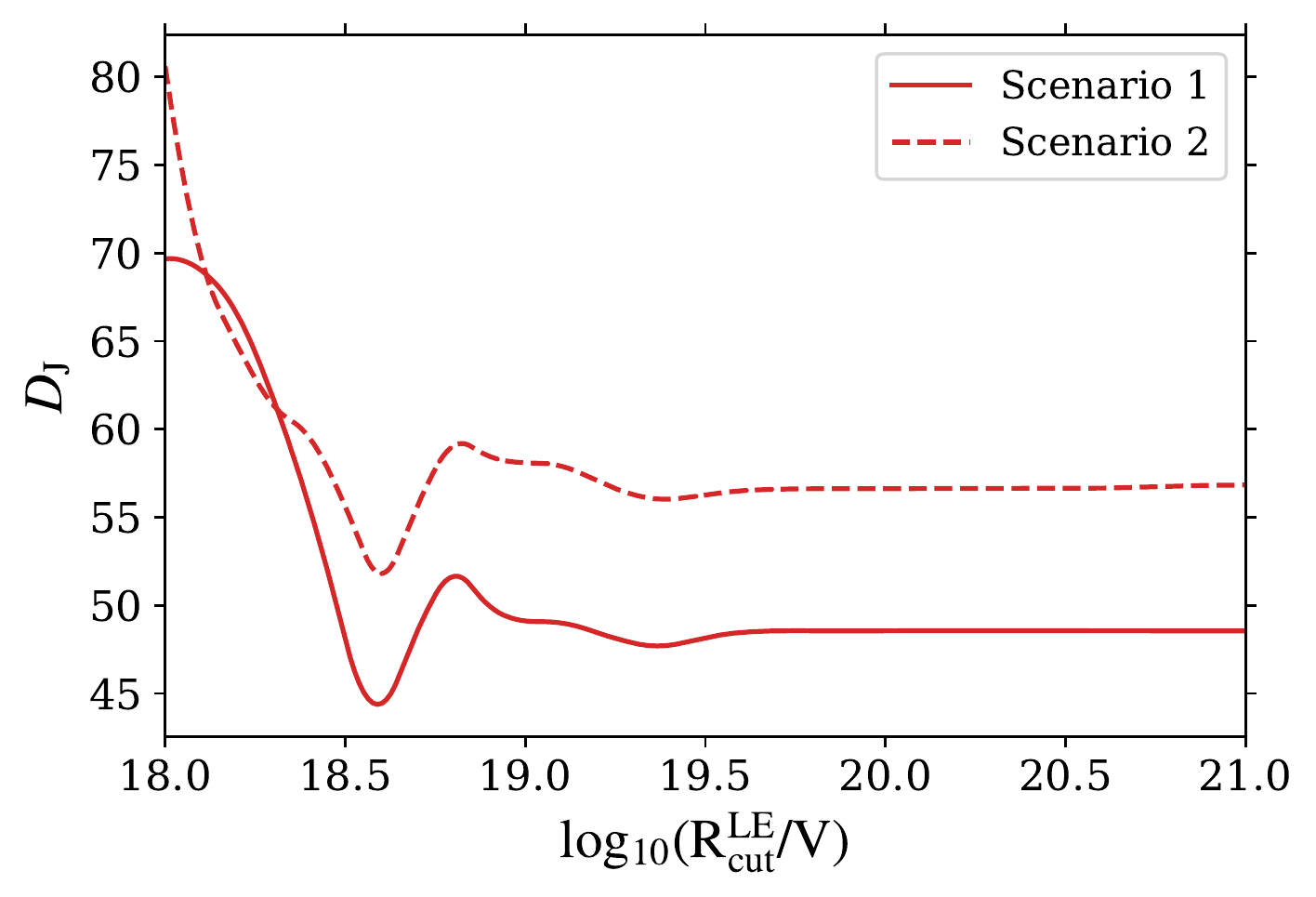}\label{scanRcut2}}\hfill
	\subfigure{\includegraphics[width=\w\linewidth]{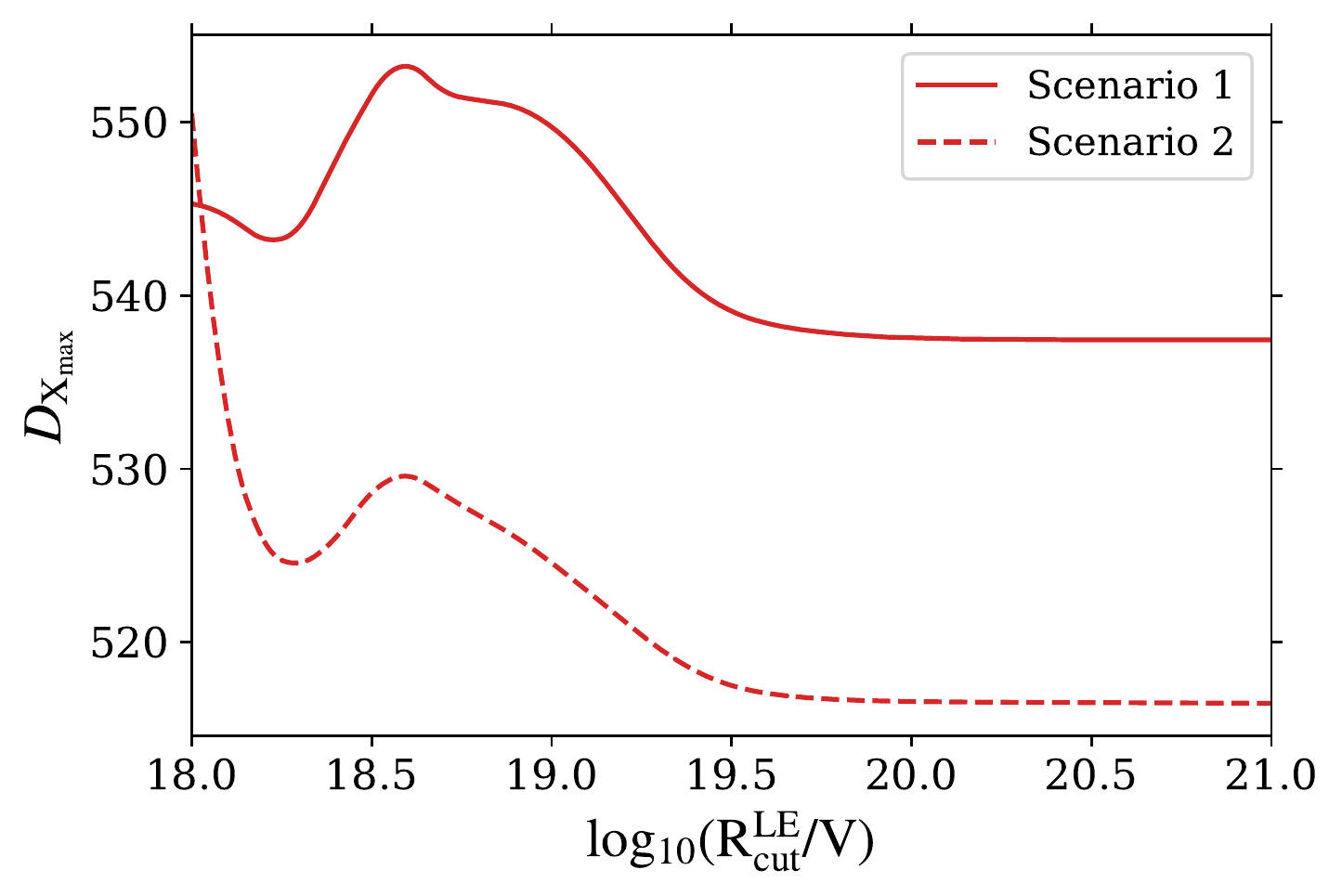}\label{scanRcut3}}
	\caption{The deviance profiles as a function of the rigidity cutoff of the \LEb\ component in the reference scenarios. The total deviance and the partial contributions from the energy spectrum fit and the \xmax fit are shown.}
	\label{scanRcut_2cases}
\end{figure}

In Fig.~\ref{scanRcut_2cases}, the values of the total deviance and of its partial contributions are shown as obtained by scanning over $R_\text{cut}^\text{\LEa}$ (re-optimizing all other parameters for each $R_\text{cut}^\text{\LEa}$ value). The deviance profiles exhibit similar trends in the two reference scenarios, despite some differences in the nominal values due to the fact that a better fit of either the energy spectrum or the \xmax distributions is provided in \textsc{Scenario 1} and \textsc{Scenario 2}, respectively.

From the total deviance profile, it is also clear that the fit is degenerate with respect to $R_\text{cut}^\text{\LEa}$ for values ${\gg}10^{19.5}$\,V, because of the very steep estimated energy spectrum of this component which is thus suppressed even in the absence of an exponential cutoff.

\section{Treatments of the \xmax scale uncertainties}
\subsection{Use of two nuisance parameters}
\label{nuisance-expl}

The probability that \xmax measurements in the 1st, \ldots, $n$-th energy bin are affected by a bias $\mathbf{x} = (x_1, \ldots, x_n)$ can be treated as a multivariate Gaussian distribution
\begin{equation}
    p(\mathbf{x}) = \frac{1}{\sqrt{\smash[b]{(2\pi)^n \det\mathbf{\Sigma}}}} \exp \left(
        -\frac{1}{2} \mathbf{x}^\mathrm{T} \mathbf{\Sigma}^{-1} \mathbf{x}
    \right), 
\end{equation}
where the covariance matrix is $\Sigma_{ii'} = \sigma_{i} \sigma_{i'} \rho_{ii'}$, in which $\sigma_i$ is the standard deviation in the $i$-th energy bin and $\rho_{ii'}$ is the correlation coefficient between the $i$-th and the $i'$-th energy bin.
Hence, if we want to model such biases by shifting \xmax values, we should add a term
\begin{equation}
    \Dsyst = -2\ln\frac{p(\mathbf{x})}{p(\mathbf{0})} = \mathbf{x}^\mathrm{T} \mathbf{\Sigma}^{-1} \mathbf{x}
\end{equation}
to the overall deviance.  
However, due to the strong bin-to-bin correlations, the matrix $\mathbf{\Sigma}$ is almost singular, so $p$ is close to $0$ ($\Dsyst$ is very large) for all values of $\mathbf{x}$ except those which do not vary very fast across neighbouring energy bins.
On the other hand, we can diagonalise $\mathbf{\Sigma}$ as $\mathbf{R}\mathbf{\Sigma}'\mathbf{R}^{-1}$, where $\mathbf{R}$ is a rotation matrix ($\mathbf{R}^{-1} = \mathbf{R}^\mathrm{T}$) and $\mathbf{\Sigma}' = \operatorname{diag}\left(\sigma_1^{\prime2}, \ldots, \sigma_n^{\prime2} \right)$ (and hence $\mathbf{\Sigma}^{-1}$ as $\mathbf{R}\operatorname{diag}\left(\sigma_1^{\prime-2}, \ldots, \sigma_n^{\prime-2} \right)\mathbf{R}^{-1}$, $\sqrt{\mathbf{\Sigma}}$ as $\mathbf{R}\operatorname{diag}\left(\sigma_1^{\prime}, \ldots, \sigma_n^{\prime} \right)\mathbf{R}^{-1}$, and so on).  The columns of $\mathbf{R}$ are the eigenvectors of $\mathbf{\Sigma}$, and $\sigma_1^{\prime2}, \ldots, \sigma_n^{\prime2}$ are its eigenvalues.  We then have
\begin{equation}
    \Dsyst = 
    \mathbf{x}^\mathrm{T} \mathbf{\Sigma}^{-1} \mathbf{x} =
    \mathbf{x}^\mathrm{T} \mathbf{R}\mathbf{\Sigma}^{\prime-1}\mathbf{R}^{-1} \mathbf{x} =
    \mathbf{x}^\mathrm{T} \mathbf{R}
    \mathbf{\Sigma}^{\prime-1/2}\mathbf{\Sigma}^{\prime-1/2}
    \mathbf{R}^\mathrm{T} \mathbf{x} =
    \left| \mathbf{\Sigma}^{\prime-1/2} \mathbf{R}^\mathrm{T} \mathbf{x} \right|^2,
\end{equation}
i.e.\ the entries of~$\mathbf{x}' = \mathbf{\Sigma}^{\prime-1/2} \mathbf{R}^\mathrm{T} \mathbf{x}$ are independent Gaussians with zero mean and unit variance, which can be converted back to $\mathbf{x} = \mathbf{R} \mathbf{\Sigma}^{\prime1/2} \mathbf{x}'$.
We can use $x'_1, \ldots, x'_n$ as the fit parameters, with $\Dsyst = x^{\prime2}_1 + \cdots + x^{\prime2}_n$, and the actual shifts are $x_i = \sum_{i'} (\mathbf{R}\mathbf{\Sigma}^{\prime1/2})_{ii'} \smash{x'\!}_{i'}$.  In practice, we have $|(\mathbf{R}\mathbf{\Sigma}^{\prime1/2})_{ii'}| \ll 1$\,g/cm$^2$ for all $i' \ge 3$, so we only use two parameters $a = x'_1$ and $b = x'_2$, as eigenvectors after the second would be unlikely to substantially improve the fit.  
The first two eigenvectors of $\mathbf{\Sigma}$ then define two functions of energy, given by $v_1(E_i) = (\mathbf{R}\mathbf{\Sigma}^{\prime1/2})_{i1}$ and $v_2(E_i) = (\mathbf{R}\mathbf{\Sigma}^{\prime1/2})_{i2}$, plotted in Fig.~\ref{fig:eigenvectors}, and all the \xmax distributions are thus shifted according to a quantity $a\,v_1(E)+b\,v_2(E)$, where $a$ and $b$ are two additional nuisance parameters of the fit, and an additional term $\Dsyst=a^2+b^2$ is added to the deviance. 

The results of adding the two parameters $a$ and $b$ to the fit are reported in Section~\ref{nuisance}.

\subsection{Fixed \xmax shifts}
\label{sysexp}

In Section~\ref{nuisance} we discussed the effect of using an approach based on nuisance parameters to treat the \xmax scale uncertainty. In order to compare with the analysis we presented in \CF, here we also show the results obtained by simultaneously shifting all \xmax distributions to higher or lower values according to their energy-dependent systematic uncertainties~$\sigma_i$, which implies a lighter or a heavier observed mass composition at all energies, respectively.

This can be justified as a first-order approximation as the systematic uncertainties on \xmax at different energies are all positively correlated with each other. Nevertheless, as already illustrated in Section~\ref{nuisance} the correlations between bins at very different energies can be rather weak, hence the approach with the nuisance parameters should be considered more complete.

The results are obtained in the \TALYS+Gilmore configuration, assuming \EPOS\ as the HIM, so they can be directly compared with the ones presented in Section~\ref{sysunc}.

\begin{table}\small
	\renewcommand\arraystretch{1.4}
	\centering
	 \begin{threeparttable}
	\begin{tabular}{|  c  c |c| c | c c c c c  | c  |}
	\hline
		$\Delta X_\text{max} $ & $\Delta E / E$ &  & $\mathcal{L}_0$~\tnote{*} & $I_\text{H}$  & $I_\text{He}$  &$I_\text{N}$  & $I_\text{Si}$  & $I_\text{Fe}$~\tnote{**}$\:$ & $D$ ($D_J$, $D_{\xmax}$) \\ \hline
		& \multirow{2}{*}{$-14\%$}& \LEb & $7.7 $ & 39.7 & 11.5 & 36.0 & 12.9 & 0.0 &\multirow{2}{*}{572.5 (50.0, 522.6)}  \\
		& &\HEb & $3.6 $ & 0.0 & 24.2 &72.5 & 0.0&3.3 & \\ \cline{3-10}
		\multirow{2}{*}{$-1\sigma_\text{syst}$}   & \multirow{2}{*}{$\phantom{+0}0\phantom{\%}$} & \LEb & $11.2$ & 33.0 & 18.5 & 20.7 & 27.8 & 0.0 &\multirow{2}{*}{597.6 (74.1, 523.5)}  \\
		& &\HEb & $4.6$ & 0.0 & 15.5 &79.6 & 0.0&5.0 & \\ \cline{3-10}
		& \multirow{2}{*}{$+14\%$}& \LEb & $15.6$ & 28.9  & 22.8 & 11.0 & 35.2 & 2.2 &\multirow{2}{*}{612.9 (92.1, 520.8)}  \\
		& &\HEb & $5.5$ & 0.0 & 5.9 & 84.5 & 3.5 &6.1 & \\ \hline \hline
		
		& \multirow{2}{*}{$-14\%$}& \LEb & $7.6$ & 47.5 & 24.5  &28.0  & 0.0 &0.0  &\multirow{2}{*}{604.9 (46.5, 558.4)}  \\
		& &\HEb & $3.9$ & 1.1 &30.0  &66.6 &0.0 &2.3 & \\ \cline{3-10}
		\multirow{2}{*}{$\phantom{+}0\phantom{\sigma_\mathrm{syst}}$}  & \multirow{2}{*}{$\phantom{+0}0\phantom{\%}$}& \LEb & $11.4$ & 48.7 & 7.3 & 44.0 & 0.0 & 0.0 &\multirow{2}{*}{573.1 (56.6, 516.5)}  \\
		& &\HEb & $5.1$ & 0.0 & 23.6 & 72.1 & 1.3 &3.1 & \\ \cline{3-10}
		& \multirow{2}{*}{$+14\%$}& \LEb & $15.8$ & 47.5 & 0.3 & 48.5 & 3.00 & 0.8 &\multirow{2}{*}{577.1 (70.3, 506.8)}  \\
		& &\HEb & $6.2$ & 0.0 &17.8  &74.5 &4.00 &3.8 & \\ \hline \hline

		& \multirow{2}{*}{$-14\%$}& \LEb & $ 7.4$ & 52.5 & 42.1 & 5.4 & 0.0 & 0.0 &\multirow{2}{*}{788.7 (68.7, 720.0)}  \\
		& &\HEb & $4.3$ & 7.5 & 29.3 & 62.1 & 0.0 & 1.1 & \\ \cline{3-10}

		\multirow{2}{*}{$+1\sigma_\text{syst}$}  & \multirow{2}{*}{$\phantom{+0}0\phantom{\%}$}& \LEb & $11.1$ & 50.2 & 31.2  & 18.7 & 0.0 & 0.0 &\multirow{2}{*}{729.7 (73.4, 656.3)}  \\
		& &\HEb & $ 5.5$ & 3.6 & 25.4 &68.5 &0.2 &2.3 & \\ \cline{3-10}
		& \multirow{2}{*}{$+14\%$}& \LEb & $15.7$ &  50.5 & 18.1 & 31.4 & 0.0 & 0.0 &\multirow{2}{*}{686.6 (78.5, 608.1)}  \\
		& &\HEb & $6.7$ & 0.3 & 21.9 & 71.8 & 3.6 & 2.4 & \\ \cline{3-10} \hline
	\end{tabular}
	\smallskip
	 \begin{tablenotes}
      \item[*] in units of $10^{44}\,\text{erg}\,\text{Mpc}^{-3}\,\text{yr}^{-1}$, from $E_\text{min}=10^{17.8}$\,eV.
      \item[**] in percentage.
      \end{tablenotes}
       \end{threeparttable}
	\caption{The effect on the deviance, the emissivities and the mass fractions of the $\pm1\sigma_\text{syst}$ shifts in the energy and \xmax scales.}
	\label{table:allsysexp}
	\vspace{-15pt}
\end{table}

We take into account the uncertainty on the energy scale and on the \xmax scale by shifting all the measured energies and \xmax values by one systematic standard deviation in each direction and consider all the possible combinations of these shifts. Their effect on the estimated emissivities, on the mass fractions at the sources and on the deviance value is summarised in Table~\ref{table:allsysexp}.
The dominant effect in terms of predictions at Earth is the one arising from the \xmax uncertainty, with the inferred composition becoming heavier as \xmax gets a negative shift. As for the remaining best fit parameters, they are not modified significantly when the experimental systematic uncertainties are considered.

\begin{figure}
	\footnotesize
	\centering
	\def\w{0.49}
	\subfigure{\includegraphics[width=\w\linewidth]{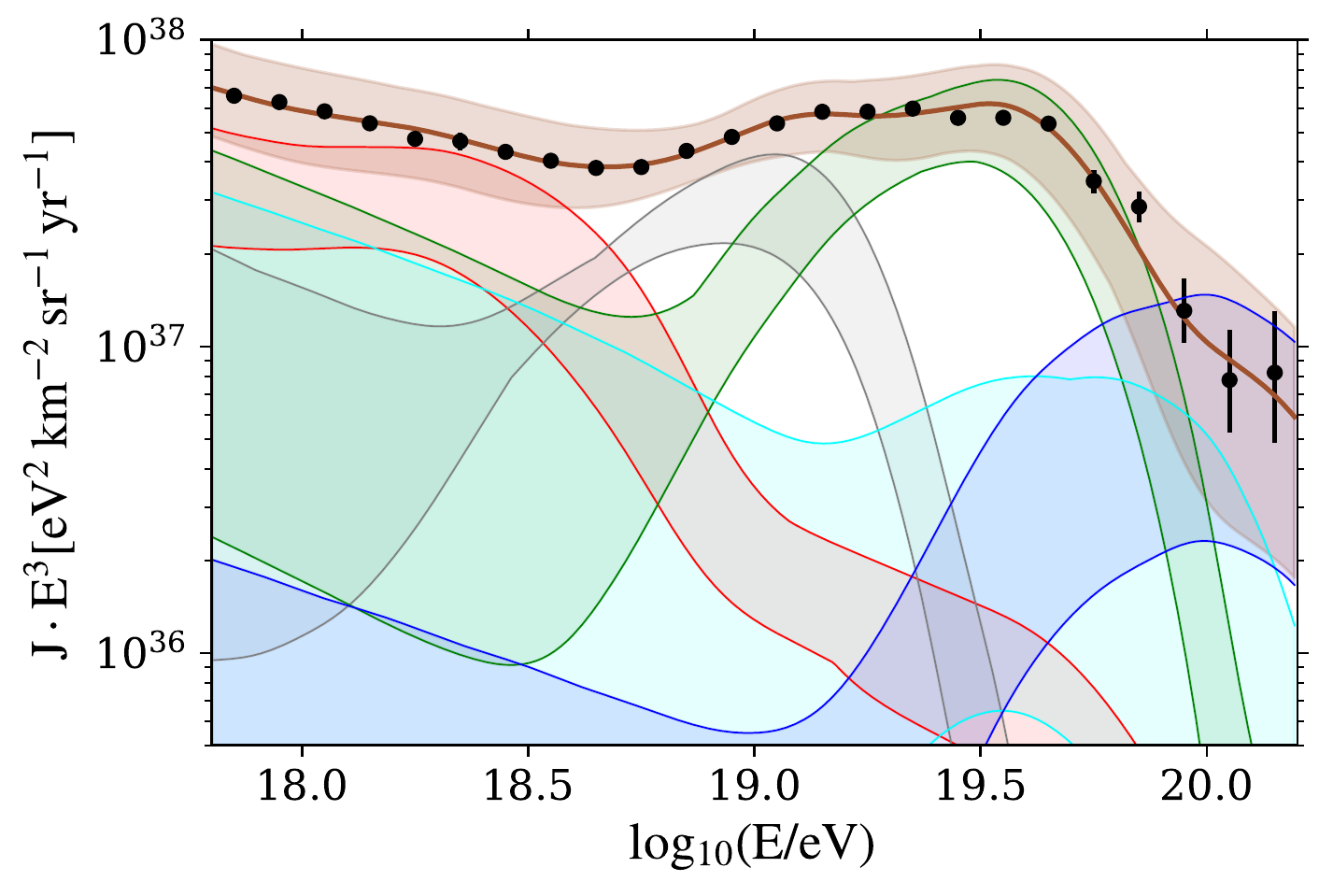}}\hfill
	\subfigure{\includegraphics[width=\w\linewidth]{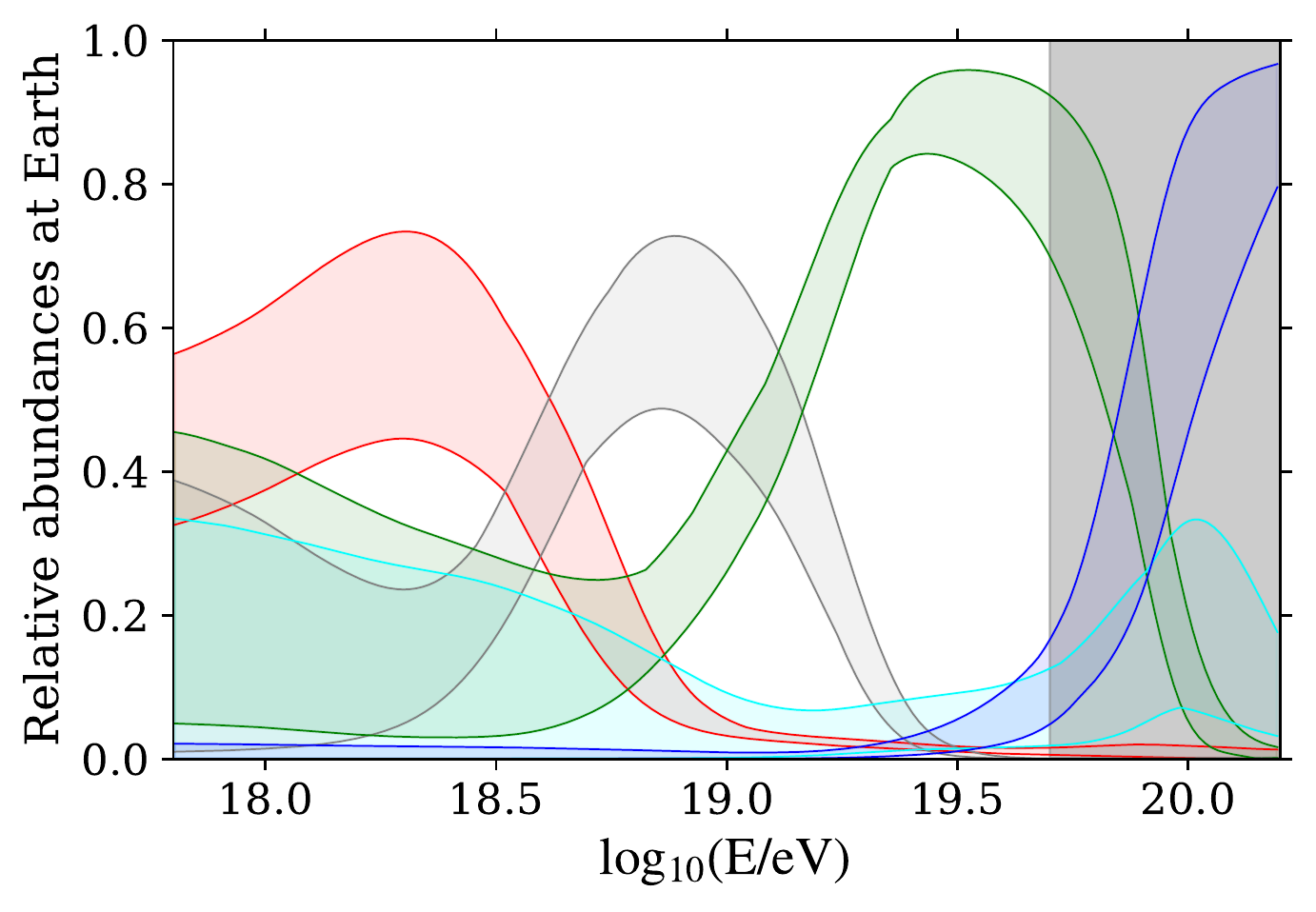}}
	\caption{\emph{Left:} the combined effect of the experimental uncertainties on the energy spectrum. \emph{Right:} the effect on the relative abundances at the top of the atmosphere. The bands represent the maximal variations induced by considering all the possible combinations of shifts. \lastXmax}
	\label{xmaxE_comparison_2comp}
\end{figure}

The maximal variations on the predicted fluxes at Earth, obtained by considering all the configurations of Table~\ref{table:allsysexp}, are shown in Fig.~\ref{xmaxE_comparison_2comp}.  The rather large uncertainty on the predicted total fluxes (brown band) is mainly due to the $\pm14\%$ shifts in the energy scale, which significantly affects only the estimated source emissivities, whereas the description of the energy spectrum and the mass composition data is very similar; on the other hand, the largest modifications of the predicted abundances at Earth are induced by the shifts in the \xmax scale, which also strongly affect the deviance value.

The main effect of the shift in the energy scale is to increase (in the case of a positive shift) or decrease (in the case of negative shift) the fraction of the heaviest masses.
This is because the observed cutoff at Earth is mainly due to the photodisintegration cutoff, which is proportional to the mass number, so a higher observed cutoff energy requires a heavier composition.  
The spectral index is also slightly changed. 
As concerns the \xmax scale uncertainty, a positive shift imposes a larger contribution of light masses, which naturally enhances the superposition of the \xmax distributions and therefore the fit requires a very negative spectral index to contrast this effect, in agreement with what predicted also in~\cite{PierreAuger:2013xim}.
    
When shifting the \xmax and energy values as in Table~\ref{table:allsysexp}, the emissivities $\mathcal{L}_0$ of the \LEb\ and \HEb\ components span the ranges~$(1.2\text{ to }2.7){\times}10^{46}$ and $(3.5\text{ to}6.5){\times}10^{44}\,\text{erg}\,\text{Mpc}^{-3}\,\text{yr}^{-1}$, respectively. The maximum decrease in $\mathcal{L}_0$ is of ${\sim}30\%$ for both components, which is given by a negative shift in both the energy scale and the \xmax scale; conversely, a positive shift in both measurements makes the $\mathcal{L}_0$ of the \LEb\ component increase by ${\sim}50\%$ and that of the \HEb\ component by ${\sim}30\%$.

\section{Distributions of sources}
\subsection{Models of local overdensity}
\label{app:overdensity}
\label{sec:overdensity}

At large distances, we assume in the benchmark model that the sources of each extragalactic population are uniformly distributed in the comoving volume.  Conversely, on small scales, since our Galaxy belongs to a group of galaxies, itself embedded in the Local Sheet~\cite{mccall2014}, and thus the density of nearby sources is greater than the average one in the Universe, we apply a correction based on the distribution of the SFR. A good approximation of the density of closer sources is important since Auger data at the highest energies are found to correlate with the flux mainly originating from nearby galaxies~\cite{PierreAuger:2018qvk,HEFD}.

To this end, we used the catalogue from Ref.~\cite{biteau}, which lists over 500\,000 galaxies from a variety of surveys (including fake galaxies in the Zone of Avoidance along the Galactic Plane where surveys are incomplete due to the Galactic foreground, obtained by cloning galaxies in zones immediately above and below it).  For each such galaxy, this flux-limited catalogue lists the luminosity distance $d$, the star formation rate SFR, the stellar mass $M_*$, and two correction factors $c$ (one for SFR and one for $M_*$) to take into account the catalogue incompleteness.

\begin{figure}
	\centering
	\includegraphics[width=0.8\linewidth]{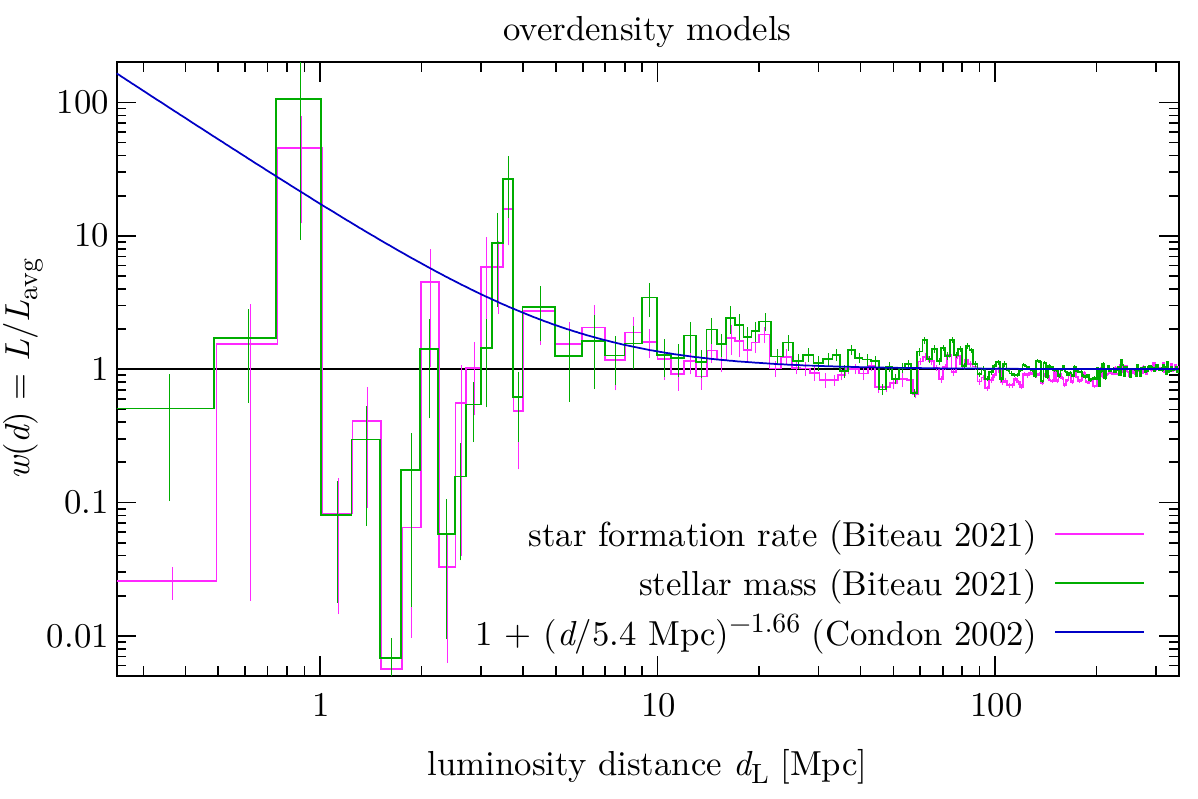}
	\caption{The three models of overdensity we tried in this work.  In the models based on Biteau (2021)~\cite{biteau}, peaks corresponding to the Andromeda Galaxy's satellite system ($d \approx 0.8$\,Mpc), the Council of Giants ($d \approx 3.6$\,Mpc) and the Virgo Cluster ($d \approx 16$\,Mpc) can be seen. At large distances, thicker bins are used in the plot than in the calculations in order to not show small-scale fluctuations in $w(d)$.}
	\label{distribution_sources_mass}
\end{figure}

We computed an overdensity correction factor $w(d)$ in each distance bin of 0.25\,Mpc thickness, proportional to the sum of SFR/$(c\,d^2)$ over galaxies in the bin, and normalised so that $w(d)$ averages to 1 between 250\,Mpc and 350\,Mpc.  
As shown in Fig.~\ref{distribution_sources_mass}, this correction produces an overdensity for distances below ${\sim}30$\,Mpc, and becomes approximately constant at larger distances.  All events in the simulations where the source of the primary particle is at a luminosity distance $d < 350$\,Mpc are then weighted by $w(d)$.  Furthermore, to avoid potential problems due to the finite statistics of simulations, compared to \CF\ we ran SimProp simulations with a thinner binning in source redshift, splitting the $[0, 0.01)$ bin into five bins $[0, 0.002), \ldots, [0.008, 0.01)$ each with as many events as previously in $[0, 0.01)$.

\begin{table}[!h]
	\small
	\renewcommand\arraystretch{1.4}
	\centering
	\begin{threeparttable}
	\begin{tabular}{|  c | c | c  c  c | c  |}
		\hline
		Overdensity model   &    &   $\mathcal{L}_0$~\tnote{*}   &   $\gamma$    &  $\log_{10}(R_\text{cut}/\text{V})$  & $D$  \\
		\hline
		\hline
		\multirow{2}{*}{No overdensity} & \LEb & $11.4$ & $\phantom{+}3.51 \pm 0.03$ & $>19.5$  & 	\multirow{2}{*}{575.1 }  \\
		                              & \HEb & $5.1$ & $-2.24 \pm 0.11$ &  $18.12 \pm 0.01$  & \\
	    \hline
		\hline
		\multirow{2}{*}{SFR~\cite{biteau}} & \LEb & $11.4$ & $\phantom{+}3.52 \pm 0.03$ &  $>19.4$ & 	\multirow{2}{*}{573.1 }  \\
		                              & \HEb & $5.1$ & $-1.99 \pm 0.11$ &  $18.15 \pm 0.01$  & \\
	    \hline
	    \hline
		\multirow{2}{*}{$M_{*}$~\cite{biteau}} & \LEb & $11.4$ & $\phantom{+}3.49 \pm 0.03$ &   $>19.4$  & 	\multirow{2}{*}{575.9 }   \\
		                              & \HEb & $4.8$ & $-2.07 \pm 0.11$ &  $18.14 \pm 0.01$ &\\
		                              \hline
		                              \hline
		\multirow{2}{*}{Infrared galaxies~\cite{condon2002, condon}} & \LEb & $11.4$ & $\phantom{+}3.49 \pm 0.03$ &  $>19.5$  & 	\multirow{2}{*}{570.8 }   \\
		                              & \HEb & $4.8$ & $-2.08 \pm 0.11$ &  $18.14 \pm 0.01$ &\\
		 \hline

	\end{tabular}
	\begin{tablenotes}
      \item[*] in units of $10^{44}\,\text{erg}\,\text{Mpc}^{-3}\,\text{yr}^{-1}$.
      \end{tablenotes}
      \end{threeparttable}
	\smallskip
    \caption{Comparison between the fit results obtained by using different models for the local overdensity correction and without applying the overdensity correction (see the text).
    }
	\label{overdensity_effect}		
\end{table}

In addition to this, we also tried using a model based on the stellar mass $M_{*}$ rather than formation rate, and a power-law approximation based on radio sources from~\cite{condon2002, condon}. These three models of overdensity are shown in Fig.~\ref{distribution_sources_mass}. The corresponding results and the ones obtained without using any correction are compared in Table~\ref{overdensity_effect} and, as expected, no significant differences are observed. 

\subsection{Minimum distance}
\begin{figure}
	\footnotesize
	\centering
	\def\w{0.49}
	\subfigure{\includegraphics[width=\w\linewidth]{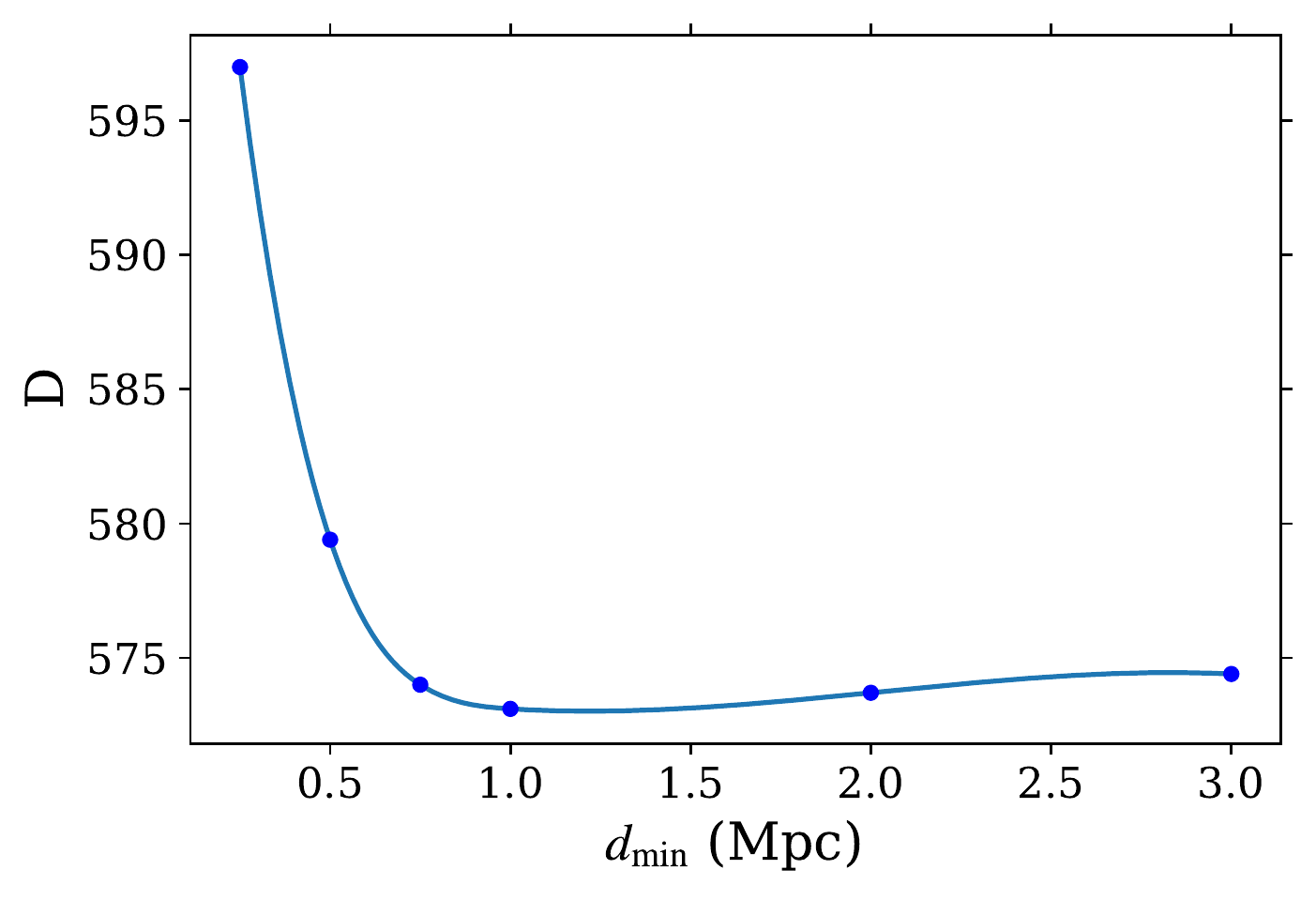}\label{dmin_effect1}}\hfill
	\subfigure{\includegraphics[width=\w\linewidth]{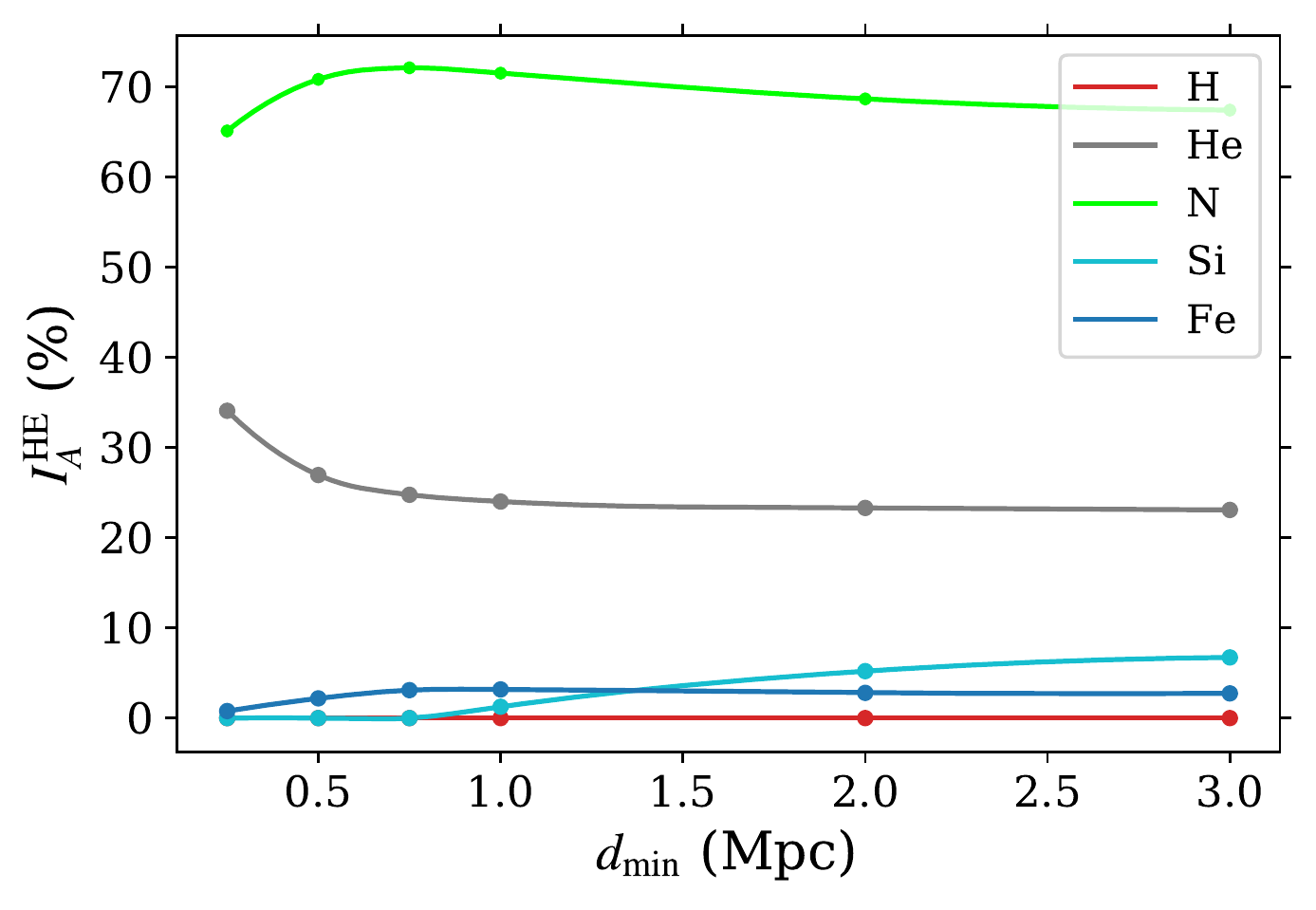}\label{dmin_effect3}}
	\\
	\subfigure{\includegraphics[width=\w\linewidth]{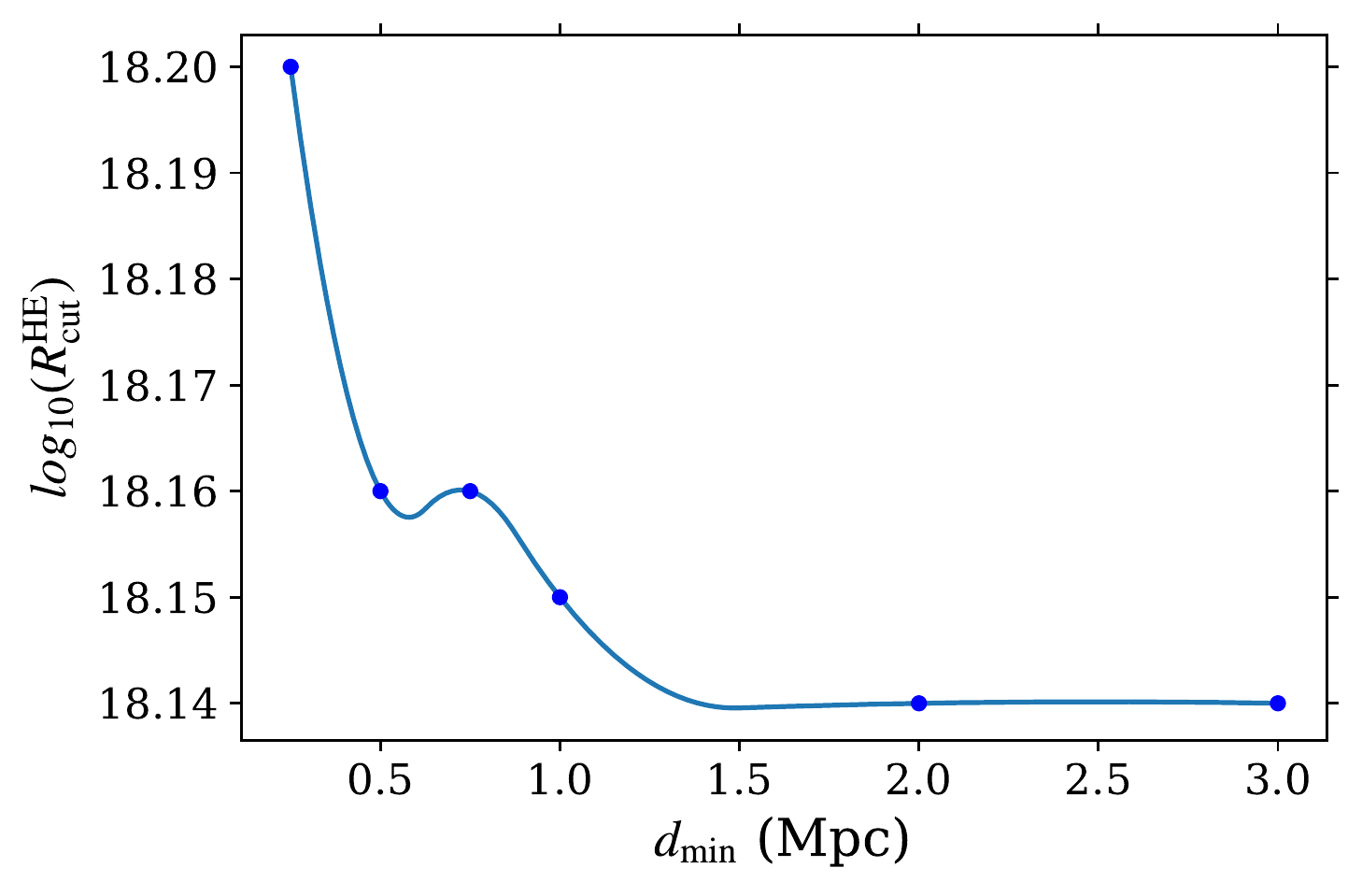}\label{dmin_effect1b}}\hfill
	\subfigure{\includegraphics[width=\w\linewidth]{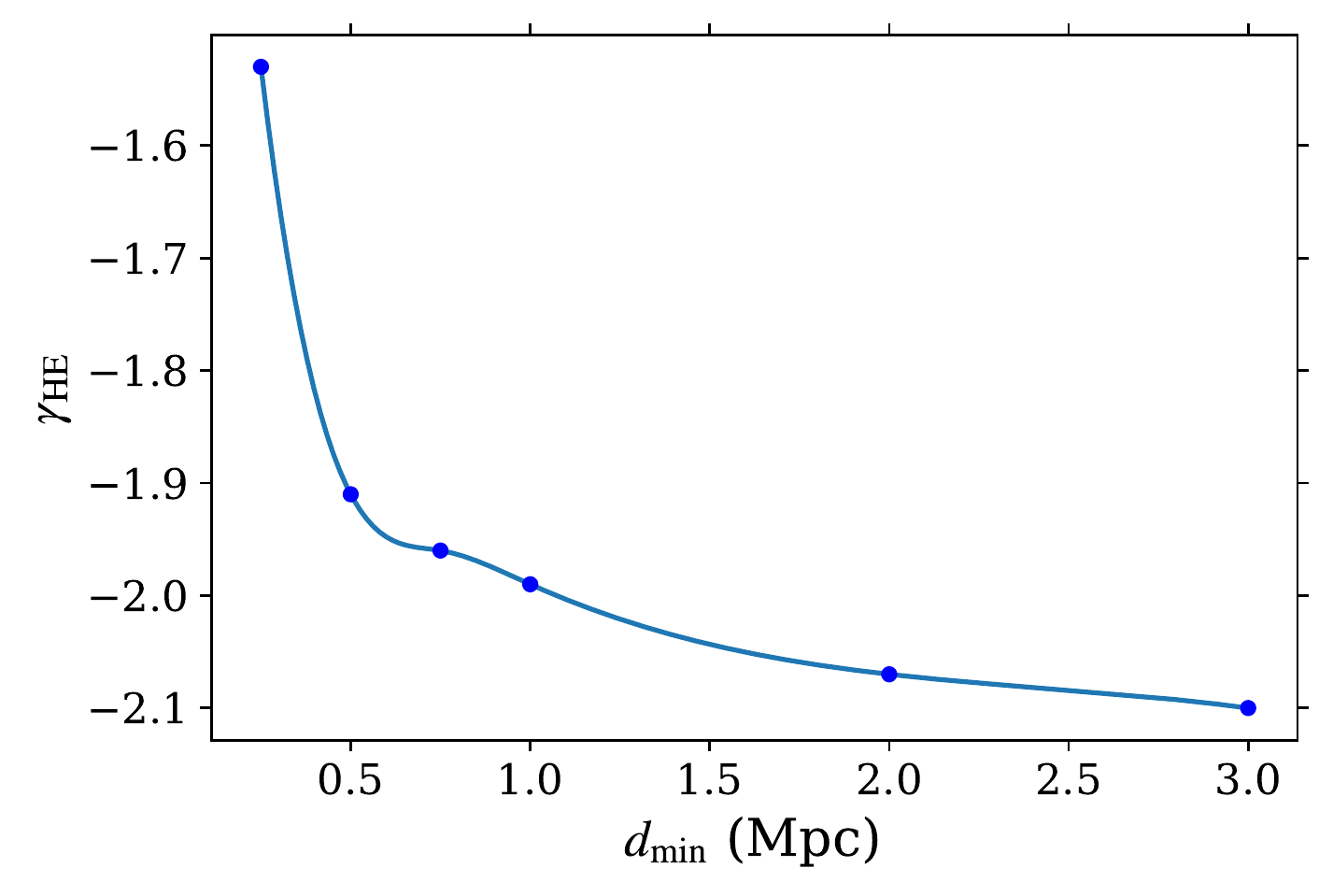}\label{dmin_effect2}}
	\caption{\emph{Top:} The effect of the minimum distance $d_\text{min}$ on the deviance values (left) and on the estimated mass fractions of the \HEb\ component (right). \emph{Bottom:} The estimated spectral parameters of the \HEb\ component as a function of the minimum distance $d_\text{min}$.}
	\label{dmin_effect}
\end{figure}

In our reference scenario the UHECR emissivity is proportional to the SFR, but if this is assumed to apply to arbitrarily small distances, the flux ($\propto \text{emissivity}/\text{distance}^2$) would be completely dominated by the Local Group (in particular the Magellanic Clouds).  To avoid this,
in all the cases illustrated in this study the minimum distance $d_\text{min}$ beyond which cosmic rays are ejected was set to 1\,Mpc.\footnote{A possible reason for the UHECR emissivity to be approximated as vanishing within 1\,Mpc and proportional to the SFR beyond would be if UHECRs are accelerated in transient events with a Poisson rate proportional to the SFR, with a proportionality constant such that most of the time there are zero events in the Local Group but several events from the Council of Giants.} However, investigating the impact of such a parameter on the deviance and the estimated fit parameters can possibly provide information about which distances are dominant for the ejection of UHECRs. We tested six different values: $d_\text{min}=\{0.25, 0.5, 0.75, 1, 2, 3\}$\,Mpc, and the corresponding results in terms of deviance are summarised in the top left plot of Fig.~\ref{dmin_effect}. 

The estimated spectral parameters and mass fractions of the \HEb\ component are also shown in the other plots of Fig.~\ref{dmin_effect}. On the other hand, the \LEb\ component parameters are almost not affected by the minimum distance, since the contribution at low energy is dominated by distant sources, thus they are not shown here.

The best fit is provided by choosing $d_\text{min}=1$\,Mpc. In general, the impact of the minimum distance on the fit results is small, with the deviance increasing for smaller $d_\text{min}$ values.

\section{Shape of the ejection cutoff function}
\label{spectrum_shape}
\label{cutoff_shape}
One of the a-priori assumptions that we made on the ejected energy spectrum is the shape of the rigidity-dependent cutoff, which is a broken exponential function in our reference fit. With such a choice we are implying that the energy spectrum is a pure power law, exponentially suppressed only at the highest energies, following the same approach introduced in \CF.

\begin{table}
	\small
	\renewcommand\arraystretch{1.4}
	\centering

	\begin{tabular}{|  c | c |  c  c | c c c  |}
		\hline
		Energy cutoff   &       &   $\gamma$    &  $\log_{10}(R_\text{cut}/\text{V})$  & $D$ &$D_J$& $D_{\xmax}$ \\
		\hline
		\hline
		\multirow{2}{*}{Broken exponential}  & \LEb &  $\phantom{+}3.52 \pm 0.03$ &  $>19.4$ & 	\multirow{2}{*}{573.1} & \multirow{2}{*}{56.6} & \multirow{2}{*}{516.5}  \\
		                              & \HEb  & $-1.99 \pm 0.11$ &  $18.15 \pm 0.01$  & & &\\
		\hline
		\hline
		\multirow{2}{*}{Exponential}  & \LEb &  $\phantom{+}3.53 \pm 0.03$ &  $>20.2$ & 	\multirow{2}{*}{575.2} & \multirow{2}{*}{58.9} & \multirow{2}{*}{516.2}  \\
		                              & \HEb  & $-2.06 \pm 0.10$ &  $18.15 \pm 0.01$  & & &\\
		
		\hline
		\hline
		Hyperbolic secant  & \LEb  & $\phantom{+}3.41 \pm 0.07$ &  $18.29 \pm 0.25$ & 	\multirow{2}{*}{595.9} & \multirow{2}{*}{70.0} & \multirow{2}{*}{525.9}  \\
		                           $\Delta=0.5$   & \HEb  & $-6.23 \pm 0.18$ &  $16.33 \pm 0.02$  & & &\\
	    \hline
	    \hline
		 Hyperbolic secant & \LEb  & $\phantom{+}3.53 \pm 0.03$ &   $>19.7$  & 	\multirow{2}{*}{575.3} & 	\multirow{2}{*}{59.3} & 	\multirow{2}{*}{516.0}   \\
		                           $\Delta=1.0$   & \HEb  & $-2.02 \pm 0.10$ &  $18.15 \pm 0.01$ & & &\\
		                              \hline
		                              \hline
		Hyperbolic secant & \LEb  & $\phantom{+}3.65 \pm 0.03$ &  $>19.7$  & 	\multirow{2}{*}{618.6} & 	\multirow{2}{*}{83.3} & 	\multirow{2}{*}{535.3}   \\
		                           $\Delta=2.0$    & \HEb  & $\phantom{+}0.32 \pm 0.05$ &  $18.78 \pm 0.01$ & & &\\
		 \hline
	\end{tabular}
	\bigskip
    \caption{Best fit results obtained by assuming that the energy cutoff is shaped as a broken exponential function (reference case), a simple exponential function and  a hyperbolic secant function with $\Delta=0.5$, $1.0$, $2.0$. 
  }
	\label{cutoff_effect}		
\end{table}

In this appendix we aim at testing the impact on the fit results of choosing some alternative energy spectrum, with a simple exponential and a hyperbolic secant as the cutoff function,
\begin{align}
                J(E) &= \sum_A J_{0A} \, \left( \frac{E}{E_0} \right)^{-\gamma}
                \,
                \exp\left(-\frac{E}{Z_A \, R_\text{cut}}\right)
\label{simpleexpo}\\
                J(E) &= \sum_A J_{0A} \, \left( \frac{E}{E_0} \right)^{-\gamma}
                \,
                \sech\left[\left(\frac{E}{Z_A \, R_\text{cut}}\right)^\Delta\right]
\label{hyperbolic}
\end{align}
Both these functions have a smooth shape and a continuous derivative. The parameter $\Delta$ in Eq.~\ref{hyperbolic} is related to the steepness and the width of the energy cutoff, and we tested three different values $\Delta=0.5$, $1.0$, $2.0$. The effect on the fit results of the cutoff function choice and of the value of $\Delta$, i.e.\ the cutoff steepness of the hyperbolic secant function, is shown in Table~\ref{cutoff_effect}. 

Both the hyperbolic secant with $\Delta=1$ and the simple exponential cutoff have very similar shape with respect to the broken exponential function and, as expected, they provide compatible fit results in terms of estimated parameters and deviance value. On the other hand, the hyperbolic secant with $\Delta=2$ produces a steeper cutoff and the one with $\Delta=0.5$ a more gradual cutoff, with a significant impact on the fit results.

First of all, the impact on the low-energy component is generally much smaller than on the high-energy component, since the cutoff at the sources of the former plays a minor role in shaping the observed energy spectrum.
As concerns the effect on the high-energy component, a steeper cutoff ($\Delta=2$) requires a softer energy spectrum with a positive spectral index and a slightly larger cutoff rigidity, and a more gradual cutoff ($\Delta=0.5$) needs to be compensated by a very low cutoff rigidity and spectral index; however, in these two cases the fit of both the energy spectrum and the \xmax distributions appears worsened.

\begin{figure}
	\centering
	\def\w{0.49}
	\subfigure{\includegraphics[width=\w\linewidth]{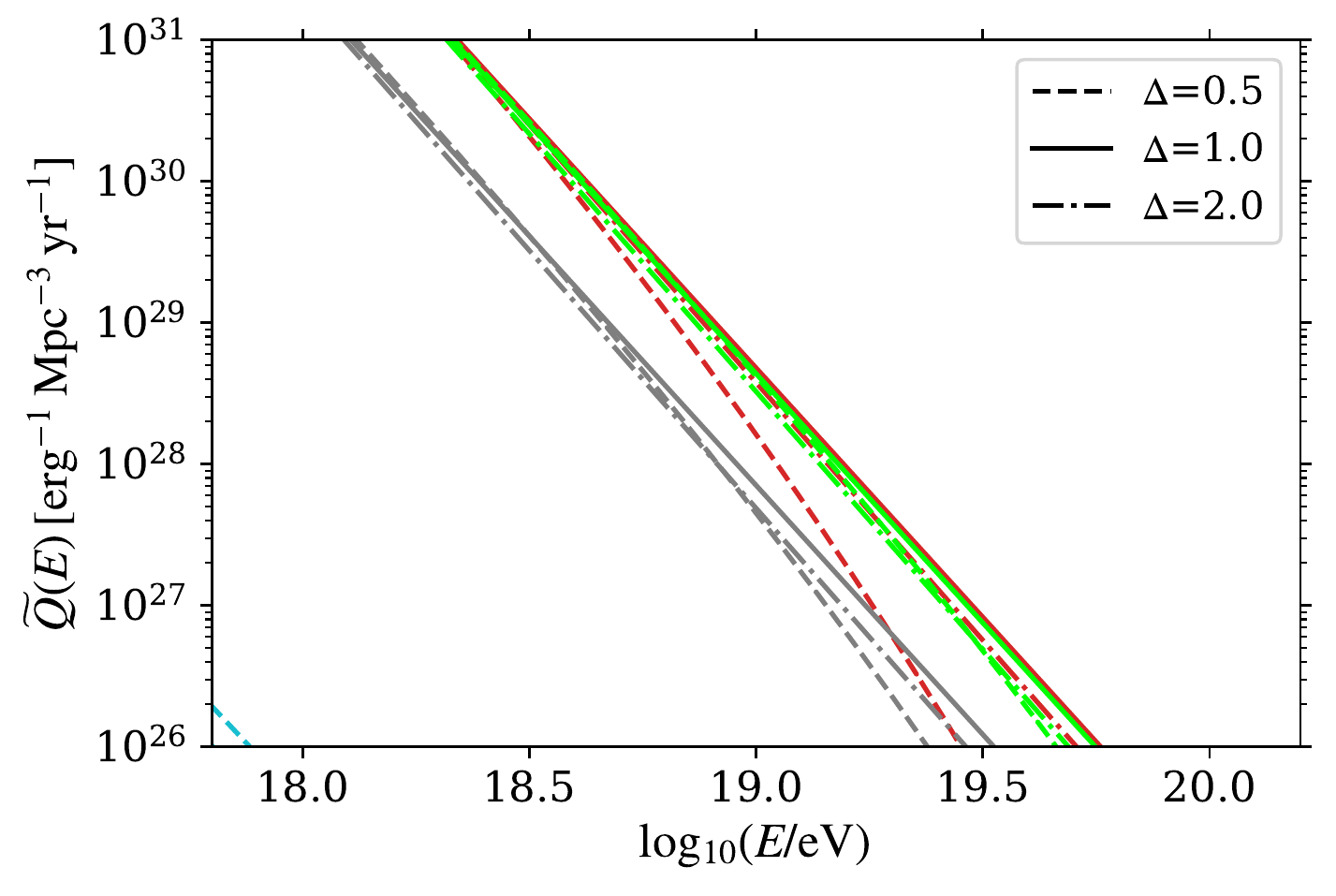}}
	\subfigure{\includegraphics[width=\w\linewidth]{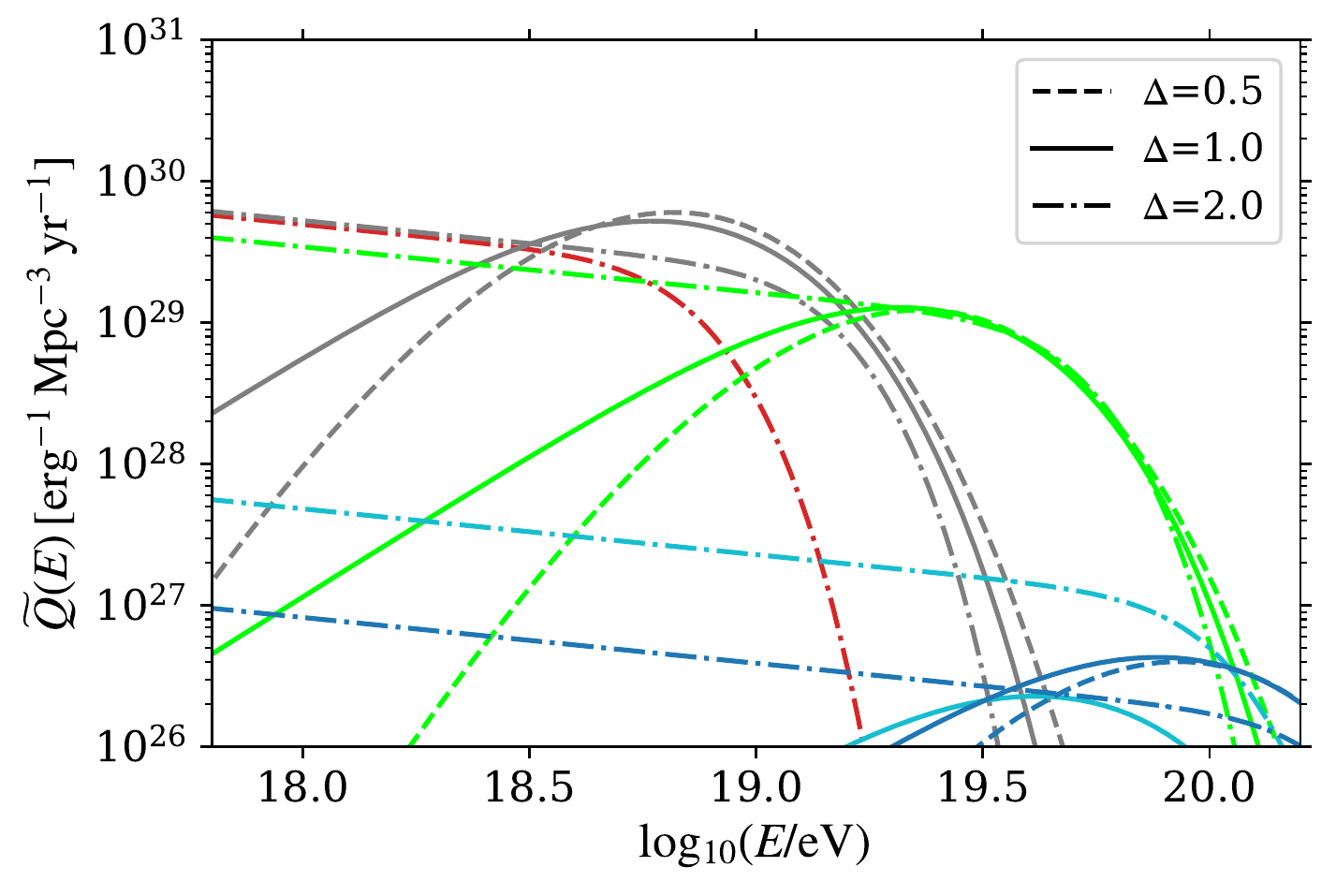}}
	\caption{Comparison between the generation rate of the \LEb\ component (left) and the \HEb\ one (right) given by the best fit parameters obtained by assuming a hyperbolic secant cutoff with different $\Delta$ values. Note that the fit results for $\Delta=1.0$ are very similar to the ones obtained with a broken exponential cutoff and a simple exponential cutoff, which are thus not shown. 
	}
	\label{fig:cutoff}
\end{figure} 

It is important to stress that in presence of a hyperbolic secant or a simple exponential cutoff the power law function may be significantly modified even at energies much lower than the estimated energy cutoff, so that the actual slope of the ejected energy spectrum is not simply the one given by the nominal value of the spectral index reported in the table. For example, in the case of a hyperbolic secant function with $\Delta=0.5$, the extremely low estimated $\gamma$ value translates into a much softer energy spectrum because of the more gradual cutoff shape.

As a consequence, note that the best-fit spectral parameters adjust to compensate the effect of the cutoff shape, so that the ejection spectra in the various cutoff shape hypotheses are actually much more similar to each other than a naive comparison of the parameter values might suggest, as shown in Fig.~\ref{fig:cutoff}.  Still, the cases $\Delta = 0.5$ and $\Delta = 2.0$ result in a noticeably larger total deviance, showing that the fit moderately disfavours excessively gradual or excessively sudden cutoffs, though the size of the deviance increases are comparable to those seen in the studies of effects of systematic uncertainties.

\section*{Acknowledgments}

\begin{sloppypar}
The successful installation, commissioning, and operation of the Pierre
Auger Observatory would not have been possible without the strong
commitment and effort from the technical and administrative staff in
Malarg\"ue. We are very grateful to the following agencies and
organizations for financial support:
\end{sloppypar}

\begin{sloppypar}
Argentina -- Comisi\'on Nacional de Energ\'\i{}a At\'omica; Agencia Nacional de
Promoci\'on Cient\'\i{}fica y Tecnol\'ogica (ANPCyT); Consejo Nacional de
Investigaciones Cient\'\i{}ficas y T\'ecnicas (CONICET); Gobierno de la
Provincia de Mendoza; Municipalidad de Malarg\"ue; NDM Holdings and Valle
Las Le\~nas; in gratitude for their continuing cooperation over land
access; Australia -- the Australian Research Council; Belgium -- Fonds
de la Recherche Scientifique (FNRS); Research Foundation Flanders (FWO);
Brazil -- Conselho Nacional de Desenvolvimento Cient\'\i{}fico e Tecnol\'ogico
(CNPq); Financiadora de Estudos e Projetos (FINEP); Funda\c{c}\~ao de Amparo \`a
Pesquisa do Estado de Rio de Janeiro (FAPERJ); S\~ao Paulo Research
Foundation (FAPESP) Grants No.~2019/10151-2, No.~2010/07359-6 and
No.~1999/05404-3; Minist\'erio da Ci\^encia, Tecnologia, Inova\c{c}\~oes e
Comunica\c{c}\~oes (MCTIC); Czech Republic -- Grant No.~MSMT CR LTT18004,
LM2015038, LM2018102, CZ.02.1.01/0.0/0.0/16{\textunderscore}013/0001402,
CZ.02.1.01/0.0/0.0/18{\textunderscore}046/0016010 and
CZ.02.1.01/0.0/0.0/17{\textunderscore}049/0008422; France -- Centre de Calcul
IN2P3/CNRS; Centre National de la Recherche Scientifique (CNRS); Conseil
R\'egional Ile-de-France; D\'epartement Physique Nucl\'eaire et Corpusculaire
(PNC-IN2P3/CNRS); D\'epartement Sciences de l'Univers (SDU-INSU/CNRS);
Institut Lagrange de Paris (ILP) Grant No.~LABEX ANR-10-LABX-63 within
the Investissements d'Avenir Programme Grant No.~ANR-11-IDEX-0004-02;
Germany -- Bundesministerium f\"ur Bildung und Forschung (BMBF); Deutsche
Forschungsgemeinschaft (DFG); Finanzministerium Baden-W\"urttemberg;
Helmholtz Alliance for Astroparticle Physics (HAP);
Helmholtz-Gemeinschaft Deutscher Forschungszentren (HGF); Ministerium
f\"ur Kultur und Wissenschaft des Landes Nordrhein-Westfalen; Ministerium
f\"ur Wissenschaft, Forschung und Kunst des Landes Baden-W\"urttemberg;
Italy -- Istituto Nazionale di Fisica Nucleare (INFN); Istituto
Nazionale di Astrofisica (INAF); Ministero dell'Istruzione,
dell'Universit\'a e della Ricerca (MIUR); CETEMPS Center of Excellence;
Ministero degli Affari Esteri (MAE); M\'exico -- Consejo Nacional de
Ciencia y Tecnolog\'\i{}a (CONACYT) No.~167733; Universidad Nacional Aut\'onoma
de M\'exico (UNAM); PAPIIT DGAPA-UNAM; The Netherlands -- Ministry of
Education, Culture and Science; Netherlands Organisation for Scientific
Research (NWO); Dutch national e-infrastructure with the support of SURF
Cooperative; Poland -- Ministry of Education and Science, grant
No.~DIR/WK/2018/11; National Science Centre, Grants
No.~2016/22/M/ST9/00198, 2016/23/B/ST9/01635, and 2020/39/B/ST9/01398;
Portugal -- Portuguese national funds and FEDER funds within Programa
Operacional Factores de Competitividade through Funda\c{c}\~ao para a Ci\^encia
e a Tecnologia (COMPETE); Romania -- Ministry of Research, Innovation
and Digitization, CNCS/CCCDI UEFISCDI, grant no. PN19150201/16N/2019 and
PN1906010 within the National Nucleus Program, and projects number
TE128, PN-III-P1-1.1-TE-2021-0924/TE57/2022 and PED289, within PNCDI
III; Slovenia -- Slovenian Research Agency, grants P1-0031, P1-0385,
I0-0033, N1-0111; Spain -- Ministerio de Econom\'\i{}a, Industria y
Competitividad (FPA2017-85114-P and PID2019-104676GB-C32), Xunta de
Galicia (ED431C 2017/07), Junta de Andaluc\'\i{}a (SOMM17/6104/UGR,
P18-FR-4314) Feder Funds, RENATA Red Nacional Tem\'atica de
Astropart\'\i{}culas (FPA2015-68783-REDT) and Mar\'\i{}a de Maeztu Unit of
Excellence (MDM-2016-0692); USA -- Department of Energy, Contracts
No.~DE-AC02-07CH11359, No.~DE-FR02-04ER41300, No.~DE-FG02-99ER41107 and
No.~DE-SC0011689; National Science Foundation, Grant No.~0450696; The
Grainger Foundation; Marie Curie-IRSES/EPLANET; European Particle
Physics Latin American Network; and UNESCO.
\end{sloppypar}

\newcommand{\vyp}[3]{\textbf{#1}\ (#2)\ #3}
\newcommand{\etal}{et~al.}

\end{document}